\documentclass{aa}

\usepackage{xcolor}
\definecolor{darkblue}{RGB}{12,94,176}
\usepackage{hyperref}
\hypersetup{
    pdfnewwindow=true,     % links in new window
    colorlinks=true,       % false: boxed links; true: colored links
    linkcolor=darkblue,    % color of internal links
    citecolor=darkblue,    % color of links to bibliography
    filecolor=darkblue,    % color of file links
    urlcolor=blue          % color of external links
}
\newcommand{\orcid}[1]{\unskip\protect\href{https://orcid.org/#1}{\protect\includegraphics[width=8pt,clip]{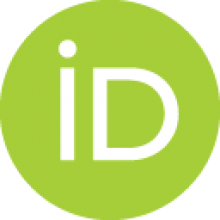}}}

\defcitealias{jiang2021ApJS}{J21}

\usepackage{CJKutf8}

\usepackage{enumitem}
\setlist{nolistsep}
\setlist[itemize]{align=parleft,left=0pt..1.5em}
\setlist[enumerate]{align=parleft,left=0pt..1.5em}
\usepackage{multirow}
\usepackage{amsmath}
\usepackage{cuted}
\usepackage{subcaption}
\usepackage{listings}

\newcommand\tensoraas[1]{\mathbf{#1}}
\newcommand\vectoraas[1]{{\boldsymbol{#1}}}

\usepackage{graphicx}
\usepackage{txfonts}
\usepackage{lipsum}
\usepackage{subcaption}         % necessary for continued figures, example in section 3
\usepackage{lscape}             % to rotate a single page table, example in appendix.
\usepackage{placeins}           % useful with \FloatBarrier, to keep 
\begin{document}
% \begin{CJK*}{UTF8}{gbsn}

   \title{AREPO-IDORT: Implicit Discrete Ordinates Radiation Transport}%\\ for Radiation Magnetohydrodynamics on an Unstructured Moving Mesh}

    \subtitle{for Radiation Magnetohydrodynamics on an Unstructured Moving Mesh}

   \titlerunning{AREPO-IDORT: Implicit Discrete Ordinates Radiation Transport on a Moving Mesh}
%\authorrunning{Ma et al.}

   \author{
   Jing-Ze Ma \begin{CJK*}{UTF8}{gbsn}(马竟泽)\end{CJK*}\thanks{jingze@mpa-garching.mpg.de} \fnmsep  \orcid{0000-0002-9911-8767}
    \and 
    Rüdiger Pakmor  \orcid{0000-0003-3308-2420}
    \and 
    Stephen Justham  \orcid{0000-0001-7969-1569}   
    \and 
    Selma E. de Mink  \orcid{0000-0001-9336-2825}
        }

   \institute{Max Planck Institute for Astrophysics, Karl-Schwarzschild-Str. 1, 85748 Garching, Germany}

   % \date{Received September 30, 20XX}

% \abstract{}{}{}{}{}
% 5 {} token are mandatory
 
  \abstract
  {Radiation is crucial not only for observing astrophysical objects, but also for transporting energy and momentum.
However, accurate on-the-fly radiation transport in astrophysical simulations is challenging and computationally expensive.
Here we introduce \texttt{AREPO-IDORT} (Implicit Discrete Ordinates Radiation Transport), a scheme coupled to the explicit magnetohydrodynamic (MHD) solver in the 3D moving-mesh code \texttt{AREPO}.
The discrete ordinates scheme means we directly solve for the specific intensities along discrete directions.
We solve the time-dependent relativistic radiation transport equation via an implicit Jacobi-like iterative finite-volume solver, which overcomes the small radiation time-steps needed by explicit methods.
Compared to commonly-used moment-based methods, e.g.\ flux-limited diffusion or M1 closure, this scheme has the advantage of correctly capturing the direction of radiation in both optically-thick and thin regions.
It is based on the scheme by Jiang 2021 for the adaptive mesh refinement code \texttt{ATHENA++}, but we generalize the scheme to support (1) an unstructured moving-mesh, (2) local time-stepping, and (3) general equations of state.
% To demonstrate those new features, 
We show various test problems that commonly-used moment-based methods fail to reproduce accurately.
To apply the scheme to a real astrophysics problem, we show the first global 3D radiation hydrodynamic simulation of the entire convective envelope of a red supergiant star. 
We even marginally resolve the photosphere, which is a known challenge for global 3D simulations of stars.
For this problem, the radiation module only takes less than half of the total computational cost.
Our current scheme assumes grey radiation, is first-order accurate in both time and space, and is memory intensive especially for large cosmological simulations, but we discuss potential avenues for future improvements.
We expect our scheme will enable more accurate multi-scale radiation MHD simulations involving supersonic bulk motions, ranging from planet formation in protoplanetary disks, stars and associated transients, to accretion flows near black holes.}
%   % context heading (optional)
%   % {} leave it empty if necessary  
%    {Optional, leave empty if necessary.  The heading “Context” is used when needed to
% give background information on the research conducted in the paper}
%   % aims heading (mandatory)
%    {Mandatory. The objectives of the paper are defined here.} 
%   % methods heading (mandatory)
%    {Mandatory. The methods of the investigation are outlined here}
%   % results heading (mandatory)
%    {Mandatory. The results are summarized here.}
%   % conclusions heading (optional), leave it empty if necessary
%    {Optional, leave empty if necessary.  “Conclusions” can be used to
% explicit the general conclusions that can be drawn from the paper.}


   \maketitle

\begin{nolinenumbers}

% \begin{widetext}

% \onecolumn

\tableofcontents

\section{Introduction}
\label{sec:intro}

Radiation plays an essential role in most astrophysical problems.
Even in this era of multi-messenger astronomy, electromagnetic radiation is still the dominant way for us to observe astrophysical objects via photometry, spectroscopy, polarimetry, or interferometry.
More importantly, radiation interferes with the dynamics of matter through emission, absorption, and scattering.
Photons exchange energy with matter in the form of heating, cooling, photoionization and photodissociation, and exchange momentum with matter via radiation pressure or radiation force \citep{mihalas1984}.
For example, radiative heating and cooling determines the thermodynamics and thickness of most astrophysical disks and streams, and significantly alters, e.g. substructures in protoplanetary disks \citep{zhang2024ApJ}, accretion onto protostars or mass-transferring binary stars \citep{geroux2016A&A}, circularization of the stellar debris in tidal disruption events \citep{hayasaki2016MNRAS}, and circumbinary-disk accretion onto supermassive black hole binaries \citep{tiwari2025arXive-prints}.
Ionizing radiation regulates star formation and thermodynamics of the intergalactic medium, thereby controlling galaxy formation, especially at high redshifts \citep{borrow2023MNRAS}.
Under certain conditions, radiation forces dominate the dynamics, driving outflows in cool \citep{habing1996A&AR} and hot stars \citep{castor1975ApJ}, active galactic nuclei \citep[AGN;][]{murray1995ApJ}, and super-Eddington accretion flows around compact objects \citep{jiang2024arXive-prints}.
For many complex astrophysical scenarios, it is therefore crucial to perform multi-dimensional numerical simulations with (magneto)hydrodynamics (MHD) fully coupled with accurate radiation transport to understand the physical processes and provide accurate predictions for the observables.

\subsection{Overview: Challenges and Available Methods for Radiation Transport}

\renewcommand{\arraystretch}{1.3} % Adjust row spacing (1.5x default)
\begin{table*}[htb!] 
\footnotesize 
\begin{center}
 \caption{Summary and comparison of radiation transport methods. 
 See the main text for details.
 For a summary of codes using different methods, see e.g. \citet{wunsch2024Front.Astron.SpaceSci.}.
}
 \label{tab:rt}
 \begin{tabular}{ccccc}
\hline
\hline
\multirow{2}{*}{Category} & Particle & Ray & Fluid & Ray-fluid hybrid \\
& \begin{minipage}{0.1\textwidth}
      \includegraphics[width=\linewidth]{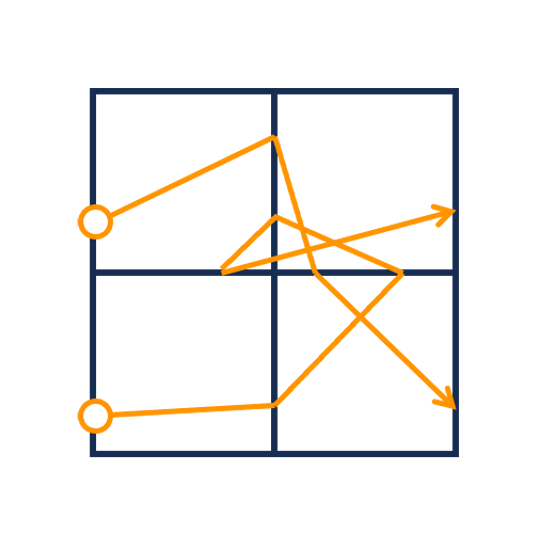}
    \end{minipage} 
& \begin{minipage}{0.1\textwidth}
      \includegraphics[width=\linewidth]{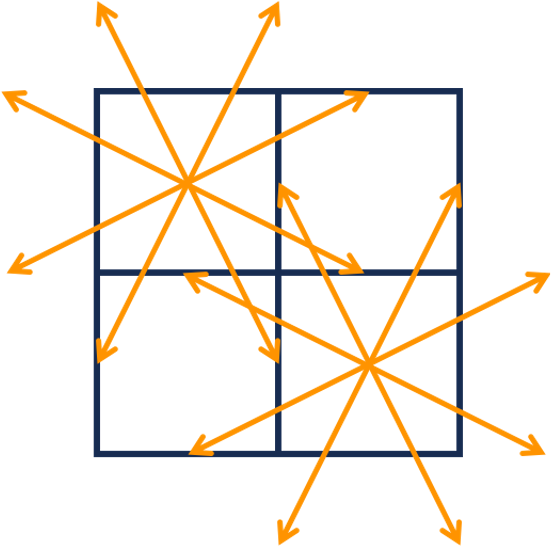}
    \end{minipage} 
& \begin{minipage}{0.1\textwidth}
      \includegraphics[width=\linewidth]{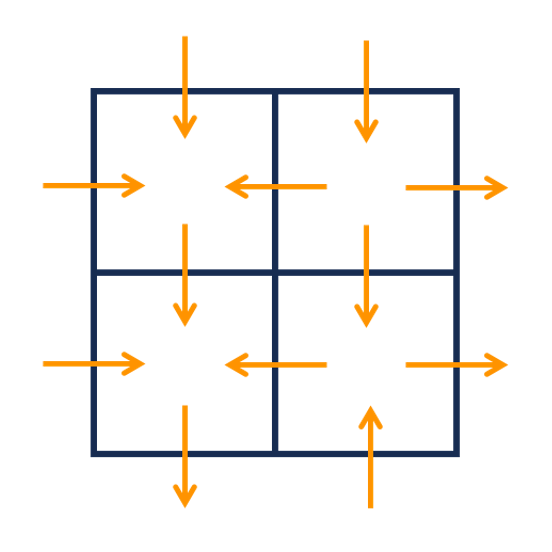}
    \end{minipage}
& \begin{minipage}{0.1\textwidth}
      \includegraphics[width=\linewidth]{ray.png}
    \end{minipage} 
    $\Longrightarrow$
    \begin{minipage}{0.1\textwidth}
      \includegraphics[width=\linewidth]{fluid.png}
    \end{minipage}
    \\
\hline
\multirow{2}{*}{Method name} & \multirow{2}{*}{Monte Carlo} & Ray-tracing & Flux-limited diffusion & Variable Eddington tensor \\
& & Characteristics & M1 closure & Discrete ordinates \\
\hline
\multicolumn{5}{l}{\textit{Physics}}\\
\hline
Accurate direction & $\checkmark$ & $\checkmark$ & $\times$ & $\checkmark$ \\
of radiation? & & & ($\tau \lesssim 1$) & \\
\hline
Accurate at different & $\times$ & $\checkmark$ & $\times$ & $\checkmark$ \\
 optical depths $\tau$? & ($\tau \gg 1$) & & ($\tau \lesssim 1$) & \\
\hline
Time-and- & $\checkmark$ & $\times$ & $\checkmark$ & $\checkmark$ \\
velocity-dependent? & & & & \\
\hline
Detailed microphysics & $\checkmark$ & $\times$ & $\times$ & $\times$ \\
(e.g. anisotropic scattering)? & & & & \\
\hline
\multirow{3}{*}{Known difficulties} & Poisson noise; &  & Filling shadows; &  \\
& Slow transport & Ray-effect\tablefootmark{*} & Merging crossing beams; & Ray-effect\tablefootmark{*} \\
 & ($\tau \gg 1$) & ($\tau \ll 1$) & Colliding radiation fronts & ($\tau \ll 1$)\\
\hline
\multicolumn{5}{l}{\textit{Computational cost for a sophisticated grey 3D simulation ($\textit{1=}$ comparable to hydro, order-of-magnitude estimate)}}\\
\hline
Relative cost & $\times 100$ & $\times 10$ & $\times 1$ & $\times 10$ \\
\hline
\multicolumn{5}{l}{\textit{Implementation in \texttt{AREPO}}} \\
\hline
\texttt{AREPO} module name & \texttt{AREPO-MCRT}\tablefootmark{a} & \texttt{SPRAI}\tablefootmark{b} & \texttt{AREPO-RT}\tablefootmark{c} & \texttt{AREPO-IDORT} (This work)\tablefootmark{d} \\
\hline
 \end{tabular}
  \end{center}
  \tablefoot{
\tablefoottext{*}{Artificial radiation peaks along discrete directions if the number of discretized angles is insufficient, see e.g. Figure~\ref{fig:test:rsg:ray}.}
\tablefoottext{a} {\citet{smith2020ApJ}.}
\tablefoottext{b}{\citet{jaura2018MNRAS, jaura2020MNRAS}; see also \citet{greif2014MNRAS}.}
\tablefoottext{c}{\citet{kannan2019MNRAS, zier2024MNRAS}; see also \citet{bauer2015MNRAS}.}
\tablefoottext{d}{see also \citet{petkova2011MNRAS} for a semi-implicit source-dependent variance of the discrete ordinates method.}
}
\end{table*}

Radiation transport in radiation MHD (RMHD) simulations is difficult for three reasons caused by the vast range of scales involved: (1) dimension, (2) length-scale, and (3) time-scale.
% These are general issues for performing any multi-dimensional simulation on modern computers, but they become exceedingly challenging, sometimes by orders of magnitude more, when radiation comes into play.
(1) Dimension: The radiation field is not only a function of space and time, but also direction and frequency \citep{mihalas1984}.
This means that if we aim to extend any existing MHD simulation to full RMHD simulation, the number of numerical quantities will expand by one to two orders of magnitude.
This poses a major challenge not only to calculate those quantities, but also save them with limited computational power and memory storage \citep{jiang2014ApJS}.
(2) Length-scale: The mean free path of photon generally differs from the length-scale of MHD processes by many orders of magnitude.
When the photon mean free path is much larger than the length-scale of gas, matter is transparent to light (`optically thin'), and photons can travel freely at the speed of light.
At the other extreme, when the photon mean free path is much smaller than the length-scale of gas, matter is opaque to light (`optically thick'), and photons are frequently absorbed and re-emitted or scattered, resulting in a diffusive random-walk of photons.
% Inside our Sun, the dense ionized plasma is highly optically-thick with an optical depth $\tau\sim 10^{11}$ \citep{krumholz2007ApJa}.
An even more challenging regime is the transition zone from optically-thick to optically-thin, from which the observed light escapes.
An accurate radiation transport scheme is expected to capture the physics correctly in all those three regimes, despite the significant differences in physics and mathematical properties of the equation \citep{stone1992ApJS, teyssier2015ARA&A, jiang2021ApJS}.
(3) Time-scale: 
In the optically-thin regime, radiation travels at the speed of light, which is orders of magnitude larger than the gas speed for non-relativistic simulations.
% This reduces the allowed explicit timestep by orders of magnitude, which renders a full explicit method infeasible for RMHD simulations.
Furthermore, typically in the optically thick regime, the heating and cooling can change the energy content of the plasma at a timescale faster than the normal MHD processes, which poses a numerical challenge in coupling the radiation and gas \citep{teyssier2015ARA&A}.

Given the three challenges, one has to make a tradeoff between speed and accuracy for radiation transport methods in RMHD simulations.
Motivated by the wave-particle duality of photons and geometrical optics, we broadly separate those methods into three categories \citep[see also][]{wunsch2024Front.Astron.SpaceSci.}, namely to treat radiation as (1) particles, (2) rays, or (3) fluids, as summarized in Table~\ref{tab:rt}.

The particle description mainly stands for the Monte Carlo radiation transport \citep[see e.g.][for a review]{noebauer2019LRCA}, which generates randomly-sampled photon packets and follows them as they propagate through matter.
It is often used for post-processing simulations to make predictions for observables, where detailed microphysics such as scattering and line transport are important \citep{noebauer2015MNRAS}.
But it has not been the main stream for multi-dimensional RMHD simulations due to Poisson noise and slow propagation of photon packets in optically-thick regions \citep{nayakshin2009MNRAS, camps2018ApJ}.

The ray description integrates intensities along straight lines, analogous to the rays of light.
This is done by interpolating between cell centers and their projections on the rays \citep{stone1992ApJS, freytag2012JCoPh}.
In typical forward ray-tracing methods, the rays originate from each radiation source and end at the simulation boundaries or at the points where the radiation is completely absorbed \citep{abel2002MNRAS, wise2011MNRAS}.
For a small number of radiation sources (e.g., some cases in star formation), forward ray-tracing is efficient to calculate heating, cooling, ionization, etc. \citep{abel2002MNRAS, wise2011MNRAS}.
However, for a large number of radiation sources, or when every cell is effectively a radiation source, other methods are more efficient.
In contrast to forward ray-tracing, short characteristics creates rays from active cell points and traces them only to nearby active mesh points \citep{stone1992ApJS, mellema2006NewAstron., davis2012ApJS, freytag2012JCoPh}.
This method is more diffusive than others and trades accuracy for stability \citep{stone1992ApJS, davis2012ApJS, freytag2012JCoPh}, because multiple interpolations are needed between cell centers and their projections on the rays, but the fact that the calculations are completely local makes the scheme faster than others \citep{stone1992ApJS, mellema2006NewAstron., davis2012ApJS}.
These ray-based methods perform reasonably well in both optically-thick and thin media, but the huge number of radiation quantities that need to be calculated is still an issue \citep{abel2002MNRAS}.

The fluid description of radiation transport, also known as the moment-based methods, does not calculate the intensities but the radiation energy density $E_\mathrm{r}$, radiation flux $\vectoraas{F}_\mathrm{r}$, and radiation pressure tensor $\tensoraas{P}_\mathrm{r}$.
This drastically reduces the number of radiation quantities to be calculated and saved, albeit at the price of making assumptions of the relations between $E_\mathrm{r}, \vectoraas{F}_\mathrm{r}, \tensoraas{P}_\mathrm{r}$.
The simplest method is the flux-limited diffusion \citep[hereafter FLD, e.g.][]{levermore1981ApJ}, which treats the radiation as a diffusion problem, where the radiation flux is always directed along the gradient of radiation energy density.
A more sophisticated method is the M1 closure \citep{levermore1984JQSRT}, which solves for $E_\mathrm{r}$ and $\vectoraas{F}_\mathrm{r}$ simultaneously, but still uses analytical approximations for the Eddington tensor $\tensoraas{f}_\mathrm{Edd} \equiv \tensoraas{P}_\mathrm{r}/E_\mathrm{r}$ to reproduce the asymptotic limits in both optically-thick and thin regimes.
% The most sophisticated moment-based method is the Variable Eddington Tensor scheme \citep[hereafter VET, e.g.][]{stone1992ApJS, jiang2012ApJS, menon2022MNRAS}, which solves for both $E_\mathrm{r}$ and $\vectoraas{F}_\mathrm{r}$ with $\tensoraas{f}_\mathrm{Edd}$ provided by another radiation transport method (short characteristics in practice).
% Generally, the more complex the method is, the more accurate and also more computationally expensive it becomes.
However, due to the diffusion description, FLD cannot cast shadows since the radiation field will diffuse into the shadow \citep{menon2022MNRAS}.
The approximated Eddington tensor of M1 causes the light to behave like a fluid, resulting in nonphysical phenomena such as merging two crossing radiation beams into one because the radiation is forced to only have one distinguished direction at one location \citep{sadowski2013MNRAS}, creating artificial shocks when two radiation fronts meet \citep{thomas2022MNRAS}, or generating unsteady vortices resembling the Karman vortex street in the shadows \citep{menon2022MNRAS}.
It is also not clear whether FLD and M1 can correctly capture the radiation field in the transition zone between optically-thick and thin regimes \citep{mihalas1984}.
In particular for M1, it has been shown that the Eddington tensor near the photosphere is always overestimated in the radiating homogeneous sphere test \citep{smit1997A&A, oconnor2015ApJS, murchikova2017MNRAS, anninos2020ApJ}.
% Direct comparisons of different methods in test problems can be found in e.g. Figure 7 in \citet{davis2012ApJS}, Figure 15, 16 in \citet{smith2020ApJ}, Figure 13 in \citet{menon2022MNRAS} and in more complex simulations \citep[e.g.][]{asahina2022ApJ}.

A way to combine the advantages of different methods is to solve the radiation transport problem twice using two methods.
This is demonstrated using the Variable Eddington Tensor scheme \citep[hereafter VET, e.g.][]{stone1992ApJS, jiang2012ApJS, menon2022MNRAS}, which solves for $\tensoraas{f}_\mathrm{Edd}$ using another radiation transport method (ray-based short characteristics in practice), and then solves for both $E_\mathrm{r}$ and $\vectoraas{F}_\mathrm{r}$ as in the moment-based methods.

Along this line, in recent years, a novel discrete ordinates radiation transport method has emerged in computational astrophysics, credited to the continuous efforts by \citet{jiang2014ApJS} and \citet{jiang2021ApJS, jiang2022ApJS}.
This method resembles the ray-based short characteristics: It discretizes the radiation field in angular space and solves for intensities along each direction.
The calculations are locally confined within neighbouring cells during each iteration.
It also resembles the moment-based methods, as it solves the time-dependent radiation transport equation taking into account the Lorentz transformation and isotropic scattering.
The intensities can be calculated using the same mesh as MHD in a finite-volume manner as one solves for energy in the mesh-based MHD simulation, which avoids the interpolation-induced diffusion in the ray-based methods.
This method shares the same features of ray-based methods and moment-based VET scheme, that (1) it performs well in all optical depths, and (2) it is relatively computationally expensive and memory intensive.
Compared to the VET scheme, it does not solve the radiation transport twice but only once, therefore is more physically self-consistent.
The semi-implicit version of this method \citep{jiang2014ApJS} has been generalized for moving-mesh codes \citep{chang2020MNRAS}, and the fully implicit version \citep{jiang2021ApJS} (hereafter \citetalias{jiang2021ApJS}) has been generalized for general relativity \citep{white2023ApJ}.

All of those methods face a similar time-scale challenge, that (1) the light propagates much faster than the plasma in optically-thin regime, and (2) the energy exchange between radiation and gas happens much faster than the propagation of fast MHD waves in optically-thick regime.
These two time-scale differences correspond to the transport terms and source terms in the radiation transport equation, respectively.
Therefore, a fully explicit time-integration where the radiation module and MHD module share the same timestep is often not adequate in RMHD simulations.
A way to circumvent this for the transport terms is to adopt a `Reduced Speed of Light Approximation' with subcycling.
This approach reduces the speed of light in radiation transport to increase the radiation timestep allowed by explicit time-integration, while keeping the hierarchy of different timescales \citep[e.g.][]{skinner2013ApJS}.
However, great caution is needed to use this approximation \citep{gnedin2016ApJ}, which sometimes leads to significant errors \citep{deparis2019A&A, ocvirk2019A&A, cain2024arXive-prints}.
Alternatively, one can adopt an implicit scheme for the transport terms, which leaps backwards in time and attempts to predict the quantities at the end of the timestep.
Besides the transport terms, the source terms also need special treatments, since the temperature field or thermochemistry often needs to be solved simultaneously with the radiation field because they are strongly coupled together \citep{sekora2010JCoPh, rosdahl2013MNRAS, jiang2012ApJS, jiang2014ApJS, kannan2019MNRAS, jiang2021ApJS}.
Those source terms can also be dealt with either explicitly with subcycling \citep[e.g.][]{rosdahl2013MNRAS, zier2024MNRAS} or implicitly \citep[e.g.][]{jiang2014ApJS, kannan2019MNRAS}.
% Putting the time-integration for transport terms and source terms together, we refer to the method as fully explicit if both terms are calculated explicitly \citep[e.g.][]{rosdahl2013MNRAS, zier2024MNRAS}, semi-implicit if some terms are calculated explicitly while the others implicitly \citep[e.g.][]{jiang2014ApJS, kannan2019MNRAS}, and fully implicit if both are calculated implicitly \citep[e.g.][]{jiang2012ApJS, jiang2021ApJS}.
Generally, implicit methods are conceptually more complex and require a more sophisticated scheme to perform well in parallel computing.

\subsection{Objective: Towards a More Accurate General-purpose 3D Moving-mesh Radiation Magnetohydrodynamic Code}

Besides accurate radiation transport, most complex astrophysical systems demand a wide range of length-scale and time-scale to be covered in one single simulation, which is often very challenging.
In this work, we focus on the 3D moving-mesh MHD code \texttt{AREPO} \citep{springel2010MNRAS, pakmor2011MNRAS, pakmor2016MNRAS, weinberger2020ApJS}.
It was initially designed for cosmological simulations of galaxy formation \citep[e.g. IllustrisTNG; ][]{springel2018MNRAS, pillepich2018MNRASa, nelson2018MNRAS, naiman2018MNRAS, marinacci2018MNRAS}.
This was made possible largely due to the unique features of \texttt{AREPO}: 
(1) Voronoi mesh construction with arbitrary refinement and derefinement criterion; (2) Local time-stepping, where different parts of the simulation can take different timesteps according to local conditions, while conserving the total energy and momentum; (3) Moving-mesh, where the mesh follows the gas velocity in a quasi-Lagrangian way.
These three features facilitate global 3D (M)HD simulations for astrophysical systems of arbitrary geometry involving highly supersonic flows and multi-scale resolutions both in space and time.

Therefore, besides galaxy formation, the \texttt{AREPO} code has also shown unique potentials in simulating other multi-scale systems sometimes with supersonic speeds, such as planet-disk interactions \citep{munoz2014MNRAS}, disk instability and circumbinary disks \citep{zier2022MNRASa, zier2023MNRAS, munoz2019ApJ, siwek2023MNRAS}, binary star interactions \citep{ohlmann2016ApJ, schneider2019Nature, ryu2024MNRAS}, neutron star mergers \citep{lioutas2024MNRAS}, type Ia and core-collapse supernovae \citep{pakmor2022MNRAS, chan2018ApJ}, tidal disruption events \citep{goicovic2019MNRAS, ryu2024MNRASa, vynatheya2024A&A}, star formation \citep{mocz2017ApJ, mayer2024arXive-prints}, multiphase gas \citep{sparre2019MNRAS, das2024arXive-prints}, and AGN jets \citep{bourne2017MNRAS, weinberger2017MNRASa, talbot2024MNRAS}.
An extreme case of the vast length-scale \texttt{AREPO} can simulate within one single simulation is presented in Morán-Fraile et al. in prep., for double-detonated white dwarfs, covering a length-scale hierarchy of more than 6 orders of magnitude, comparable to the most radical simulations by \texttt{GIZMO} \citep{hopkins2024OJAp} or \texttt{AthenaK} \citep{guo2023ApJ}.
It is therefore promising to use \texttt{AREPO} as a general-purpose RMHD code to simulate the aforementioned problems, where radiation influences the multi-scale dynamics.

So far, there are multiple radiation transport modules implemented in \texttt{AREPO}, as summarized in the last row of Table~\ref{tab:rt}.
These include the semi-implicit source-dependent discrete ordinates method \citep{petkova2011MNRAS}, explicit source-dependent ray-tracing method \citep{greif2014MNRAS}, explicit short characteristics method \citep[\texttt{SPRAI};][]{jaura2018MNRAS, jaura2020MNRAS}, explicit moment-based M1 closure \citep{bauer2015MNRAS}, semi-implicit moment-based M1 closure \citep[\texttt{AREPO-RT};][]{kannan2019MNRAS, zier2024MNRAS}, and semi-implicit Monte-Carlo method \citep[\texttt{AREPO-MCRT};][]{smith2020ApJ}.
However, all of the current radiation transport modules in \texttt{AREPO} are specialized in ionizing radiation for either star formation or cosmological simulations, and therefore with thermochemistry often hard-coded to couple with radiation transport.
This makes it difficult to use existing radiation modules to deal with general problems.
Also, current implementations are not applicable for some problems, e.g. when there are too many radiation sources, intersecting radiation fronts, or radiation-dominated flows.

It is therefore clear that a new implementation is needed to push forward \texttt{AREPO} towards a general-purpose 3D moving-mesh RMHD code, where radiation transport is treated accurately in all regimes with minimal assumptions.
In this paper, we aim to bridge this gap by introducing \texttt{AREPO-IDORT}, an implicit discrete ordinates radiation transport fully compatible with all the main features of \texttt{AREPO}.
To this end, we generalize the implicit discrete ordinates radiation transport scheme developed by \citetalias{jiang2021ApJS} for the 3D adaptive mesh refinement finite-volume code \texttt{ATHENA++} \citep{stone2020ApJS}.
The reasons we choose this method are: (1) It produces accurate results across all optical depths; (2) It is fully implicit, and therefore does not suffer from artifacts introduced by the reduced speed of light approximation; (3) It exploits a finite-volume approach, which can be easily generalized to an unstructured mesh compared to e.g. ray-based methods; (4) It has been successfully applied to a wide range of problems, which aligns with our intention for a general-purpose code. Examples using the \citetalias{jiang2021ApJS} method include, but are not limited to, planet formation \citep{zhu2021MNRAS}, protoplanetary disks \citep{zhang2024ApJ}, stellar envelopes \citep{goldberg2022ApJ}, supernovae \citep{goldberg2022ApJa}, neutron star accretion \citep{zhang2022MNRAS}, tidal disruption events \citep{huang2023ApJ}, AGN disks \citep{chen2023ApJ}, and circumbinary disks \citep{tiwari2025arXive-prints}.

We improve the scheme by \citetalias{jiang2021ApJS} in multiple aspects by supporting:
(1) an unstructured moving mesh, (2) local time-stepping, and (3) general equations of state.
Our scheme also achieves fast performance.
In Section~\ref{sec:eq}, we derive the radiation transport equations from first principles with minimal assumptions.
We describe the grey radiation transport scheme in Section~\ref{sec:solver}, where we summarize the flow of our scheme in Section~\ref{sec:solver:scheme} for direct usage.
In Section~\ref{sec:test}, we present extensive test problems covering all regimes of optical thickness and dominant energy content, with a special focus on the performance of the scheme on an unstructured moving mesh with local time-stepping.
We then show a global 3D radiation hydrodynamic model of a red supergiant star in Section~\ref{sec:test:rhd:rsg}, which for the first time covers the entire convective envelope while resolving the radiation flux near the photosphere.
Finally, we discuss the caveats of our scheme and future plans to improve and generalize it further in Section~\ref{sec:disc:code}.
We summarize and conclude in Section~\ref{sec:conc} and point out additional possible applications of our new scheme in different fields of astrophysics that only very few existing codes can simulate efficiently.

\section{Basic Equations}
\label{sec:eq}

In this section, we review the equations involved in the radiation transport processes.
We first state the general forms of radiation transport equations in Section~\ref{sec:eq:general}.
Then in Section~\ref{sec:eq:special}, assuming isotropic scattering and a Planck source function, we derive the radiation transport equations used to develop our scheme.
Radiation is coupled to the total energy and momentum of the system, along with the MHD and gravitational processes.
Therefore, we describe the equations of RMHD in Section~\ref{sec:eq:rmhd}.

\subsection{General Equations of Radiation Transport}
\label{sec:eq:general}

The general form of radiation transport equation is
\begin{equation}
    \frac{\partial I_\nu}{\partial t} + c \vectoraas{n}\cdot \vectoraas{\nabla} I_\nu = c(\eta_\nu - \chi_\nu I_\nu) \, ,
\end{equation}
where $I_\nu$ is the specific intensity at the frequency $\nu$ along the direction given by the unit vector $\vectoraas{n}$, $t$ the time, and $c$ the speed of light. The emmisivity $\eta_\nu$ and the extinction coefficient $\chi_\nu$ here are defined in the rest frame.
Adopting the convention that the source function is $S_\nu \equiv \eta_\nu/\chi_\nu$ and the opacity is $\kappa_\nu \equiv \chi_\nu/\rho$ where $\rho$ is the fluid density, this equation can be written as
\begin{equation}
    \frac{\partial I_\nu}{\partial t} + c \vectoraas{n}\cdot \vectoraas{\nabla} I_\nu = c\rho\kappa_\nu (S_\nu - I_\nu) \, .
\end{equation}
Since the source function and the opacity are often given in the frame comoving with the fluid at a velocity $\vectoraas{v}$, it is natural to convert quantities to the comoving frame (denoted with a prime $\prime$) by the Lorentz transformation \citep{mihalas1984}
\begin{align}
    & I_{\nu^\prime}^\prime(\vectoraas{n^\prime}) = \Gamma^3 I_\nu(\vectoraas{n})\, ,\\
    & S_{\nu^\prime}^\prime = \Gamma^3 S_\nu\, ,\\
    & \kappa_{\nu^\prime}^\prime = \Gamma^{-1}\kappa_\nu\, ,\\
    & \nu^\prime = \Gamma\nu\, ,\\
    & \vectoraas{n^\prime} = \Gamma^{-1}\left(\vectoraas{n}-\frac{1+\Gamma}{1+\gamma}\gamma\vectoraas{\beta}\right)\, ,
\end{align}
where
\begin{align}
    & \Gamma(\vectoraas{n},\vectoraas{v}) \equiv \gamma\left(1-\vectoraas{n}\cdot\vectoraas{\beta}\right)\, ,\\
    & \gamma = 1/\sqrt{1-\beta^2}\, ,\\
    & \vectoraas{\beta} = \vectoraas{v}/c\, .
\end{align}
Therefore, the radiation transport equation in a mixed-frame expression reads
\begin{equation}
    \frac{\partial I_\nu}{\partial t} + c \vectoraas{n}\cdot \vectoraas{\nabla} I_\nu = c\rho\kappa_{\nu^\prime}^\prime\Gamma^{-2} (S_{\nu^\prime}^\prime - I_{\nu^\prime}^\prime) \, .
    \label{eq:rt:mix}
\end{equation}

The zeroth-, first-, and second-moment of $I_\nu$ are related to the radiation energy density, radiation flux, and radiation pressure tensor per frequency via
\begin{align}
    & E_\nu \equiv \frac{1}{c}\int I_\nu d\Omega\, , \\
    & \vectoraas{F_\nu} \equiv \int I_\nu \vectoraas{n}d\Omega\, , \\
    & \tensoraas{P_\nu} \equiv \frac{1}{c}\int I_\nu \vectoraas{n}\vectoraas{n}d\Omega\, ,
\end{align}
where $\Omega$ is the solid angle.
Their counterparts in the comoving frame are
\begin{align}
    & E_{\nu^\prime}^\prime \equiv \frac{1}{c}\int I_{\nu^\prime}^\prime d\Omega^\prime\, , \\
    & \vectoraas{F_{\nu^\prime}^\prime} \equiv \int I_{\nu^\prime}^\prime \vectoraas{n^\prime}d\Omega^\prime\, , \\
    & \tensoraas{P_{\nu^\prime}^\prime} \equiv \frac{1}{c}\int I_{\nu^\prime}^\prime \vectoraas{n^\prime}\vectoraas{n^\prime}d\Omega^\prime\, ,
\end{align}
where
\begin{equation}
    d\Omega^\prime = \Gamma^{-2}d\Omega\, .
\end{equation}
Taking the zeroth- and first-moment of the radiation transport equation~\eqref{eq:rt:mix}, we obtain
\begin{align}
    & \frac{\partial}{\partial t} E_\nu + c\vectoraas{\nabla}\cdot \frac{\vectoraas{F_\nu}}{c} = \rho \int\left[\kappa_{\nu^\prime}^\prime (S_{\nu^\prime}^\prime - I_{\nu^\prime}^\prime)\right]d\Omega^\prime \, ,\\
    & \frac{\partial}{\partial t}\frac{\vectoraas{F_\nu}}{c} + c\vectoraas{\nabla}\cdot \tensoraas{P_\nu} = \rho \int\left[\kappa_{\nu^\prime}^\prime (S_{\nu^\prime}^\prime - I_{\nu^\prime}^\prime)\right]\vectoraas{n}d\Omega^\prime \, .
\end{align}

The frequency-integrated quantities are defined as
\begin{align}
    & I = \int I_\nu d\nu\, ,\\
    & E_\mathrm{r} = \int E_\nu d\nu\, ,\\
    & \vectoraas{F}_\mathrm{r} = \int \vectoraas{F_\nu} d\nu\, ,\\
    & \tensoraas{P}_\mathrm{r} = \int \tensoraas{P_\nu} d\nu\, .
\end{align}
By integrating over the frequency, the radiation transport equation and the moment equations become
\begin{align}
    & \frac{\partial I}{\partial t} + c \vectoraas{n}\cdot \vectoraas{\nabla} I = c\rho\Gamma^{-3} \int \left[\kappa_{\nu^\prime}^\prime(S_{\nu^\prime}^\prime - I_{\nu^\prime}^\prime)\right]d\nu^\prime \, ,\\
    & \frac{\partial}{\partial t} E_\mathrm{r} + c\vectoraas{\nabla}\cdot \frac{\vectoraas{F}_\mathrm{r}}{c} = \rho \int\left\{\int\left[\kappa_{\nu^\prime}^\prime (S_{\nu^\prime}^\prime - I_{\nu^\prime}^\prime)\right]d\nu^\prime \right\}\Gamma^{-1}d\Omega^\prime \, ,\\
    & \frac{\partial}{\partial t}\frac{\vectoraas{F}_\mathrm{r}}{c} + c\vectoraas{\nabla}\cdot \tensoraas{P}_\mathrm{r} = \rho \int\left\{\int\left[\kappa_{\nu^\prime}^\prime (S_{\nu^\prime}^\prime - I_{\nu^\prime}^\prime)\right]d\nu^\prime \right\}\Gamma^{-1}\vectoraas{n}d\Omega^\prime \, .
\end{align}

\subsection{Radiation Transport Equations under Assumptions}
\label{sec:eq:special}

For the rest of this work, we assume:
\begin{enumerate}%[noitemsep]
    \item Isotropic and coherent scattering in the comoving frame;
    \item The source function is the Planck function;
    \item The fluid is moving at non-relativistic speed ($\beta \ll 1$).
\end{enumerate}
% \begin{itemize}
%     \item[$\bullet$] Isotropic and coherent scattering in the comoving frame;
%     \item[$\bullet$] The source function is the Planck function;
%     \item[$\bullet$] The fluid is moving at non-relativistic speed ($\beta \ll 1$).
% \end{itemize}
Under the first two assumptions, the source terms on the right hand sides of the equations can be further expressed in terms of scattering opacity $\kappa_{\nu^\prime,\mathrm{s}}^\prime$, absorption opacity $\kappa_{\nu^\prime,\mathrm{a}}^\prime$, Planck function $B_{\nu^\prime}^\prime$, and the zeroth moment 
\begin{equation}
    J_{\nu^\prime}^\prime = \int I_{\nu^\prime}^\prime d\Omega^\prime / (4\pi) = cE_{\nu^\prime}^\prime / (4\pi)\, ,
\end{equation}
which gives
\begin{align}
    & \frac{\partial I}{\partial t} + c \vectoraas{n}\cdot \vectoraas{\nabla} I = c\rho\Gamma^{-3} \int \begin{aligned}[t]
    & \bigg[\kappa_{\nu^\prime,\mathrm{s}}^\prime(J_{\nu^\prime}^\prime - I_{\nu^\prime}^\prime) \\
    & + \kappa_{\nu^\prime,\mathrm{a}}^\prime(B_{\nu^\prime}^\prime - I_{\nu^\prime}^\prime)\bigg]d\nu^\prime \, ,
    \end{aligned} \\
    & \frac{\partial}{\partial t} E_\mathrm{r} + c\vectoraas{\nabla}\cdot \frac{\vectoraas{F}_\mathrm{r}}{c} = \rho \int
    \begin{aligned}[t]
        & \bigg[(\kappa_{\nu^\prime,\mathrm{s}}^\prime J_{\nu^\prime}^\prime + \kappa_{\nu^\prime,\mathrm{a}}^\prime B_{\nu^\prime}^\prime )\int\Gamma^{-1}d\Omega^\prime \\
        & - (\kappa_{\nu^\prime,\mathrm{s}}^\prime+\kappa_{\nu^\prime,\mathrm{a}}^\prime)\left(\int I_{\nu^\prime}^\prime\Gamma^{-1}d\Omega^\prime\right)\bigg]d\nu^\prime\, ,
    \end{aligned}\\
    & \frac{\partial}{\partial t}\frac{\vectoraas{F}_\mathrm{r}}{c} + c\vectoraas{\nabla}\cdot \tensoraas{P}_\mathrm{r} = \rho  \int
    \begin{aligned}[t]
    & \bigg[(\kappa_{\nu^\prime,\mathrm{s}}^\prime J_{\nu^\prime}^\prime + \kappa_{\nu^\prime,\mathrm{a}}^\prime B_{\nu^\prime}^\prime )\int\Gamma^{-1}\vectoraas{n}d\Omega^\prime \\
    & - (\kappa_{\nu^\prime,\mathrm{s}}^\prime+\kappa_{\nu^\prime,\mathrm{a}}^\prime)\left(\int I_{\nu^\prime}^\prime\Gamma^{-1}\vectoraas{n}d\Omega^\prime\right)\bigg]d\nu^\prime\, .
    \end{aligned}
\end{align}
Here, the Planck function is
\begin{equation}
    B_{\nu^\prime}^\prime = \frac{2h\nu^{\prime 3}}{c^2}\frac{1}{\exp{(h\nu^\prime/k_\mathrm{B}T)}-1}\, ,
\end{equation}
where $h$ is the Planck constant, $k_\mathrm{B}$ is the Boltzmann constant, and $T$ is the local gas temperature.
Assuming that the fluid is moving at non-relativistic speed ($\beta \ll 1$), we can expand the integral over directions as
\begin{align}
    & \int\Gamma^{-1}d\Omega^\prime \begin{aligned}[t]
    & = \int\Gamma^{-3}d\Omega \\
    & = \int\left[1+3\vectoraas{n}\cdot\vectoraas{\beta}+6(\vectoraas{n}\cdot\vectoraas{\beta})^2-3\beta^2/2+\mathcal{O}(\beta^3)\right]d\Omega \\
    & = 4\pi (1+\beta^2/2)+\mathcal{O}(\beta^3)\, ,
    \end{aligned}\\
    & \int\Gamma^{-1}\vectoraas{n}d\Omega^\prime = \int\Gamma^{-3}\vectoraas{n}d\Omega = 4\pi\vectoraas{\beta}+\mathcal{O}(\beta^3)\, ,\\
    & \int I_{\nu^\prime}^\prime\Gamma^{-1}d\Omega^\prime = 4\pi J_{\nu^\prime}^\prime (1+\beta^2/2) + \vectoraas{\beta}\cdot \vectoraas{F_{\nu^\prime}^\prime} + \mathcal{O}(J_{\nu^\prime}^\prime\beta^3)\, ,\\
    & \int I_{\nu^\prime}^\prime\Gamma^{-1}\vectoraas{n}d\Omega^\prime = \vectoraas{F_{\nu^\prime}^\prime} + 4\pi J_{\nu^\prime}^\prime\vectoraas{\beta} + \mathcal{O}(J_{\nu^\prime}^\prime\beta^2)\, .
\end{align}
To the second order of $\beta$ for the energy, these yield
\begin{align}
    & \frac{\partial}{\partial t} E_\mathrm{r} + c\vectoraas{\nabla}\cdot \frac{\vectoraas{F}_\mathrm{r}}{c} = \rho \int \begin{aligned}[t]
    & \bigg[\kappa_{\nu^\prime,a}^\prime 4\pi(B_{\nu^\prime}^\prime - J_{\nu^\prime}^\prime)(1+\beta^2/2) \\
    & - (\kappa_{\nu^\prime,s}^\prime+\kappa_{\nu^\prime,a}^\prime)\vectoraas{\beta}\cdot \vectoraas{F_{\nu^\prime}^\prime}\bigg]d\nu^\prime\, ,
    \end{aligned}\\
    & \frac{\partial}{\partial t}\frac{\vectoraas{F}_\mathrm{r}}{c} + c\vectoraas{\nabla}\cdot \tensoraas{P}_\mathrm{r} = \rho \int \begin{aligned}[t]
    & \bigg[\kappa_{\nu^\prime,a}^\prime 4\pi(B_{\nu^\prime}^\prime - J_{\nu^\prime}^\prime)\vectoraas{\beta} \\
    & - (\kappa_{\nu^\prime,s}^\prime+\kappa_{\nu^\prime,a}^\prime)\vectoraas{F_{\nu^\prime}^\prime}\bigg]d\nu^\prime\, ,
    \end{aligned}
\end{align}
and the Lorentz transformations of the radiation quantities give
\begin{align}
    & E_\mathrm{r}^\prime \begin{aligned}[t]
    & = \frac{1}{c}\int\int I_{\nu^\prime}^\prime d\Omega^\prime d\nu^\prime = \frac{1}{c}\int\int \Gamma^2 I_\nu d\Omega d\nu \\
    & = (1+\beta^2)E_\mathrm{r} - 2\vectoraas{\beta}\cdot\frac{\vectoraas{F}_\mathrm{r}}{c} + \vectoraas{\beta}\cdot\vectoraas{\beta}\cdot\tensoraas{P}_\mathrm{r} + \mathcal{O}(E_\mathrm{r}\beta^3)\, ,
    \end{aligned}\\
    & \frac{\vectoraas{F}_\mathrm{r}^\prime}{c} \begin{aligned}[t]
    & = \frac{1}{c}\int\int I_{\nu^\prime}^\prime \vectoraas{n^\prime}d\Omega^\prime d\nu^\prime = \frac{1}{c}\int\int \Gamma I_\nu \left(\vectoraas{n}-\frac{1+\Gamma}{1+\gamma}\gamma\vectoraas{\beta}\right)d\Omega d\nu \\
    & = \frac{\vectoraas{F}_\mathrm{r}}{c} - \vectoraas{\beta} E_\mathrm{r} - \vectoraas{\beta}\cdot\tensoraas{P}_\mathrm{r} + \mathcal{O}(E_\mathrm{r}\beta^2)\, .
    \end{aligned}
    % & \tensoraas{P_\mathrm{r}^\prime} = \frac{1}{c}\int\int I_{\nu^\prime}^\prime \vectoraas{n^\prime}\vectoraas{n^\prime}d\Omega^\prime d\nu^\prime = \frac{1}{c}\int\int I_\nu \left(\vectoraas{n}-\frac{1+\Gamma}{1+\gamma}\gamma\vectoraas{\beta}\right)\left(\vectoraas{n}-\frac{1+\Gamma}{1+\gamma}\gamma\vectoraas{\beta}\right)d\Omega d\nu = \tensoraas{P_\mathrm{r}} - \left(\vectoraas{\beta}\frac{\vectoraas{F_\mathrm{r}}}{c} + \frac{\vectoraas{F_\mathrm{r}}}{c}\vectoraas{\beta}\right) - \vectoraas{\beta} E_\mathrm{r} - \vectoraas{\beta}\cdot\tensoraas{P_\mathrm{r}} + O(E_\mathrm{r}\beta^2)\, .
\end{align}
By defining the frequency-integrated opacities
\begin{align}
    & \kappa_\mathrm{P} \equiv \frac{\int\kappa_{\nu^\prime,\mathrm{a}}^\prime B_{\nu^\prime}^\prime d\nu^\prime}{\int B_{\nu^\prime}^\prime d\nu^\prime}\, ,\\
    & \kappa_\mathrm{E} \equiv \frac{\int\kappa_{\nu^\prime,\mathrm{a}}^\prime E_{\nu^\prime}^\prime d\nu^\prime}{\int E_{\nu^\prime}^\prime d\nu^\prime}\, ,\\
    & \kappa_{\mathrm{F},i} \equiv \frac{\int\kappa_{\nu^\prime,\mathrm{a}}^\prime F_{\nu^\prime,i}^\prime d\nu^\prime}{\int F_{\nu^\prime,i}^\prime d\nu^\prime},\ i=1,2,3\, ,\\
    & \kappa_{\mathrm{F,s},i} \equiv \frac{\int\kappa_{\nu^\prime,\mathrm{s}}^\prime F_{\nu^\prime,i}^\prime d\nu^\prime}{\int F_{\nu^\prime,i}^\prime d\nu^\prime},\ i=1,2,3\, ,
\end{align}
the moment equations become
\begin{align}
    & \frac{\partial}{\partial t} E_\mathrm{r} + c\vectoraas{\nabla}\cdot \frac{\vectoraas{F}_\mathrm{r}}{c} = \rho \begin{aligned}[t]
    & \bigg\{(\kappa_\mathrm{P} 4\pi B^\prime - \kappa_\mathrm{E} cE_\mathrm{r}^\prime)(1+\beta^2/2) \\
    & - \vectoraas{\beta}\cdot \left[(\vectoraas{\kappa}_\mathrm{F}+\vectoraas{\kappa}_\mathrm{F,s})\odot\vectoraas{F}_\mathrm{r}^\prime\right]\bigg\}\, ,
    \end{aligned}\\
    & \frac{\partial}{\partial t}\frac{\vectoraas{F}_\mathrm{r}}{c} + c\vectoraas{\nabla}\cdot \tensoraas{P}_\mathrm{r} = \rho \begin{aligned}[t]
    & \bigg\{(\kappa_\mathrm{P} 4\pi B^\prime - \kappa_\mathrm{E} cE_\mathrm{r}^\prime)\vectoraas{\beta} \\
    & - \left[(\vectoraas{\kappa}_\mathrm{F}+\vectoraas{\kappa}_\mathrm{F,s})\odot\vectoraas{F}_\mathrm{r}^\prime\right]\bigg\}\, ,
    \end{aligned}
\end{align}
where $\vectoraas{A}\odot \vectoraas{B}$ is the element-wise Hadamard product of vectors $\vectoraas{A}$ and $\vectoraas{B}$, and
\begin{equation}
    B^\prime = \int B_{\nu^\prime}^\prime d\nu^\prime = \frac{caT^4}{4\pi}\, .
\end{equation}
Here, $a = 8\pi^5 k_\mathrm{B}^4/(15 c^3 h^3)$ is the radiation constant.

We further assume that
\begin{equation}
    \kappa_{\mathrm{F},i} \approx \kappa_\mathrm{F},\ \kappa_{\mathrm{F,s},i} \approx \kappa_\mathrm{s},\ i=1,2,3\, ,
\end{equation}
where $\kappa_\mathrm{s}$ is the scattering opacity.
These assumptions only hold when the radiative fluxes of different frequencies are pointing to the same direction.
For a general case, better treatment of opacities are needed, e.g., multigroup or direct frequency-dependent calculations.

In the optically thick regime where local thermal equilibrium (LTE) and radiative diffusion are good approximations, the energy-weighted opacity equals to the Planck mean opacity and the flux-weighted opacity equals to the Rosseland mean opacity, i.e., $\kappa_\mathrm{E}\approx \kappa_\mathrm{P}$ and $\kappa_\mathrm{F}\approx \kappa_\mathrm{R}$, where 
\begin{equation}
    \kappa_\mathrm{R} \equiv \left[\frac{\int\kappa_{\nu^\prime,\mathrm{a}}^{\prime\ -1} (\partial B_{\nu^\prime}^\prime/\partial T) d\nu^\prime}{\int (\partial B_{\nu^\prime}^\prime/\partial T) d\nu^\prime}\right]^{-1}\, ,
\end{equation}
is the Rosseland opacity.
However, we develop the scheme for a more generalized scenario here, and therefore do not make such assumption.

The moment equations then read
\begin{align}
    & \frac{\partial}{\partial t} E_\mathrm{r} \begin{aligned}[t]
    & + c\vectoraas{\nabla}\cdot \frac{\vectoraas{F}_\mathrm{r}}{c} \\
    & = \rho \left[(\kappa_\mathrm{P} 4\pi B^\prime - \kappa_\mathrm{E} cE_\mathrm{r}^\prime)(1+\beta^2/2) -  (\kappa_\mathrm{F}+\kappa_\mathrm{s})\vectoraas{\beta}\cdot\vectoraas{F}_\mathrm{r}^\prime\right]\, ,
    \end{aligned}\\
    & \frac{\partial}{\partial t}\frac{\vectoraas{F}_\mathrm{r}}{c} + c\vectoraas{\nabla}\cdot \tensoraas{P}_\mathrm{r} = \rho \left[(\kappa_\mathrm{P} 4\pi B^\prime - \kappa_\mathrm{E} cE_\mathrm{r}^\prime)\vectoraas{\beta} - (\kappa_\mathrm{F}+\kappa_\mathrm{s})\vectoraas{F}_\mathrm{r}^\prime\right]\, .
\end{align}
Based on the structure of these equations, it is then evident that in order to arrive at the same moment equations, the frequency-integrated radiation transport equation is
\begin{align}
    &\frac{\partial I}{\partial t} + c \vectoraas{n}\cdot \vectoraas{\nabla} I = c S_I\, ,
    \label{eq:rt:I} \\
    & S_I = \rho\Gamma^{-3} \left[(\kappa_\mathrm{P} B^\prime - \kappa_\mathrm{E} J^\prime) +(\kappa_\mathrm{F}+\kappa_\mathrm{s})(J^\prime - I^\prime)\right] \, .
\end{align}

For numerical schemes, it is more useful to express the source terms in the rest frame.
Plugging in the transformations of radiation energy density and radiation flux, we arrive at the final form of the moment equations
\begin{align}
    & \frac{\partial}{\partial t} E_\mathrm{r} + c\vectoraas{\nabla}\cdot \frac{\vectoraas{F}_\mathrm{r}}{c} = S_\mathrm{r,E}\, ,
    \label{eq:rt:E} \\
    & \frac{\partial}{\partial t}\frac{\vectoraas{F}_\mathrm{r}}{c} + c\vectoraas{\nabla}\cdot \tensoraas{P}_\mathrm{r} = c\vectoraas{S}_\mathrm{r,p}\, ,
    \label{eq:rt:F} 
\end{align}
where
% \begin{equation}
% \begin{split}
%     S_\mathrm{r} = \int S_I d\Omega = & c\rho \bigg\{ \left[\left(\kappa_\mathrm{P}\frac{4\pi B^\prime}{c} - \kappa_\mathrm{E} E_\mathrm{r}\right)(1+\beta^2/2) + \kappa_\mathrm{E}\vectoraas{\beta}\cdot\frac{\vectoraas{F}_\mathrm{r}}{c}\right] \\
%     & -  (\kappa_\mathrm{F}+\kappa_\mathrm{s}-\kappa_\mathrm{E})\vectoraas{\beta}\cdot\left(\frac{\vectoraas{F}_\mathrm{r}}{c}- \vectoraas{\beta} E_\mathrm{r} - \vectoraas{\beta}\cdot\tensoraas{P}_\mathrm{r}\right)+ \mathcal{O}(\kappa E_\mathrm{r}\beta^3)\bigg\}\, ,
% \end{split}
% \end{equation}
% \begin{equation}
%     \vectoraas{S}_\mathrm{r} = \frac{1}{c}\int S_I \vectoraas{n}d\Omega  = \rho \left[ (\kappa_\mathrm{P}\frac{4\pi B^\prime}{c} - \kappa_\mathrm{E} E_\mathrm{r})\vectoraas{\beta} - (\kappa_\mathrm{F}+\kappa_\mathrm{s})\left(\frac{\vectoraas{F}_\mathrm{r}}{c}- \vectoraas{\beta} E_\mathrm{r} - \vectoraas{\beta}\cdot\tensoraas{P}_\mathrm{r}\right) + \mathcal{O}(\kappa E_\mathrm{r} \beta^2)\right]\, .
% \end{equation}
\begin{align}
    & S_\mathrm{r,E} \begin{aligned}[t]
    & = \int S_I d\Omega \\
    & = c\rho \bigg\{ \begin{aligned}[t]
        & \left[\left(\kappa_\mathrm{P}\frac{4\pi B^\prime}{c} - \kappa_\mathrm{E} E_\mathrm{r}\right)(1+\beta^2/2) + \kappa_\mathrm{E}\vectoraas{\beta}\cdot\frac{\vectoraas{F}_\mathrm{r}}{c}\right] \\
        & -  (\kappa_\mathrm{F}+\kappa_\mathrm{s}-\kappa_\mathrm{E})\vectoraas{\beta}\cdot\left(\frac{\vectoraas{F}_\mathrm{r}}{c}- \vectoraas{\beta} E_\mathrm{r} - \vectoraas{\beta}\cdot\tensoraas{P}_\mathrm{r}\right)+ \mathcal{O}(\kappa E_\mathrm{r}\beta^3)\bigg\}\, ,
        \end{aligned}
    \end{aligned}
    \label{eq:Sr}\\
    & \vectoraas{S}_\mathrm{r,p} \begin{aligned}[t]
    & = \frac{1}{c}\int S_I \vectoraas{n}d\Omega \\
    & = \rho \bigg[ \begin{aligned}[t]
        &(\kappa_\mathrm{P}\frac{4\pi B^\prime}{c} - \kappa_\mathrm{E} E_\mathrm{r})\vectoraas{\beta} \\
        & - (\kappa_\mathrm{F}+\kappa_\mathrm{s})\left(\frac{\vectoraas{F}_\mathrm{r}}{c}- \vectoraas{\beta} E_\mathrm{r} - \vectoraas{\beta}\cdot\tensoraas{P}_\mathrm{r}\right) + \mathcal{O}(\kappa E_\mathrm{r} \beta^2)\bigg]\, .
        \end{aligned}
    \end{aligned}
    \label{eq:Srvec}
\end{align}
% \begin{align}
%     & S_\mathrm{r} = \int S_I d\Omega = c\rho \left\{\kappa_\mathrm{P} \left[\left(\frac{4\pi B^\prime}{c} - E_\mathrm{r}\right)(1+\beta^2/2) + \vectoraas{\beta}\cdot\frac{\vectoraas{F}_\mathrm{r}}{c}\right] -  (\kappa_\mathrm{s}+\kappa_\mathrm{R}-\kappa_\mathrm{P})\vectoraas{\beta}\cdot\left(\frac{\vectoraas{F}_\mathrm{r}}{c}- \vectoraas{\beta} E_\mathrm{r} - \vectoraas{\beta}\cdot\tensoraas{P}_\mathrm{r}\right)+ \mathcal{O}(\kappa E_\mathrm{r}\beta^3)\right\}\, ,\\
%     & \vectoraas{S}_\mathrm{r} = \frac{1}{c}\int S_I \vectoraas{n}d\Omega  = \rho \left[\kappa_\mathrm{P} (\frac{4\pi B^\prime}{c} - E_\mathrm{r})\vectoraas{\beta} - (\kappa_\mathrm{R}+\kappa_\mathrm{s})\left(\frac{\vectoraas{F}_\mathrm{r}}{c}- \vectoraas{\beta} E_\mathrm{r} - \vectoraas{\beta}\cdot\tensoraas{P}_\mathrm{r}\right) + \mathcal{O}(\kappa E_\mathrm{r} \beta^2)\right]\, .
% \end{align}
We note that those two moment equations cannot be solved alone, since there are three variables in only two equations.
A closure relation between $E_\mathrm{r}$ and $\tensoraas{P}_\mathrm{r}$ (or an effective `equation of state' for radiation) must be provided to solve the moment equations, which is the essence of moment-based radiation transport methods.
The radiation source terms derived here are consistent with those derived in \citet{krumholz2007ApJa}, but are slightly different from those documented in VET schemes \citep{jiang2012ApJS, menon2022MNRAS}.
Based on the scaling argument in \citet{krumholz2007ApJa}, it is important to keep all the second order terms, and we recommend to use these source terms in the future development of the VET scheme.

\subsection{Equations of Radiation Ideal Magnetohydrodynamics}
\label{sec:eq:rmhd}

The Euler equations of radiation ideal MHD can be written as
\begin{equation}
    \frac{\partial}{\partial t}\vectoraas{U} + \vectoraas{\nabla}\cdot \tensoraas{F} = \vectoraas{S}\, ,
\end{equation}
where the conserved quantities, their associated fluxes, and source terms are
\begin{equation}
\begin{split}
    \vectoraas{U} = \begin{pmatrix}
    \rho\\ \rho\vectoraas{v}+\vectoraas{F}_\mathrm{r}/c^2\\ E^*+E_\mathrm{r}
    \end{pmatrix}\, ,\ \
    & \tensoraas{F} = \begin{pmatrix}
    \rho\vectoraas{v}\\ \rho\vectoraas{v}\vectoraas{v}+\tensoraas{P^*}-\vectoraas{B}\vectoraas{B}+\tensoraas{P}_\mathrm{r}\\ (E^*+P^*)\vectoraas{v}-\vectoraas{B}(\vectoraas{v}\cdot\vectoraas{B})+\vectoraas{F}_\mathrm{r}
    \end{pmatrix}\, ,\ \ \\
    & \vectoraas{S} = \begin{pmatrix}
    0\\ -\rho\vectoraas{\nabla}\Phi\\ -\rho\vectoraas{v}\cdot\vectoraas{\nabla}\Phi
    \end{pmatrix}\, .
\end{split}
\end{equation}
Here, $\vectoraas{B}$ is the magnetic field, $\Phi$ is the gravitational potential, $E^* = E_\mathrm{g} + \rho v^2/2 + B^2/2$ is the total MHD energy density, $\tensoraas{P^*} = (P_\mathrm{g} + B^2/2)\tensoraas{I}$ is the total MHD pressure, where $E_\mathrm{g}$ and $P_\mathrm{g}$ are the internal energy density and gas pressure.
This set of equations, coupled with the induction equation 
\begin{equation}
    \frac{\partial}{\partial t}\vectoraas{B} - \vectoraas{\nabla}\times (\vectoraas{v} \times \vectoraas{B}) = 0
\end{equation}
and the radiation transport equation, form the system of radiation ideal MHD.

For the numerical methods, it is more useful to separate the radiation terms as radiation source terms $S_\mathrm{r,E}$ and $\vectoraas{S}_\mathrm{r,p}$ (equations~\eqref{eq:Sr} and \eqref{eq:Srvec}), such that the conserved quantities, fluxes, and source terms are
\begin{equation}
    \vectoraas{U} = \begin{pmatrix}
    \rho\\ \rho\vectoraas{v}\\ E^*
    \end{pmatrix}\, ,\ \
    \tensoraas{F} = \begin{pmatrix}
    \rho\vectoraas{v}\\ \rho\vectoraas{v}\vectoraas{v}+\tensoraas{P^*}-\vectoraas{B}\vectoraas{B}\\ (E^*+P^*)\vectoraas{v}-\vectoraas{B}(\vectoraas{v}\cdot\vectoraas{B})
    \end{pmatrix}\, ,\ \
    \vectoraas{S} = \begin{pmatrix}
    0\\ -\rho\vectoraas{\nabla}\Phi - \vectoraas{S}_\mathrm{r,p}\\ -\rho\vectoraas{v}\cdot\vectoraas{\nabla}\Phi - S_\mathrm{r,E}
    \end{pmatrix}\, .
\end{equation}

\section{Implicit Discrete Ordinates Solver}
\label{sec:solver}

Implicit time integration allows the radiation solver to evolve the radiation field at a timestep not limited by the speed of light.
This is at the expense of losing the short-timescale physics between the large timesteps, which is typically acceptable if the radiation varies at a timescale at least comparable to the dynamical timescale.
The radiation transport equations, either in the angle-dependent form~\eqref{eq:rt:I} or the moment form~\eqref{eq:rt:E}, \eqref{eq:rt:F}, are linear in radiation quantities.
This suggests that the radiation quantities in all the cells can be grouped into a vector, and the radiation transport equations form a linear solver involving a large sparse matrix.
However, the radiation source term in the energy equation can be stiff \citep{sekora2010JCoPh, jiang2012ApJS}, such that the gas temperature has to be evolved simultaneously with the radiation quantities by calculating the change in the internal energy density due to the radiation source terms
\begin{equation}
\begin{split}
    % \frac{\partial}{\partial t}E_\mathrm{g} = - S_\mathrm{r} + \vectoraas{v}\cdot \vectoraas{S}_\mathrm{r} = -\frac{1}{\gamma} \int S_I  \Gamma d\Omega = - c\rho\kappa_\mathrm{P}\left(\frac{4\pi B^\prime}{c} - E_\mathrm{r}^\prime\right)(1-\beta^2/2) + O(\kappa E_\mathrm{r} \beta^3)\, .
    \frac{\partial}{\partial t}E_\mathrm{g} & = - S_\mathrm{r} + \vectoraas{v}\cdot \vectoraas{S}_\mathrm{r} - \frac{1}{2}\vectoraas{S}_\mathrm{r}\cdot\vectoraas{S}_\mathrm{r} = -\frac{1}{\gamma} \int S_I  \Gamma d\Omega - \frac{1}{2}\vectoraas{S}_\mathrm{r}\cdot\vectoraas{S}_\mathrm{r} \\
    & = - c\rho\left(\kappa_\mathrm{P}\frac{4\pi B^\prime}{c} - \kappa_\mathrm{E}E_\mathrm{r}^\prime\right) + \mathcal{O}(\kappa E_\mathrm{r} \beta^3)\, .
\end{split}
\end{equation}
% \begin{equation}
%     % \frac{\partial}{\partial t}E_\mathrm{g} = - S_\mathrm{r} + \vectoraas{v}\cdot \vectoraas{S}_\mathrm{r} = -\frac{1}{\gamma} \int S_I  \Gamma d\Omega = - c\rho\kappa_\mathrm{P}\left(\frac{4\pi B^\prime}{c} - E_\mathrm{r}^\prime\right)(1-\beta^2/2) + O(\kappa E_\mathrm{r} \beta^3)\, .
%     \frac{\partial}{\partial t}E_\mathrm{g} = - S_\mathrm{r} + \vectoraas{v}\cdot \vectoraas{S}_\mathrm{r} - \frac{1}{2}\vectoraas{S}_\mathrm{r}\cdot\vectoraas{S}_\mathrm{r} = -\frac{1}{\gamma} \int S_I  \Gamma d\Omega - \frac{1}{2}\vectoraas{S}_\mathrm{r}\cdot\vectoraas{S}_\mathrm{r} = - c\rho\kappa_\mathrm{P}\left(\frac{4\pi B^\prime}{c} - E_\mathrm{r}^\prime\right) + \mathcal{O}(\kappa E_\mathrm{r} \beta^3)\, .
% \end{equation}
Both the internal energy density of the gas and the Planck function are non-linear functions of the gas temperature.
Therefore, the equations are linear in radiation quantities but non-linear in gas temperature.
For this work, we adopt an iterative solver to solve for the intensities based on \citetalias{jiang2021ApJS}, and use the Newton-Raphson method to solve for the gas temperature.
The scheme described here is only first-order accurate both in time and space, as in \citepalias{jiang2021ApJS}.
This is because higher-order spatial reconstruction requires flux limiters to avoid oscillatory behavior, but flux limiters render the transport term non-linear, which is very difficult to solve implicitly \citepalias{jiang2021ApJS}.
Also, stiff source terms can be difficult to treat with higher-order implicit time integration \citep{sekora2010JCoPh, jiang2012ApJS, jiang2021ApJS}.

The implicit discrete-ordinate radiation transport described in this paper is based on the method in \citetalias{jiang2021ApJS} implemented in \texttt{ATHENA++} \citep{stone2020ApJS}.
This method solves for the specific intensities along discrete angles.
It exploits the finite-volume approach, which makes the method particularly suitable for unstructured meshes.
We extend the scheme to be fully compatible with the features of \texttt{AREPO} \citep{springel2010MNRAS, pakmor2016MNRAS, weinberger2020ApJS}: unstructured moving mesh with local time-stepping and general EOS.
We first review the discretization (Section~\ref{sec:solver:ray} and \ref{sec:solver:eq}) and the derivation of the equations for the iterative solver with local time-stepping (Section~\ref{sec:solver:derivation}).
We present the detailed scheme in Section~\ref{sec:solver:scheme}.
Then we describe the source terms to be coupled back to the \texttt{AREPO} MHD solver (Section~\ref{sec:solver:couple}).
We also briefly describe some optional features for the users to choose and experiment with (Section~\ref{sec:solver:option}).

\subsection{Discretization in Space, Time, and Angular Space}
\label{sec:solver:ray}

To numerically integrate the equations, all the radiation transport equations are discretized in space, time, and solid angle.
We perform the radiation transport calculations on the same Voronoi mesh as the \texttt{AREPO} MHD solver uses, but the method applies to arbitrary meshes.
For the Voronoi mesh, the cells are labeled in a 1D sequence.
The specific intensities are defined at the volume center of each cell.
The numbers of active cells and of directions are denoted as $N_\mathrm{p}$ and $N_\Omega$, respectively.
For the spatial discretization, the subscript $i$ indicates the quantities of the active cell $i \in [0, N_\mathrm{p}-1]$.
The $i^\mathrm{th}$ cell has $N_i$ neighboring cells.
Among those neighboring cells, the $j$th cell is denoted by the subscript $j \in [0, N_i-1]$.
For local time-stepping, we further divide those $N_i$ neighboring cells into $\Bar{N_i}$ active cells and $M_i$ passive cells, such that $N_i = \Bar{N_i} + M_i$.
The active cells are cells holding the local timestep.
The passive cells are cells with larger local timesteps and therefore the primitive variables are not updated in those cells.
We use the subscript $ij$ to indicate the quantities at the face between cell $i$ and $j$.
For angular discretization, the subscript $n \in [0, N_\Omega-1]$ indicates the $n$th direction among the $4\pi$ solid angles.
For temporal discretization, we use the superscript $m$ to indicate the primitive variables at the end of the last active timestep, and $m+1$ as the primitive variables at the end of the current active timestep.
For local time-stepping, the last active timestep is different for different cells.

We use the same angular discretization as the implementation in \texttt{ATHENA++} \citep{davis2012ApJS, jiang2014ApJS, jiang2021ApJS}.
The unit vector along each direction $\vectoraas{n}_n = ((n_x)_n,(n_y)_n,(n_z)_n)$ is defined with respect to the $(x,y,z)$ coordinates.
Each direction is assigned a weight $w_n$ for integrated quantities.
We keep the weights fixed for all the cells and throughout the simulation.
The angular discretization scheme is based on \citet{bruls1999A&A}.
The scheme guarantees that when the radiation field is isotropic, the radiation flux is strictly zero and the Eddington tensor $\tensoraas{f}_\mathrm{Edd} \equiv \tensoraas{P}_\mathrm{r}/E_\mathrm{r}$ is strictly $1/3$ the unit matrix, namely
\begin{align}
    & \sum_{n=0}^{N_\Omega-1} w_n = 1\, ,\\
    & \sum_{n=0}^{N_\Omega-1} (n_x)_n w_n = \sum_{n=0}^{N_\Omega-1} (n_y)_n w_n = \sum_{n=0}^{N_\Omega-1} (n_z)_n w_n = 0\, ,\\
    & \sum_{n=0}^{N_\Omega-1} (n_x)_n^2 w_n = \sum_{n=0}^{N_\Omega-1} (n_y)_n^2 w_n = \sum_{n=0}^{N_\Omega-1} (n_z)_n^2 w_n = \frac{1}{3}\, .
\end{align}
This angular discretization ensures that each octant has $k(k+1)/2$ directions, where $k$ is an arbitrary integer.
The total number of angles can only take specific values as $4k(k+1)$, that is $8,24,48,80,120$, etc.
Therefore, although the Voronoi mesh does not have preferred directions, the discretized radiation field has, especially if only a small number of angles is chosen.
It is generally recommended to choose a number of angles above $48$ for 3D simulations to obtain a decent coverage of the sphere and to avoid ray-effects \citep[i.e., peaks along the discretized directions in radiation and temperature fields due to the small number of directions selected;][see also our Figure~\ref{fig:test:rsg:ray}]{huang2023ApJ, zhang2024ApJ}.
Based on the Lorentz transformation of the solid angle, we have $(w^\prime)_{n,i}^m = (\Gamma_{n,i}^m)^{-2} w_n$.
However, as noted in \citetalias{jiang2021ApJS}, the normalization $\sum_{n=0}^{N_\Omega-1} (w^\prime)_{n,i}^m = 1$ is typically not satisfied to machine precision.
This is corrected by
\begin{equation}
    (w^\prime)_{n,i}^m = \frac{(\Gamma_{n,i}^m)^{-2} w_n}{\sum_{n=0}^{N_\Omega-1} (\Gamma_{n,i}^m)^{-2} w_n}\, .
\end{equation}
We note that although the weight $w_n$ in the lab frame does not depend on space and time, the weight $(w^\prime)_{n,i}^m$ in the comoving frame does, since the Lorentz factor depends on space and time.

We calculate the radiation energy density, radiation flux, and radiation pressure tensor by
\begin{align}
    & (E_\mathrm{r})_i^{m+1} = \frac{4\pi}{c}\sum_{n=0}^{N_\Omega-1} I_{n,i}^{m+1} w_n\, , \\
    & (\vectoraas{F}_\mathrm{r})_i^{m+1} = 4\pi\sum_{n=0}^{N_\Omega-1} I_{n,i}^{m+1} \vectoraas{n}_n w_n\, ,\\
    & (\tensoraas{P}_\mathrm{r})_i^{m+1} = \frac{4\pi}{c}\sum_{n=0}^{N_\Omega-1} I_{n,i}^{m+1} \vectoraas{n}_n \vectoraas{n}_n w_n\, .
\end{align}

\subsection{Discretized Equations}
\label{sec:solver:eq}

In a moving-mesh code with refinement and local time-stepping such as \texttt{AREPO}, it is essential to distinguish between the conserved quantities and primitive variables.
We only update the primitive variables of the cell when the cell is active (i.e. synchronised), but the conserved quantities are always updated when they exchange fluxes with the neighboring cells.
Here, we use $(IV)$ as the conserved quantity corresponding to the primitive variable $I$.
The conserved form of the discretized radiation transport equation can be written as
% \begin{equation}
%     [I_n^{m+1}V^m - (IV)_n^m]_i = -\sum_{j=0}^{N_i-1}\left[I_n (c\vectoraas{n}_n-\vectoraas{v})\right]_{ij}^{m+1}\cdot \vectoraas{A}_{ij}^m \Delta t_{ij}^m - \sum_{j=0}^{N_i-1}\left[I_n (\vectoraas{v}-\vectoraas{w})\right]_{ij}^m\cdot \vectoraas{A}_{ij}^m \Delta t_{ij}^m + \left[(S_I)_{n}^{m+1} V^m c\Delta t^m\right]_i\, ,
% \end{equation}
\begin{equation}
\begin{split}
    [I_n^{m+1}V^m - (IV)_n^m]_i = & -\sum_{j=0}^{N_i-1}\left[I_n (c\vectoraas{n}_n-\vectoraas{u})\right]_{ij}^{m+1}\cdot \vectoraas{A}_{ij}^m \Delta t_{ij}^m \\
    & + \left[(S_I)_{n}^{m+1} V^m c\Delta t^m\right]_i\, ,
\end{split}
    \label{eq:discretizedrt}
\end{equation}
where $\vectoraas{A}_{ij}$, $\Delta t_{ij}=\min(\Delta t_i,\Delta t_j)$ and $\vectoraas{u}_{ij}$ are the normal area vector, minimum timestep and the face velocity between cell $i$ and $j$.
The local timesteps are limited by the local MHD CFL condition and grouped into a power-of-two hierarchy \citep{springel2010MNRAS}.
For a fixed-mesh code, $\vectoraas{u}_{ij} = 0$.
For a code that uses global timesteps, we have $\Delta t_{ij}^m = \Delta t_i^m = \Delta t^m$.
Here, $V_i$ is the volume of the cell $i$, and the implicit source term is
\begin{equation}
\begin{split}
    (S_{I})_{n,i}^{m+1} = (\rho \Gamma_n^{-3})_i^m \bigg[ & (\kappa_{\mathrm{P}})_i^m (B^\prime)_i^{m+1} - (\kappa_{\mathrm{E}})_i^m (J^\prime)_i^{m+1} \\
    & +(\kappa_\mathrm{F}+\kappa_\mathrm{s})_i^m (J^\prime - I^\prime_n)_i^{m+1}\bigg] \, .
\end{split}
\end{equation}
% \begin{equation}
%     (S_{I})_{n,i}^{m+1} = (\rho \Gamma_n^{-3})_i^m \left[(\kappa_{\mathrm{P}})_i^m (B^\prime - J^\prime)_i^{m+1} +(\kappa_\mathrm{R}+\kappa_\mathrm{s})_i^m (J^\prime - I^\prime_n)_i^{m+1}\right] \, .
% \end{equation}
The first moment of the specific intensities is discretized as
\begin{equation}
    J^\prime = \sum_{n=0}^{N_\Omega-1} I^\prime_n w^\prime_n\, ,
\end{equation}
where the specific intensities in the comoving frame are related to their counterparts in the lab frame via the Lorentz transformation
\begin{equation}
    I^\prime_n = \Gamma_n^4 I_n\, .
\end{equation}
% In this equation, we separate the gas velocity advection term $I (\vectoraas{v}-\vectoraas{w})$ from the total speed-of-light transport term $I c\vectoraas{n}$ following \citetalias{jiang2021ApJS}.
% We evolve the gas velocity advection term explicitly with the hydrodynamic finite-volume solver of \texttt{AREPO}, treating $I_n$ as passive scalars.
% This step is to ensure the accuracy in the dynamic diffusion regime, where the numerical truncation errors arising from the gas velocity advection can easily dominate over the physical diffusion \citep{jiang2013ApJ, jiang2014ApJS, jiang2021ApJS}.
We do not separate the gas velocity advection term from the total transport term as in \citetalias{jiang2021ApJS}, because we find that the separation leads to unstable behavior with local time-stepping.

The discretized update to the specific gas internal energy is
\begin{equation}
    (e_\mathrm{g}^{m+1} - e_\mathrm{g}^m)_i = -4\pi  \left[\Delta t^m  (\kappa_\mathrm{P}^m B^{\prime, m+1} - \kappa_\mathrm{E}^m J^{\prime, m+1})\right]_i\, ,
\end{equation}
% \begin{equation}
%     (e_\mathrm{g}^{m+1} - e_\mathrm{g}^m)_i = -4\pi  \left[\Delta t^m \kappa_\mathrm{P}^m (B^\prime - J^\prime)^{m+1}\right]_i\, ,
% \end{equation}
where $e_\mathrm{g}^{m+1} = E_\mathrm{g}^{m+1}/\rho^m = e_\mathrm{g} (\rho^m, T^{m+1})$ and $B^{\prime, m+1} = B^\prime (T^{m+1})$.
By coupling these two implicit equations together, we solve for the radiation field and the temperature field simultaneously.
All the other physical quantities are kept unchanged from the last timestep.
Specifically, the opacities are not solved implicitly.
This is to ensure convergence and to avoid oscillatory behavior from the highly non-linear dependence of opacities on temperature.
For passive cells, only the conserved radiation quantities $(IV)$ are changed, and all the other MHD quantities (e.g. temperature, internal energy, total energy, total momentum) are only updated by the next time the passive cells become active again.

\subsection{Derivation of the Equations for the Iterative Solver}
\label{sec:solver:derivation}

Let us focus on the specific intensity along one direction in one cell.
It only depends on the other physical quantities of the cell, the specific intensities of the cell along other directions, and the specific intensities in neighboring cells along this direction.
The iterative solver is constructed such that the specific intensities along this direction in neighboring cells are kept unchanged after the last iteration, in order to update the specific intensities along all directions in the target cell in the current iteration.

Specifically, the iterated equation is derived as follows.
First, we consider the transport terms.
% We put all the terms that only depend on specific instensities from last timestep into one term, namely
% \begin{equation}
%     (I_\mathrm{c1})_{n,i}^m = \left\{(IV)_{n,i}^m - \sum_{j=0}^{N_i-1}\left[I_n (\vectoraas{v}-\vectoraas{w})\right]_{ij}^m\cdot \vectoraas{A}_{ij}^m \Delta t_{ij}^m\right\}/V^m_i \, .
%     \label{eq:gasadv}
% \end{equation}
% This term represents the updated specific intensity owning to the gas velocity advection, which does not change during the iterations.
The implicit transport term in the radiation transport equation is a linear combination of the specific intensities in the target cell and in neighboring cells.
Therefore, it can be written as
% \begin{equation}
%     \left\{\sum_{j=0}^{N_i-1}\left[I_n (c\vectoraas{n}_n-\vectoraas{v})\right]_{ij}^{m+1}\cdot \vectoraas{A}_{ij}^m \Delta t_{ij}^m\right\}/V^m_i = \left(\sum_{j=0}^{N_i-1} C_{n,ij}^m \right) I_{n,i}^{m+1} + \left(\sum_{j=0}^{\Bar{N_i}-1} D_{n,ij}^m I_{n,j}^{m+1} \right) + \left(\sum_{j=0}^{M_i-1} D_{n,ij}^m I_{n,j}^m \right)\, ,
% \end{equation}
\begin{equation}
\begin{split}
    \bigg\{\sum_{j=0}^{N_i-1} & \left[I_n (c\vectoraas{n}_n-\vectoraas{u})\right]_{ij}^{m+1}\cdot \vectoraas{A}_{ij}^m \Delta t_{ij}^m\bigg\}/V^m_i \\
    & = \left(\sum_{j=0}^{N_i-1} C_{n,ij}^m \right) I_{n,i}^{m+1} + \left(\sum_{j=0}^{\Bar{N_i}-1} D_{n,ij}^m I_{n,j}^{m+1} \right) + \left(\sum_{j=0}^{M_i-1} D_{n,ij}^m I_{n,j}^m \right)\, ,
\end{split}
\end{equation}
where the coefficients $C$ and $D$ only depend on the grid configuration and the Riemann solver.
The last two terms on the right hand side are separated according to whether the neighboring cell is active or not.

These coefficients in the transport terms are calculated by a modified HLLE solver as proposed by \citetalias{jiang2021ApJS} based on \citet{sekora2010JCoPh}, where
% \begin{equation}
% \begin{split}
%     \left[I_n (c\vectoraas{n}_n-\vectoraas{v})\right]_{ij}^{m+1}\cdot \vectoraas{\mu}_{ij}^m = 
%     & \left(\frac{S^+}{S^+-S^-}\right)_{n,ij}^m I_{n,i}^{m+1}(c\vectoraas{n}_n-\vectoraas{v}_{ij}^m)\cdot \vectoraas{\mu}_{ij}^m
%     - \left(\frac{S^-}{S^+-S^-}\right)_{n,ij}^m I_{n,j}^{m+1}(c\vectoraas{n}_n-\vectoraas{v}_{ij}^m)\cdot \vectoraas{\mu}_{ij}^m \\
%     & + \left(\frac{S^+ S^-}{S^+-S^-}\right)_{n,ij}^m \left[I_j - I_i\right]_n^{m+1}\, .
% \end{split}
% \end{equation}
\begin{equation}
\begin{split}
    \left[I_n (c\vectoraas{n}_n-\vectoraas{u})\right]_{ij}^{m+1}\cdot \vectoraas{\mu}_{ij}^m = &
    \left(\frac{S^+}{S^+-S^-}\right)_{n,ij}^m I_{n,i}^{m+1}(c\vectoraas{n}_n-\vectoraas{u}_{ij}^m)\cdot \vectoraas{\mu}_{ij}^m \\
    & - \left(\frac{S^-}{S^+-S^-}\right)_{n,ij}^m I_{n,j}^{m+1}(c\vectoraas{n}_n-\vectoraas{u}_{ij}^m)\cdot \vectoraas{\mu}_{ij}^m \\
    & + \left(\frac{S^+ S^-}{S^+-S^-}\right)_{n,ij}^m \left[I_j - I_i\right]_n^{m+1}\, .
\end{split}
\end{equation}
% \begin{equation}
% \begin{split}
%     \left[I_n (c\vectoraas{n}_n-\vectoraas{w})\right]_{ij}^{m+1}\cdot \vectoraas{\mu}_{ij}^m = 
%     & \left(\frac{S^+}{S^+-S^-}\right)_{n,ij}^m I_{n,i}^{m+1}(c\vectoraas{n}_n-\vectoraas{w}_{ij}^m)\cdot \vectoraas{\mu}_{ij}^m
%     - \left(\frac{S^-}{S^+-S^-}\right)_{n,ij}^m I_{n,j}^{m+1}(c\vectoraas{n}_n-\vectoraas{w}_{ij}^m)\cdot \vectoraas{\mu}_{ij}^m \\
%     & + \left(\frac{S^+ S^-}{S^+-S^-}\right)_{n,ij}^m \left[I_j - I_i\right]_n^{m+1}\, .
% \end{split}
% \end{equation}
% Here, $\vectoraas{v}_{ij}^m$ is the fluid velocity at the face calculated from the \texttt{AREPO} MHD solver taking into account of the time and spatial extrapolation \citep{pakmor2016MNRAS}.
Here, $\vectoraas{u}_{ij}^m$ is the face velocity of the moving mesh calculated in the \texttt{AREPO} MHD finite-volume solver.
The unit vector $\vectoraas{\mu}_{ij}^m$ is normal to the face pointing from cell $i$ to cell $j$, obtained from the Voronoi mesh construction.
The maximum and minimum signal speeds are 
\begin{align}
    & (S^+)_{n,ij}^m = 
    \begin{cases}
    c\left|\vectoraas{n}_n\cdot\vectoraas{\mu}_{ij}^m\right|\sqrt{\left\{1-\exp\left[-(\tau_{ij}^m)^2\right]\right\}/(\tau_{ij}^m)^2}, & \vectoraas{n}_n\cdot\vectoraas{\mu}_{ij}^m \ge 0\, ,\\
    c\left|\vectoraas{n}_n\cdot\vectoraas{\mu}_{ij}^m\right|\sqrt{\left\{1-\exp\left[-(\tau_{ij}^m)^4\right]\right\}/(\tau_{ij}^m)^2}, & \vectoraas{n}_n\cdot\vectoraas{\mu}_{ij}^m < 0\, ,\\
    \end{cases} \\
    & (S^-)_{n,ij}^m = 
    \begin{cases}
    -c\left|\vectoraas{n}_n\cdot\vectoraas{\mu}_{ij}^m\right|\sqrt{\left\{1-\exp\left[-(\tau_{ij}^m)^4\right]\right\}/(\tau_{ij}^m)^2}, & \vectoraas{n}_n\cdot\vectoraas{\mu}_{ij}^m \ge 0\, ,\\
    -c\left|\vectoraas{n}_n\cdot\vectoraas{\mu}_{ij}^m\right|\sqrt{\left\{1-\exp\left[-(\tau_{ij}^m)^2\right]\right\}/(\tau_{ij}^m)^2}, & \vectoraas{n}_n\cdot\vectoraas{\mu}_{ij}^m < 0\, .
    \end{cases}
\end{align}
This definition is designed to reduce the simple signal speed $c\left|\vectoraas{n}_n\cdot\vectoraas{\mu}_{ij}^m\right|$ by a factor of $\tau_{ij}^m$ in the optically thick limit while approaching an upwind flux in the optically thin limit \citepalias{jiang2021ApJS}.
Here, the effective optical depth at the face is defined as
\begin{equation}
    \tau_{ij}^m = \alpha \rho_{ij}^m (\kappa_\mathrm{F}+\kappa_\mathrm{s})_{ij}^m \Delta R_{ij}^m\, ,
    \label{eq:alpha}
\end{equation}
% \begin{equation}
%     \tau_{ij}^m = \alpha \rho_{ij}^m (\kappa_\mathrm{R}+\kappa_\mathrm{s})_{ij}^m \Delta R_{ij}^m\, ,
%     \label{eq:alpha}
% \end{equation}
where $\alpha$ is a numerical parameter taken as $\alpha=5$ by default \citepalias{jiang2021ApJS}.
All the face values on the right hand side of this equation are calculated as the average of the two values in $i$ and $j$ cells, namely $q_{ij}^m = (q_i^m + q_j^m)/2$.
The effective cell radius is converted from the cell volume, namely $\Delta R_i^m = [3V_i^m/(4\pi)]^{1/3}$.
We note that the cell `volume' is different in 1D and 2D cases.
Therefore, the coefficients are
% \begin{align}
%     & C_{n,ij}^m = \left(\frac{S^+}{S^+-S^-}\right)_{n,ij}^m \left[(c\vectoraas{n}_n-\vectoraas{v}_{ij}^m)\cdot \vectoraas{\mu}_{ij}^m - (S^-)_{n,ij}^m\right] \left(\frac{A\Delta t}{V}\right)_{ij}^m\, ,
%     \label{eq:C}\\
%     & D_{n,ij}^m = - \left(\frac{S^-}{S^+-S^-}\right)_{n,ij}^m \left[(c\vectoraas{n}_n-\vectoraas{v}_{ij}^m)\cdot \vectoraas{\mu}_{ij}^m - (S^+)_{n,ij}^m\right] \left(\frac{A\Delta t}{V}\right)_{ij}^m\, .
%     \label{eq:D}
% \end{align}
\begin{align}
    & C_{n,ij}^m = \left(\frac{S^+}{S^+-S^-}\right)_{n,ij}^m \left[(c\vectoraas{n}_n-\vectoraas{u}_{ij}^m)\cdot \vectoraas{\mu}_{ij}^m - (S^-)_{n,ij}^m\right] \left(\frac{A\Delta t}{V}\right)_{ij}^m\, ,
    \label{eq:C}\\
    & D_{n,ij}^m = - \left(\frac{S^-}{S^+-S^-}\right)_{n,ij}^m \left[(c\vectoraas{n}_n-\vectoraas{u}_{ij}^m)\cdot \vectoraas{\mu}_{ij}^m - (S^+)_{n,ij}^m\right] \left(\frac{A\Delta t}{V}\right)_{ij}^m\, .
    \label{eq:D}
\end{align}

The radiation transport equation can be further written as
% \begin{equation}
%     I_{n,i}^{m+1} = - \left(\sum_{j=0}^{N_i-1} C_{n,ij}^m \right) I_{n,i}^{m+1} - \left(\sum_{j=0}^{\Bar{N_i}-1} D_{n,ij}^m I_{n,j}^{m+1} \right)  - \left(\sum_{j=0}^{M_i-1} D_{n,ij}^m I_{n,j}^m \right) + (I_\mathrm{c1})_{n,i}^m + \left[(S_I)_{n}^{m+1} c\Delta t^m \right]_i\, .
% \end{equation}
\begin{equation}
\begin{split}
    I_{n,i}^{m+1} = & - \left(\sum_{j=0}^{N_i-1} C_{n,ij}^m \right) I_{n,i}^{m+1} - \left(\sum_{j=0}^{\Bar{N_i}-1} D_{n,ij}^m I_{n,j}^{m+1} \right)  - \left(\sum_{j=0}^{M_i-1} D_{n,ij}^m I_{n,j}^m \right) \\
    & + (IV)_{n,i}^m/V_i^m + \left[(S_I)_{n}^{m+1} c\Delta t^m \right]_i\, .
\end{split}
\end{equation}
We rewrite this in the context of iterations.
The last iteration is denoted by a superscript $l$, while the current iteration is denoted as $l+1$:
% \begin{equation}
% \begin{split}
%     I_{n,i}^{m+1, l+1} = 
%     & - \left(\sum_{j=0}^{N_i-1} (C^+)_{n,ij}^m \right) I_{n,i}^{m+1, l+1} - \left(\sum_{j=0}^{N_i-1} (C^-)_{n,ij}^m \right) I_{n,i}^{m+1, l}\\ 
%     & - \left(\sum_{j=0}^{\Bar{N_i}-1} D_{n,ij}^m I_{n,j}^{m+1, l} \right) - \left(\sum_{j=0}^{M_i-1} D_{n,ij}^m I_{n,j}^m \right)
%      + (I_\mathrm{c1})_{n,i}^m + \left[(S_I)_{n}^{m+1, l+1} c\Delta t^m)\right]_i\, .
% \end{split}
% \end{equation}
\begin{equation}
\begin{split}
    I_{n,i}^{m+1, l+1} = 
    & - \left(\sum_{j=0}^{N_i-1} (C^+)_{n,ij}^m \right) I_{n,i}^{m+1, l+1} - \left(\sum_{j=0}^{N_i-1} (C^-)_{n,ij}^m \right) I_{n,i}^{m+1, l} \\
    & - \left(\sum_{j=0}^{\Bar{N_i}-1} D_{n,ij}^m I_{n,j}^{m+1, l} \right) - \left(\sum_{j=0}^{M_i-1} D_{n,ij}^m I_{n,j}^m \right) \\
     & + (IV)_{n,i}^m/V_i^m + \left[(S_I)_{n}^{m+1, l+1} c\Delta t^m)\right]_i\, .
\end{split}
\end{equation}
% \begin{equation}
% \begin{split}
%     I_{n,i}^{m+1, l+1} = 
%     & - \left(\sum_{j=0}^{N_i-1} (C^+)_{n,ij}^m \right) I_{n,i}^{m+1, l+1} - \left(\sum_{j=0}^{N_i-1} (C^-)_{n,ij}^m \right) I_{n,i}^{m+1, l}\\ 
%     & - \left(\sum_{j=0}^{\Bar{N_i}-1} D_{n,ij}^m I_{n,j}^{m+1, l} \right) - \left(\sum_{j=0}^{M_i-1} D_{n,ij}^m I_{n,j}^m \right)
%      + (IV)_{n,i}^m/V_i^m + \left[(S_I)_{n}^{m+1, l+1} c\Delta t^m)\right]_i\, .
% \end{split}
% \end{equation}
Here we further split the coefficients $C$ according to whether they are positive or negative, namely $(C^+)_{n,ij}^m = \max (C_{n,ij}^m,0)$ and $(C^-)_{n,ij}^m = \min (C_{n,ij}^m,0)$.
All the negative terms are grouped into the last iteration.
This is to ensure the coefficients in front of the specific intensities are always above unity during the iterations for a more robust performance \citepalias[see appendix of][]{jiang2021ApJS}.
The equation then reads
\begin{equation}
    \left[1+(g^+)_{n,i}^m \right] I_{n,i}^{m+1, l+1} = (I_\mathrm{c})_{n,i}^{m+1,l} + \left[(S_I)_{n}^{m+1, l+1} c\Delta t^m\right]_i\, ,
    \label{eq:Idiscretized}
\end{equation}
where
\begin{align}
    & (g^+)_{n,i}^m = \sum_{j=0}^{N_i-1} (C^+)_{n,ij}^m\, ,\\
    & (g^-)_{n,i}^m = \sum_{j=0}^{N_i-1} (C^-)_{n,ij}^m\, ,\\
    & (I_\mathrm{c2})_{n,i}^{m+1,l} = - (g^-)_{n,i}^m I_{n,i}^{m+1, l} - \left(\sum_{j=0}^{\Bar{N_i}-1} D_{n,ij}^m I_{n,j}^{m+1, l} \right) - \left(\sum_{j=0}^{M_i-1} D_{n,ij}^m I_{n,j}^m\right)\, ,\\
    % & (I_\mathrm{c})_{n,i}^{m+1,l} = (I_\mathrm{c2})_{n,i}^{m+1,l} + (I_\mathrm{c1})_{n,i}^m\, .
    & (I_\mathrm{c})_{n,i}^{m+1,l} = (I_\mathrm{c2})_{n,i}^{m+1,l} + (IV)_{n,i}^m/V_i^m\, .
\end{align}

Since the source terms are measured in the comoving frame, we multiply Equation~\eqref{eq:Idiscretized} by $\Gamma^4$, which yields
% \begin{equation}
% \begin{split}
%     \left[1+(g^+)_{n,i}^m \right] (I^\prime)_{n,i}^{m+1, l+1} = & (\Gamma_{n,i}^m)^4(I_\mathrm{c})_{n,i}^{m+1,l}  \\
%     & + (c\Delta t \rho \Gamma_n)_i^m \left[(\kappa_\mathrm{P})_i^m (B^\prime)_i^{m+1, l+1} - (\kappa_\mathrm{E})_i^m (J^\prime)_i^{m+1, l+1} +(\kappa_\mathrm{F}+\kappa_\mathrm{s})_i^m (J^\prime - I^\prime_n)_i^{m+1, l+1}\right]\, .
% \end{split}
% \end{equation}
\begin{equation}
\begin{split}
    \big[1 & +(g^+)_{n,i}^m \big] (I^\prime)_{n,i}^{m+1, l+1} = (\Gamma_{n,i}^m)^4(I_\mathrm{c})_{n,i}^{m+1,l}  \\
    & + (c\Delta t \rho \Gamma_n)_i^m \bigg[ \begin{aligned}[t]       
        & (\kappa_\mathrm{P})_i^m (B^\prime)_i^{m+1, l+1} - (\kappa_\mathrm{E})_i^m (J^\prime)_i^{m+1, l+1} \\
        & +(\kappa_\mathrm{F}+\kappa_\mathrm{s})_i^m (J^\prime - I^\prime_n)_i^{m+1, l+1}\bigg]\, .
    \end{aligned}
\end{split}
\end{equation}
% \begin{equation}
%     \left[1+(g^+)_{n,i}^m \right] (I^\prime)_{n,i}^{m+1, l+1} = (\Gamma_{n,i}^m)^4(I_\mathrm{c})_{n,i}^{m+1,l} + (c\Delta t \rho \Gamma_n)_i^m \left[(\kappa_\mathrm{P})_i^m (B^\prime - J^\prime)_i^{m+1, l+1} +(\kappa_\mathrm{R}+\kappa_\mathrm{s})_i^m (J^\prime - I^\prime_n)_i^{m+1, l+1}\right]\, .
% \end{equation}
This gives
% \begin{equation}
% \begin{split}
%     \left[1+g_n^+ + c\Delta t\rho (\kappa_\mathrm{F}+\kappa_\mathrm{s})\Gamma_n\right]_i^m (I^\prime)_{n,i}^{m+1, l+1} = 
%     & (\Gamma_{n,i}^m)^4(I_\mathrm{c})_{n,i}^{m+1,l} \\
%     & + (c\Delta t \rho \Gamma_n)_i^m \left[(\kappa_\mathrm{P})_i^m (B^\prime)_i^{m+1, l+1} +(\kappa_\mathrm{F}+\kappa_\mathrm{s}-\kappa_\mathrm{E})_i^m (J^\prime)_i^{m+1, l+1}\right]\, .
% \end{split}
% \end{equation}
\begin{equation}
\begin{split}
    \big[1 & +g_n^+ + c\Delta t\rho (\kappa_\mathrm{F}+\kappa_\mathrm{s})\Gamma_n\big]_i^m (I^\prime)_{n,i}^{m+1, l+1} = 
    (\Gamma_{n,i}^m)^4(I_\mathrm{c})_{n,i}^{m+1,l} \\
    & + (c\Delta t \rho \Gamma_n)_i^m \left[(\kappa_\mathrm{P})_i^m (B^\prime)_i^{m+1, l+1} +(\kappa_\mathrm{F}+\kappa_\mathrm{s}-\kappa_\mathrm{E})_i^m (J^\prime)_i^{m+1, l+1}\right]\, .
\end{split}
\end{equation}
% \begin{equation}
% \begin{split}
%     \left[1+g_n^+ + c\Delta t\rho (\kappa_\mathrm{R}+\kappa_\mathrm{s})\Gamma_n\right]_i^m (I^\prime)_{n,i}^{m+1, l+1} = 
%     & (\Gamma_{n,i}^m)^4(I_\mathrm{c})_{n,i}^{m+1,l} \\
%     & + (c\Delta t \rho \Gamma_n)_i^m \left[(\kappa_\mathrm{P})_i^m (B^\prime)_i^{m+1, l+1} +(\kappa_\mathrm{R}+\kappa_\mathrm{s}-\kappa_\mathrm{P})_i^m (J^\prime)_i^{m+1, l+1}\right]\, .
% \end{split}
% \end{equation}
We define the coefficient
\begin{equation}
    f_{n,i}^m \equiv \frac{1}{1+\left[g_n^+ + c\Delta t\rho (\kappa_\mathrm{F}+\kappa_\mathrm{s}) \Gamma_n\right]_i^m}\, ,
\end{equation}
% \begin{equation}
%     f_{n,i}^m \equiv \frac{1}{1+\left[g_n^+ + c\Delta t\rho (\kappa_\mathrm{R}+\kappa_\mathrm{s}) \Gamma_n\right]_i^m}\, ,
% \end{equation}
such that the equation becomes
\begin{equation}
\begin{split}
    ( & I^\prime)_{n,i}^{m+1, l+1} = (f\Gamma^4)_{n,i}^m(I_\mathrm{c})_{n,i}^{m+1,l} \\
    & + (c\Delta t \rho \Gamma_n f_n)_i^m \left[(\kappa_\mathrm{P})_i^m (B^\prime)_i^{m+1, l+1} +(\kappa_\mathrm{F}+\kappa_\mathrm{s}-\kappa_\mathrm{E})_i^m (J^\prime)_i^{m+1, l+1}\right]\, .
\end{split}
    \label{eq:iprime}
\end{equation}
% \begin{equation}
%     (I^\prime)_{n,i}^{m+1, l+1} = (f\Gamma^4)_{n,i}^m(I_\mathrm{c})_{n,i}^{m+1,l} + (c\Delta t \rho \Gamma_n f_n)_i^m \left[(\kappa_\mathrm{P})_i^m (B^\prime)_i^{m+1, l+1} +(\kappa_\mathrm{R}+\kappa_\mathrm{s}-\kappa_\mathrm{P})_i^m (J^\prime)_i^{m+1, l+1}\right]\, .
%     \label{eq:iprime}
% \end{equation}
Summing it up over the solid angle yields
% \begin{equation}
% \begin{split}
%     (J^\prime)_i^{m+1, l+1} = & \left[\sum_{n=0}^{N_\Omega-1}(w_n^\prime f_n\Gamma_n^4)^m(I_\mathrm{c})_n^{m+1,l}\right]_{i} + \left(c\Delta t \rho \kappa_\mathrm{P}\sum_{n=0}^{N_\Omega-1}w^\prime_n f_n \Gamma_n\right)_i^m (B^\prime)_i^{m+1, l+1} \\
%     & + \left[c\Delta t \rho (\kappa_\mathrm{F}+\kappa_\mathrm{s}-\kappa_\mathrm{E})\sum_{n=0}^{N_\Omega-1}w^\prime_n f_n \Gamma_n\right]_i^m (J^\prime)_i^{m+1, l+1}\, .
% \end{split}
% \end{equation}
\begin{equation}
\begin{split}
    (J^\prime)_i^{m+1, l+1} = & \left[\sum_{n=0}^{N_\Omega-1}(w_n^\prime f_n\Gamma_n^4)^m(I_\mathrm{c})_n^{m+1,l}\right]_{i} \\
    & + \left(c\Delta t \rho \kappa_\mathrm{P}\sum_{n=0}^{N_\Omega-1}w^\prime_n f_n \Gamma_n\right)_i^m (B^\prime)_i^{m+1, l+1} \\ 
    & + \left[c\Delta t \rho (\kappa_\mathrm{F}+\kappa_\mathrm{s}-\kappa_\mathrm{E})\sum_{n=0}^{N_\Omega-1}w^\prime_n f_n \Gamma_n\right]_i^m (J^\prime)_i^{m+1, l+1}\, .
\end{split}
\end{equation}
% \begin{equation}
% \begin{split}
%     (J^\prime)_i^{m+1, l+1} = & \left[\sum_{n=0}^{N_\Omega-1}(w_n^\prime f_n\Gamma_n^4)^m(I_\mathrm{c})_n^{m+1,l}\right]_{i} + \left(c\Delta t \rho \kappa_\mathrm{P}\sum_{n=0}^{N_\Omega-1}w^\prime_n f_n \Gamma_n\right)_i^m (B^\prime)_i^{m+1, l+1} \\
%     & + \left[c\Delta t \rho (\kappa_\mathrm{R}+\kappa_\mathrm{s}-\kappa_\mathrm{P})\sum_{n=0}^{N_\Omega-1}w^\prime_n f_n \Gamma_n\right]_i^m (J^\prime)_i^{m+1, l+1}\, .
% \end{split}
% \end{equation}
We note that this summation was not written correctly in equation (25) of \citetalias{jiang2021ApJS} because it did not take into account the angular dependence of $g_n^+$.
This was later corrected in \citet{jiang2022ApJS}.
We now define another coefficient
\begin{equation}
    h_i^m \equiv \frac{1}{1-\left[c\Delta t \rho (\kappa_\mathrm{F}+\kappa_\mathrm{s}-\kappa_\mathrm{E})\sum_{n=0}^{N_\Omega-1}w^\prime_n f_n \Gamma_n\right]_i^m }\, .
\end{equation}
% \begin{equation}
%     h_i^m \equiv \frac{1}{1-\left[c\Delta t \rho (\kappa_\mathrm{R}+\kappa_\mathrm{s}-\kappa_\mathrm{P})\sum_{n=0}^{N_\Omega-1}w^\prime_n f_n \Gamma_n\right]_i^m }\, .
% \end{equation}
The equation then becomes
\begin{equation}
\begin{split}
    (J^\prime)_i^{m+1, l+1} = & h_i^m\left[\sum_{n=0}^{N_\Omega-1}(w_n^\prime f_n\Gamma_n^4)^m(I_\mathrm{c})_n^{m+1,l}\right]_{i} \\
    & + \left(h c\Delta t \rho \kappa_\mathrm{P}\sum_{n=0}^{N_\Omega-1}w^\prime_n f_n \Gamma_n\right)_i^m (B^\prime)_i^{m+1, l+1}\, .
\end{split}
    \label{eq:jprime}
\end{equation}
Plugging it into the equation for the specific gas internal energy, we get
% \begin{equation}
% \begin{split}
%     (e_\mathrm{g})_i^{m+1,l+1}  & = (e_\mathrm{g})^m_i + \left(4\pi\kappa_\mathrm{E}\Delta t h\right)_i^m\left[\sum_{n=0}^{N_\Omega-1}(w_n^\prime f_n\Gamma_n^4)^m(I_\mathrm{c})_n^{m+1,l}\right]_{i} \\
%     & - \left[4\pi \kappa_\mathrm{P}\Delta t \left(1-h c\Delta t \rho \kappa_\mathrm{E}\sum_{n=0}^{N_\Omega-1}w^\prime_n f_n \Gamma_n\right)\right]_i^m (B^\prime)^{m+1,l+1}_i\, .
% \end{split}
% \end{equation}
\begin{equation}
\begin{split}
    (e_\mathrm{g})_i^{m+1,l+1}  = & (e_\mathrm{g})^m_i + \left(4\pi\kappa_\mathrm{E}\Delta t h\right)_i^m\left[\sum_{n=0}^{N_\Omega-1}(w_n^\prime f_n\Gamma_n^4)^m(I_\mathrm{c})_n^{m+1,l}\right]_{i} \\
    & - \left[4\pi \kappa_\mathrm{P}\Delta t \left(1-h c\Delta t \rho \kappa_\mathrm{E}\sum_{n=0}^{N_\Omega-1}w^\prime_n f_n \Gamma_n\right)\right]_i^m (B^\prime)^{m+1,l+1}_i\, .
\end{split}
\end{equation}
% \begin{equation}
% \begin{split}
%     (e_\mathrm{g})_i^{m+1,l+1}  & = (e_\mathrm{g})^m_i + \left(4\pi\kappa_\mathrm{P}\Delta t h\right)_i^m\left[\sum_{n=0}^{N_\Omega-1}(w_n^\prime f_n\Gamma_n^4)^m(I_\mathrm{c})_n^{m+1,l}\right]_{i} \\
%     & - \left[4\pi \kappa_\mathrm{P}\Delta t \left(1-h c\Delta t \rho \kappa_\mathrm{P}\sum_{n=0}^{N_\Omega-1}w^\prime_n f_n \Gamma_n\right)\right]_i^m (B^\prime)^{m+1,l+1}_i\, .
% \end{split}
% \end{equation}
Taking
\begin{align}
    & b_i^{m+1,l} = (e_\mathrm{g})^m_i + \left(4\pi\kappa_\mathrm{E}\Delta t h\right)_i^m\left[\sum_{n=0}^{N_\Omega-1}(w_n^\prime f_n\Gamma_n^4)^m(I_\mathrm{c})_n^{m+1,l}\right]_{i}\, ,\\
    & k_i^m = - \left[\rho\kappa_\mathrm{P}c\Delta t \left(1-h c\Delta t \rho \kappa_\mathrm{E}\sum_{n=0}^{N_\Omega-1}w^\prime_n f_n \Gamma_n\right)\right]_i^m\, ,\\
    & (e_\mathrm{B})_i^{m+1,l+1} = \frac{a (T_i^{m+1,l+1})^4}{\rho_i^m}\, ,
\end{align}
% \begin{align}
%     & b_i^{m+1,l} = (e_\mathrm{g})^m_i + \left(4\pi\kappa_\mathrm{P}\Delta t h\right)_i^m\left[\sum_{n=0}^{N_\Omega-1}(w_n^\prime f_n\Gamma_n^4)^m(I_\mathrm{c})_n^{m+1,l}\right]_{i}\, ,\\
%     & k_i^m = - \left[\rho\kappa_\mathrm{P}c\Delta t \left(1-h c\Delta t \rho \kappa_\mathrm{P}\sum_{n=0}^{N_\Omega-1}w^\prime_n f_n \Gamma_n\right)\right]_i^m\, ,\\
%     & (e_\mathrm{B})_i^{m+1,l+1} = \frac{a (T_i^{m+1,l+1})^4}{\rho_i^m}\, ,
% \end{align}
we thus arrive at a non-linear equation for $T_i^{m+1,l+1}$ as
\begin{equation}
    y(T) = e_\mathrm{g} (T) - k e_\mathrm{B} (T) - b = 0\, ,
\end{equation}
where $e_\mathrm{g}$ and $e_\mathrm{B}$ are functions of $T_i^{m+1,l+1}$.
The solution can be obtained via the Newton-Raphson method if the specific heat capacity $c_V$ can be directly read from the EOS table to give $\mathrm{d}y/\mathrm{d}T$.
This is a small improvement upon the implementation in \texttt{ATHENA++}, where an ideal gas is assumed for the EOS and a fourth-order polynomial is solved for each iteraction \citepalias{jiang2021ApJS}.
% The derivative of this equation is 
% \begin{equation}
%     \frac{\mathrm{d}y}{\mathrm{d}T} = c_V(T) - \frac{4ak_i^m}{\rho_i^m}T^3\, .
% \end{equation}
With $T_i^{m+1,l+1}$ at hand, we can go back to Equation~\eqref{eq:jprime} to directly obtain $(J^\prime)_i^{m+1, l+1}$, and then to Equation~\eqref{eq:iprime} for $(I^\prime)_{n,i}^{m+1, l+1}$, and finally obtain $I_{n,i}^{m+1, l+1}$ via the Lorentz transformation $I = \Gamma^{-4} I^\prime$.

\subsection{Detailed Scheme for the Radiation Transport Module}
\label{sec:solver:scheme}

In this subsection, we describe the steps the scheme takes to calculate the radiation field.

At the beginning of the simulation, we discretise the angular space in the lab frame based on Section~\ref{sec:solver:ray}, which gives
\begin{equation}
    \vectoraas{n}_n,\ w_n\, ,
\end{equation}
and set the initial radiation field.

% During each hydro timestep, treat the specific intensities as passive scalars and save the updated specific intensities due to the gas velocity advection (Equation~\ref{eq:gasadv})
% \begin{equation}
%     (I_\mathrm{c1})_{n,i}^m\, .
% \end{equation}

% At the end of each hydro timestep, update the gas internal energy, and start the radiation transport solver.
At the end of each timestep, we start the radiation transport solver.
First, we loop over all the active faces between cells, and calculate all the coefficients from the modified HLLE Riemann solver that do not change during the iterations, including (Equations~\eqref{eq:C} and \eqref{eq:D})
\begin{equation}
    C_{n,ij}^m,\ D_{n,ij}^m\, ,
\end{equation}
by which we get
\begin{equation}
    (g^+)_{n,i}^m = \sum_{j=0}^{N_i-1} (C^+)_{n,ij}^m,\ (g^-)_{n,i}^m = \sum_{j=0}^{N_i-1} (C^-)_{n,ij}^m\, .\ 
\end{equation}
For the summation, we need to perform the first flux communication between tasks in the radiation module to add the coefficients to the correct cells.
We save these coefficients for later use
\begin{equation}
    (g^+)_{n,i}^m,\ (g^-)_{n,i}^m,\ C_{n,ij}^m,\ D_{n,ij}^m\, .
\end{equation}
Second, we loop over all the active cells, and calculate all the coefficients that do not change during the iterations, including
\begin{align}
    & \Gamma_{n,i}^m = (1-\vectoraas{n}_n\cdot\vectoraas{\beta}_i^m)/\sqrt{1-(\beta_i^m)^2}\, ,\\
    & (w^\prime)_{n,i}^m = \frac{(\Gamma_{n,i}^m)^{-2} w_n}{\sum_{n=0}^{N_\Omega-1} (\Gamma_{n,i}^m)^{-2} w_n}\, ,\\
    & f_{n,i}^m = \frac{1}{1+\left[g_n^+ + c\Delta t\rho (\kappa_\mathrm{F}+\kappa_\mathrm{s})\Gamma_n\right]_i^m}\, ,\\
    & h_i^m = \frac{1}{1-\left[c\Delta t \rho (\kappa_\mathrm{F}+\kappa_\mathrm{s}-\kappa_\mathrm{E})\sum_{n=0}^{N_\Omega-1}w^\prime_n f_n \Gamma_n\right]_i^m }\, ,\\
    & (f_{I\Sigma})_{n,i}^m = \left[hc\Delta t \rho (\kappa_\mathrm{F}+\kappa_\mathrm{s}-\kappa_\mathrm{E}) f_n \Gamma_n^{-3}\right]_i^m\, ,\\
    & (f_{IB})_{n,i}^m = (h c\Delta t \rho \kappa_\mathrm{P} f_n \Gamma_n^{-3})_i^m\, ,\\
    & (f_{\Sigma I})_{n,i}^m = (w_n^\prime f_n \Gamma_n^4)_i^m\, ,\\
    & (f_{b\Sigma})_i^m = 4\pi (h\Delta t \kappa_\mathrm{E})_i^m\, ,\\
    & k_i^m = \left[c\Delta t \rho\kappa_\mathrm{P}\left(h c\Delta t \rho \kappa_\mathrm{E}\sum_{n=0}^{N_\Omega-1}w^\prime_n f_n \Gamma_n - 1\right)\right]_i^m\, .
\end{align}
% \begin{align}
%     & \Gamma_{n,i}^m = (1-\vectoraas{n}_n\cdot\vectoraas{\beta}_i^m)/\sqrt{1-(\beta_i^m)^2}\, ,\\
%     & (w^\prime)_{n,i}^m = \frac{(\Gamma_{n,i}^m)^{-2} w_n}{\sum_{n=0}^{N_\Omega-1} (\Gamma_{n,i}^m)^{-2} w_n}\, ,\\
%     & f_{n,i}^m = \frac{1}{1+\left[g_n^+ + c\Delta t\rho (\kappa_\mathrm{R}+\kappa_\mathrm{s})\Gamma_n\right]_i^m}\, ,\\
%     & h_i^m = \frac{1}{1-\left[c\Delta t \rho (\kappa_\mathrm{R}+\kappa_\mathrm{s}-\kappa_\mathrm{P})\sum_{n=0}^{N_\Omega-1}w^\prime_n f_n \Gamma_n\right]_i^m }\, ,\\
%     & (f_{I\Sigma})_{n,i}^m = \left[hc\Delta t \rho (\kappa_\mathrm{R}+\kappa_\mathrm{s}-\kappa_\mathrm{P}) f_n \Gamma_n^{-3}\right]_i^m\, ,\\
%     & (f_{IB})_{n,i}^m = (h c\Delta t \rho \kappa_\mathrm{P} f_n \Gamma_n^{-3})_i^m\, ,\\
%     & (f_{\Sigma I})_{n,i}^m = (w_n^\prime f_n \Gamma_n^4)_i^m\, ,\\
%     & (f_{b\Sigma})_i^m = 4\pi (h\Delta t \kappa_\mathrm{P})_i^m\, ,\\
%     & k_i^m = \left[c\Delta t \rho\kappa_\mathrm{P}\left(h c\Delta t \rho \kappa_\mathrm{P}\sum_{n=0}^{N_\Omega-1}w^\prime_n f_n \Gamma_n - 1\right)\right]_i^m\, .
% \end{align}
We save these coefficients for later use
\begin{equation}
    f_{n,i}^m,\ (f_{I\Sigma})_{n,i}^m,\ (f_{IB})_{n,i}^m,\ (f_{\Sigma I})_{n,i}^m,\  (f_{b\Sigma})_i^m,\ k_i^m\, .
\end{equation}

Then, we enter the iteration. The initial values are set to the values at the $m$th timestep.
During each iteration, proceed as follows:
Exchange the primitive radiation variables $I_{n,j}^{m+1, l}$ between tasks.
Loop over all the active cells to get
% \begin{equation}
%     (I_\mathrm{c})_{n,i}^{m+1,l} = (I_\mathrm{c1})_{n,i}^m - (g^-)_{n,i}^m I_{n,i}^{m+1, l}\, .
% \end{equation}
\begin{equation}
    (I_\mathrm{c})_{n,i}^{m+1,l} = (IV)_{n,i}^m/V_i^m - (g^-)_{n,i}^m I_{n,i}^{m+1, l}\, .
\end{equation}
Then, we loop over all the active faces to apply the fluxes from neighboring cells
\begin{equation}
    (I_\mathrm{c})_{n,i}^{m+1,l} = (I_\mathrm{c})_{n,i}^{m+1,l} - \left(\sum_{j=0}^{\Bar{N_i}-1} D_{n,ij}^m I_{n,j}^{m+1, l} \right) - \left(\sum_{j=0}^{M_i-1} D_{n,ij}^m I_{n,j}^m \right)\, .
\end{equation}
This requires another flux communication between tasks.
Afterwards, we loop over all the active cells to calculate
\begin{align}
    & (I_\Sigma)_i^{m+1,l} = \sum_{n=0}^{N_\Omega-1} (f_{\Sigma I})_{n,i}^m(I_\mathrm{c})_{n,i}^{m+1,l}\, ,\\
    & b_i^{m+1,l} = (e_\mathrm{g})^m_i + (f_{b\Sigma})_i^m(I_\Sigma)_i^{m+1,l}\, .
\end{align}
For each active cell, we use the Newton-Raphson method to solve the non-linear equation for $T$
\begin{equation}
    y(T) = e_\mathrm{g} (\rho_i^m, T) - k_i^m e_\mathrm{B} (\rho_i^m, T) - b_i^{m+1,l} = 0\, ,
\end{equation}
which gives the solution $T = T_i^{m+1,l+1}$ and the integrated Planck function $(B^\prime)_i^{m+1,l+1} = ca(T_i^{m+1,l+1})^4/(4\pi)$.
Then calculate the specific intensities as
\begin{equation}
    I_{n,i}^{m+1,l+1} = f_{n,i}^m (I_\mathrm{c})_{n,i}^{m+1,l} + (f_{IB})_{n,i}^m (B^\prime)_i^{m+1,l+1} + (f_{I\Sigma})_{n,i}^m (I_\Sigma)_i^{m+1,l}\, .
\end{equation}
When the maximum iteration number $N_\mathrm{iter}^{\max}$ is reached ($N_\mathrm{iter}^{\max}=10$ is the default for steady-state 3D simulations), or when the convergence criterion is reached, i.e. \citepalias{jiang2021ApJS}
\begin{equation}
    \frac{\sum_{i=0}^{N_\mathrm{p}-1}\sum_{n=0}^{N_\mathrm{\Omega}-1}|I_{n,i}^{m+1,l+1} - I_{n,i}^{m+1,l}|}{\sum_{i=0}^{N_\mathrm{p}-1}\sum_{n=0}^{N_\mathrm{\Omega}-1}|I_{n,i}^{m+1,l+1}|} < \epsilon = 10^{-8}\, \mathrm{(default)}\,,
    \label{eq:criterion}
\end{equation}
we stop the iteration and set $I_{n,i}^{m+1}$.
See Section~\ref{sec:disc:code:criterion} for a detailed discussion of the criterion and alternative choices.
This global value of the convergence criterion requires another communication between tasks.

Finally after we exit the iteration, we update the conserved quantities
\begin{equation}
    (IV)_{n,k}^{m+1} = 
    \begin{cases}
    I_{n,k}^{m+1}V_k^m, & \mathrm{cell\ k\ is\ active}\, ,\\
    (IV)_{n,k}^m + \Delta(IV)_{n,k}^{m+1}, & \begin{aligned}
    & \mathrm{cell\ k\ is\ passive\ but} \\
    & \mathrm{has\ adjacent\ active\ cells}\, ,
    \end{aligned}
    \end{cases}
\end{equation}
where $\Delta(IV)_{n,k}^{m+1}$ is the change in the conserved quantity for the passive cell due to the exchanged fluxes between the passive cell and neighboring active cells.
This step requires another flux communication between tasks just to apply the fluxes to passive cells.
This way, we guarantee that the conserved quantity $IV$ and its integrated values, e.g., radiation energy and radiation flux, are indeed conserved.
But the local time-stepping may still introduce artificial energy transport or momentum transport between cells while conserving the total conserved quantities.\footnote{In particular, in the free-streaming regime, local time-stepping may yield accumulating radiation near the active-passive cell boundaries.
This is tested in Section~\ref{sec:test:rd:bubble}, where we do not see such artifacts, but more complex simulations are needed to test the free-streaming regime with local time-stepping.}
In total, within each timestep, the radiation module performs $N_\mathrm{iter}$ primitive radiation variable exchanges between tasks, $N_\mathrm{iter}+2$ flux exchanges between tasks, and $1$ convergence criterion exchange between tasks, where $N_\mathrm{iter}$ is the number of iterations.
Within each variable or flux exchange between tasks, $\mathcal{O}(N_\Omega)$ variables are exchanged per active cell or per active face.

\subsection{Coupling to the MHD Solver via the Radiation Source Terms}
\label{sec:solver:couple}

The major difference between post-processing with radiative transfer and full radiation hydrodynamics is that the radiation field needs to be coupled back to the MHD solver for the energy and momentum exchange between radiation and fluid.
There are two ways to express the radiation source terms.
One is to calculate the radiation source terms as they appear in the radiation transport equations.
The other is to express them in terms of time derivatives and the transport terms, which conserves the total energy and total momentum.

By default, we use the conservative form of the radiation source terms.
The change in $(IV)$ due to the transport terms are
\begin{equation}
    (\Delta_\mathrm{tr} IV)_{n,i}^{m+1}  = -(g^+)_{n,i}^m I_{n,i}^{m+1}V_i^m + (I_\mathrm{c})_{n,i}^{m+1}V_i^m - (IV)_{n,i}^m \, .
\end{equation}
Therefore, we have
\begin{equation}
    (IV)_{n,i}^{m+1} - (IV)_{n,i}^m - (\Delta_\mathrm{tr} IV)_{n,i}^{m+1}  = (g^+)_{n,i}^m I_{n,i}^{m+1}V_i^m - (I_\mathrm{c})_{n,i}^{m+1}V_i^m \, .
\end{equation}
This yields the radiation source terms as
\begin{align}
    & (S_\mathrm{r,E}V)_i^{m+1} \begin{aligned}[t] 
        & = (E_\mathrm{r}V)_i^{m+1} - (E_\mathrm{r}V)_i^m - (\Delta_\mathrm{tr} E_\mathrm{r}V)_i^{m+1} \\
        & = \frac{4\pi}{c}V_i^m\sum_{n=0}^{N_\Omega-1} \left[(g^+)_{n,i}^m I_{n,i}^{m+1} - (I_\mathrm{c})_{n,i}^{m+1}\right] w_n\, ,
        \end{aligned}\\
    & (\vectoraas{S}_\mathrm{r,p}V)_i^{m+1} \begin{aligned}[t] 
        & = \frac{1}{c^2}\left[(\vectoraas{F}_\mathrm{r}V)_i^{m+1} - (\vectoraas{F}_\mathrm{r}V)_i^m - (\Delta_\mathrm{tr} \vectoraas{F}_\mathrm{r}V)_i^{m+1}\right] \\
        & = \frac{4\pi}{c^2}V_i^m\sum_{n=0}^{N_\Omega-1} \left[(g^+)_{n,i}^m I_{n,i}^{m+1} - (I_\mathrm{c})_{n,i}^{m+1}\right] \vectoraas{n}_n w_n\, .
        \end{aligned}
\end{align}
At the end of each hydro timestep, we perform the radiation transport calculations, and only for the active cells do we subtract the source terms from the total gas energy and total gas momentum of the equations to get the conserved quantities.
Then we update the primitive variables for the active cells in the MHD part, and the primitive variables for the radiation module including opacities and intensities.

\subsection{Optional Features}
\label{sec:solver:option}

For the boundary conditions, our current implementation supports periodic and inflow/outflow boundaries.
To use the periodic boundary condition for the radiation module, the hydrodynamic boundary has to be set as periodic as well, which applies to most scenarios.
The inflow/outflow boundary condition for the radiation module is independent of the hydrodynamic boundary.
However, the radiation inflow boundary has to be set manually in the source code, as there is no universal way to inject radiation into the computational domain.
For an example that uses a mixture of periodic and inflow/outflow boundaries, see test~\ref{sec:test:steady:shadow}.
For another example that injects radiation in the middle of the simulation box, see Section~\ref{sec:test:rhd:rsg}.
We currently do not support reflective boundary condition, as radiation is rarely reflective and this is more difficult to realize with an angle-dependent discrete ordinates method.

We offer an optional feature to solve the \textit{time-independent} radiation transport equation with the implicit discrete ordinates solver.
This is done by neglecting the time derivative term, gas velocities, face velocities, and setting $\Delta t = 1$ in the factors and coefficients.
Specifically, we solve for
\begin{equation}
    c \vectoraas{n}\cdot \vectoraas{\nabla} I = c \rho\left[(\kappa_\mathrm{P} B - \kappa_\mathrm{E} J) +(\kappa_\mathrm{F}+\kappa_\mathrm{s})(J - I)\right] \, .
    \label{eq:rt:notime}
\end{equation}
By default, this is used to set the initial conditions for the radiation field, where we assume the radiation field starts to evolve from an initial equilibrium state.
It can also be used to directly compare with the ray-tracing results, and to provide the Eddington tensor for a VET scheme.
It can be further exploited for post-processing, in particular if multi-group radiation transport is feasible in the future.
For this time-independent version, we also implement an option to couple the radiation source terms to the MHD solver as heating/cooling terms and radiation forces.

Moreover, we implement different options to test various numerical choices for specific problems.
% For example, the gas advection term can be solved implicitly, which renders the whole scheme fully implicit.
For example, the gas advection term can be solved explicitly as done in \citetalias{jiang2021ApJS}, which is expected to suppress numerical diffusion from gas advection but only works for global time-stepping.
The coupling source terms can be solved in the non-conservative form, although caution is needed when using this with local time-stepping.
The radiation transport step can be solved without coupling it to the gas temperature, although this typically creates large errors in our tests.
Finally, it is possible to run the radiation solver only on globally synchronised timesteps despite using local timesteps for MHD.
The timesteps taken in the radiation module are replaced by the globally synchronised timestep instead of the local timestep of the cell or the face.
% However, this will likely cause problems when the radiation-dominated cells hold the small timebins.
This will become a poor approximation when gas dynamics changes the radiation field on the timescales of MHD processes.

% \end{strip}

% \end{widetext}
% \twocolumn

\section{Test Problems}
\label{sec:test}

Test problems are essential for debugging new schemes and demonstrating the capabilities and limitations of numerical schemes.
Here, we repeat most of the test problems in \citetalias{jiang2021ApJS} for direct comparisons between \texttt{AREPO} and \texttt{ATHENA++} results, with a few additional tests for broader contexts.
Most of those tests can also be found in the method papers for the VET scheme \citep{davis2012ApJS, jiang2012ApJS} and the semi-implicit discrete ordinates scheme \citep{jiang2014ApJS} for \texttt{ATHENA++}.

We perform all the tests on the moving Voronoi mesh except for test~\ref{sec:test:rhd:shock}.
For 2D tests, the mesh is initialized in a honeycomb-like configuration (Figure~\ref{fig:test:2d}).
This is done by constructing the cell centers in a Cartesian grid first, and then shifting every second row by half the cell size.
In the evolving radiation tests and radiation hydrodynamic tests, we switch on local time-stepping.
For the cases where the background medium is fixed or too homogeneous for automatic local time-stepping, we force different timesteps in different parts of the computational domain, as illustrated in Figure~\ref{fig:test:2d}.
For outflow boundary conditions, we set the incoming intensities to be zero and copy the outgoing intensities from the last active cells to the ghost cells, as in \citetalias{jiang2021ApJS}.
The radiation constant $a$ and the speed of light $c$ are sometimes set to other values for exploring different regimes easily, which are specified in different tests.
For all the tests, we do not distinguish between the flux-weighted opacity $\kappa_\mathrm{F}$ and the Rosseland opacity $\kappa_\mathrm{R}$, and also do not distinguish between the energy-weighted opacity $\kappa_\mathrm{E}$ and the Planck opacity $\kappa_\mathrm{P}$.

\begin{figure}[htb!]
\centering
\includegraphics[width=\columnwidth]{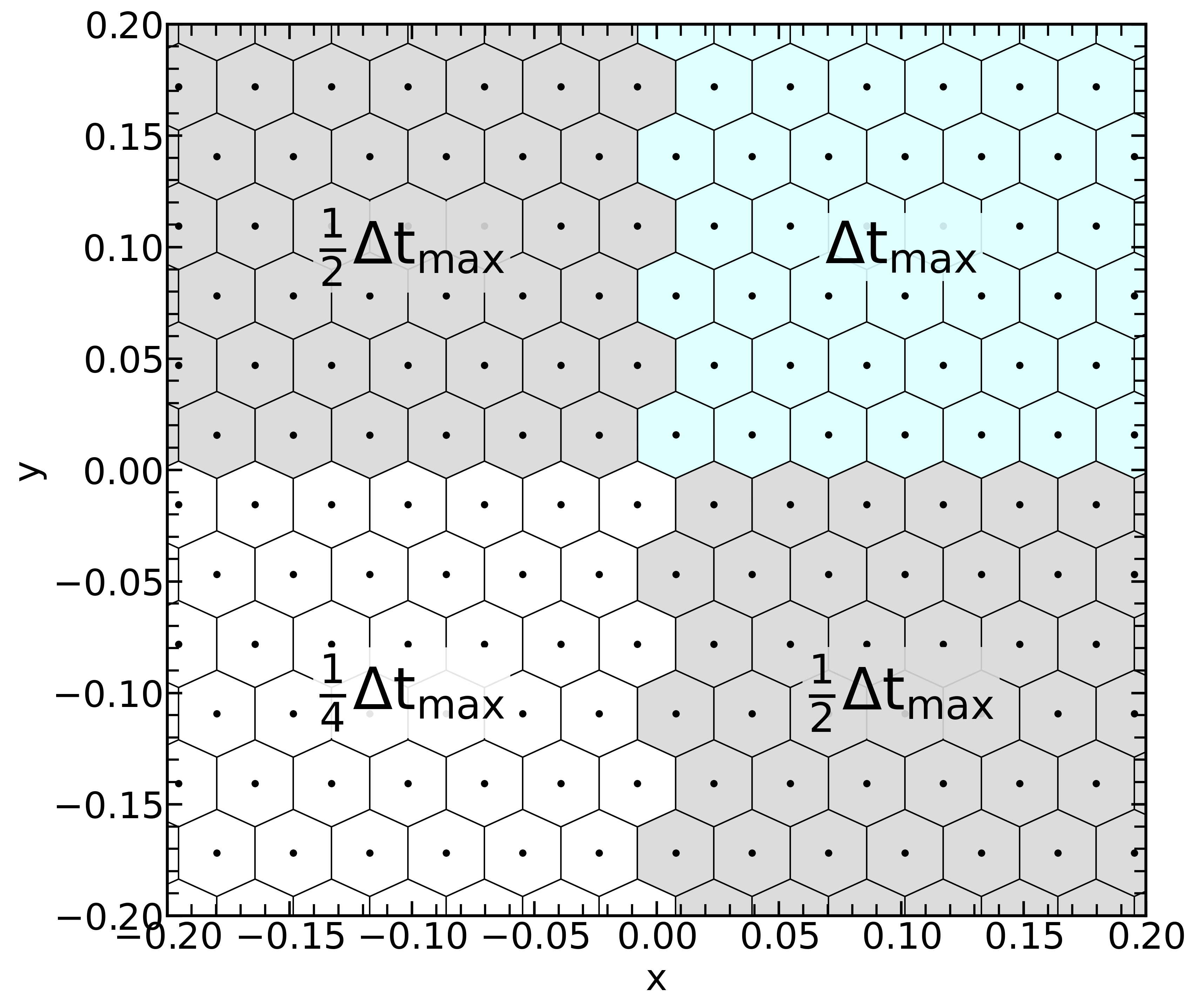}
\caption{
Illustration of the forced local time-stepping on the initial honeycomb mesh in 2D tests.
We use the hierarchical time-stepping in \texttt{AREPO}.
The top right octant holds the normal timestep $\Delta t_{\max}$ allowed by either maximum timestep limit or CFL condition for hydrodynamics.
The diagonal octants are updated with timesteps $\Delta t_{\max}/2$, and the bottom left octant is updated with a timestep $\Delta t_{\max}/4$.
}
\label{fig:test:2d}
\end{figure}

We normally adopt the angular discretization as described in Section~\ref{sec:solver:ray} for 1D and 3D tests.
However, for 2D tests, we use another set of angular bins for better coverage of the angular space.
When we also need to keep the Eddington tensor to be $1/3$ for isotropic radiation in 2D, we set up the directions in two groups.
One group is confined in the 2D plane ($n_z=0$) and uniformly distributed on a unit circle where $n_x=\cos\theta, n_y=\sin\theta$.
The other group shares the same set of $\theta$ as the first group, but has $n_x=\sqrt{1/3}\cos\theta, n_y=\sqrt{1/3}\sin\theta,n_z=\sqrt{2/3}$.
When the Eddington tensor constraint is not needed for a purely optically-thin medium (e.g., Section~\ref{sec:test:steady:thin}), we only use the first group for the angular discretization.

We present the test results based on their categories:
\begin{enumerate}%[noitemsep]
    \item We first test whether the code can capture the radiation field in a steady state with a fixed background (Section~\ref{sec:test:steady}) by solving the \textit{time-independent} radiation transport equation, where we set the transport terms to exactly balance the source terms.
    We use the time-independent flavour (Section~\ref{sec:solver:option}) of our scheme for these tests.
    The tests explore different regimes, including 2D optically-thin medium, 1D optically-thick-to-thin medium with pure absorption, pure scattering, and a standard 2D shadow test.
    \item Afterwards, we test the evolving radiation field with a fixed background (Section~\ref{sec:test:rd}) by solving the time-dependent radiation transport equation.
    We demonstrate 2D dynamic diffusion for optically-thick regime and 2D expanding multi-source radiation bubble test for optically-thin-dominated regime.
    \item Then we test the coupling between radiation and gas by performing radiation hydrodynamic tests (Section~\ref{sec:test:rhd}).
    We explore 2D thermal equilibrium, 2D linear waves, 1D radiation shocks, and 2D irradiated clouds.
\end{enumerate}

% Finally, we present more sophisticated radiation hydrodynamic simulations.
% % These include 2D irradiated clouds (Section~\ref{sec:test:rhd:cloud}), 2D dust-driven wind (Section~\ref{sec:test:rhd:wind}), and 3D convective envelope of a red supergiant star (Section~\ref{sec:test:rhd:rsg}).
% These include 2D irradiated clouds (Section~\ref{sec:test:rhd:cloud}) and 3D convective envelope of a red supergiant star (Section~\ref{sec:test:rhd:rsg}).

Most of the tests can be directly compared with the figures in \citetalias{jiang2021ApJS}, since we almost always adopt the same setup.
% Some of the tests and their variants also appear in \citet{davis2012ApJS, jiang2012ApJS, jiang2014ApJS}.
However, we use the honeycomb mesh for initial 2D mesh configuration, and switch on the moving mesh with local time-stepping for the second (evolving radiation) and third categories of tests (radiation hydrodynamics).
In particular, except for tests in Section~\ref{sec:test:rd:diff}, \ref{sec:test:rhd:te} and \ref{sec:test:rhd:wave}, FLD and M1 will not produce accurate results for all the other tests due to the inaccurate representation for the direction of radiation.
These are demonstrated in e.g. \citet{sadowski2013MNRAS}, \citet{murchikova2017MNRAS},  \citet{thomas2022MNRAS} and \citet{menon2022MNRAS} for some of the tests.

\subsection{Steady-state Tests with Fixed Gas Background}
\label{sec:test:steady}

\subsubsection{2D Crossing Beams in Optically-thin Vacuum}
\label{sec:test:steady:thin}

\begin{figure*}[htb!]
\centering
% First row
    \begin{subfigure}{0.3\textwidth}
        \centering
        \includegraphics[width=\linewidth]{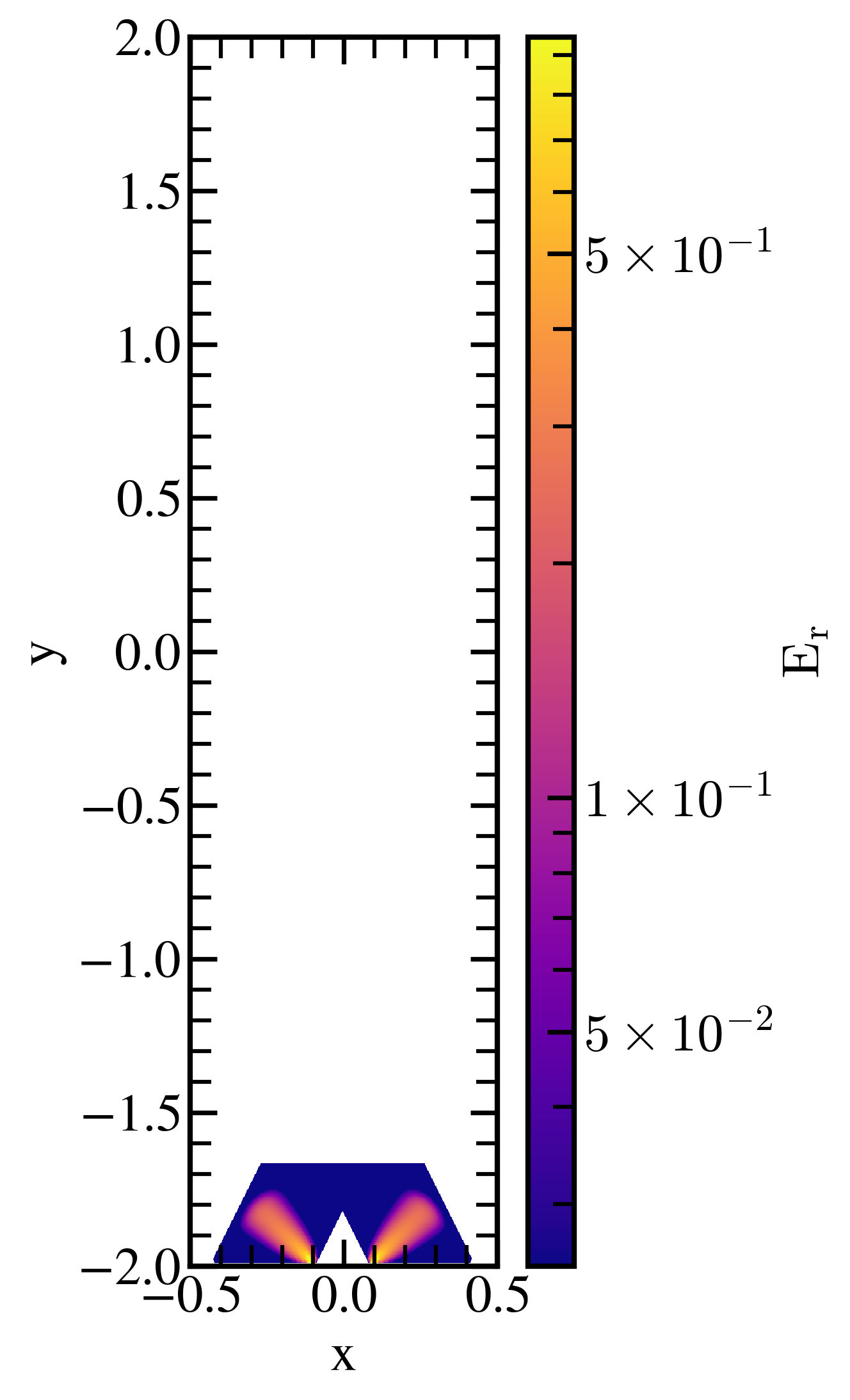}
        \caption{After 20 iterations with resolution\\ $64\times 256$.}
    \end{subfigure}
    \hfill
    \begin{subfigure}{0.3\textwidth}
        \centering
        \includegraphics[width=\linewidth]{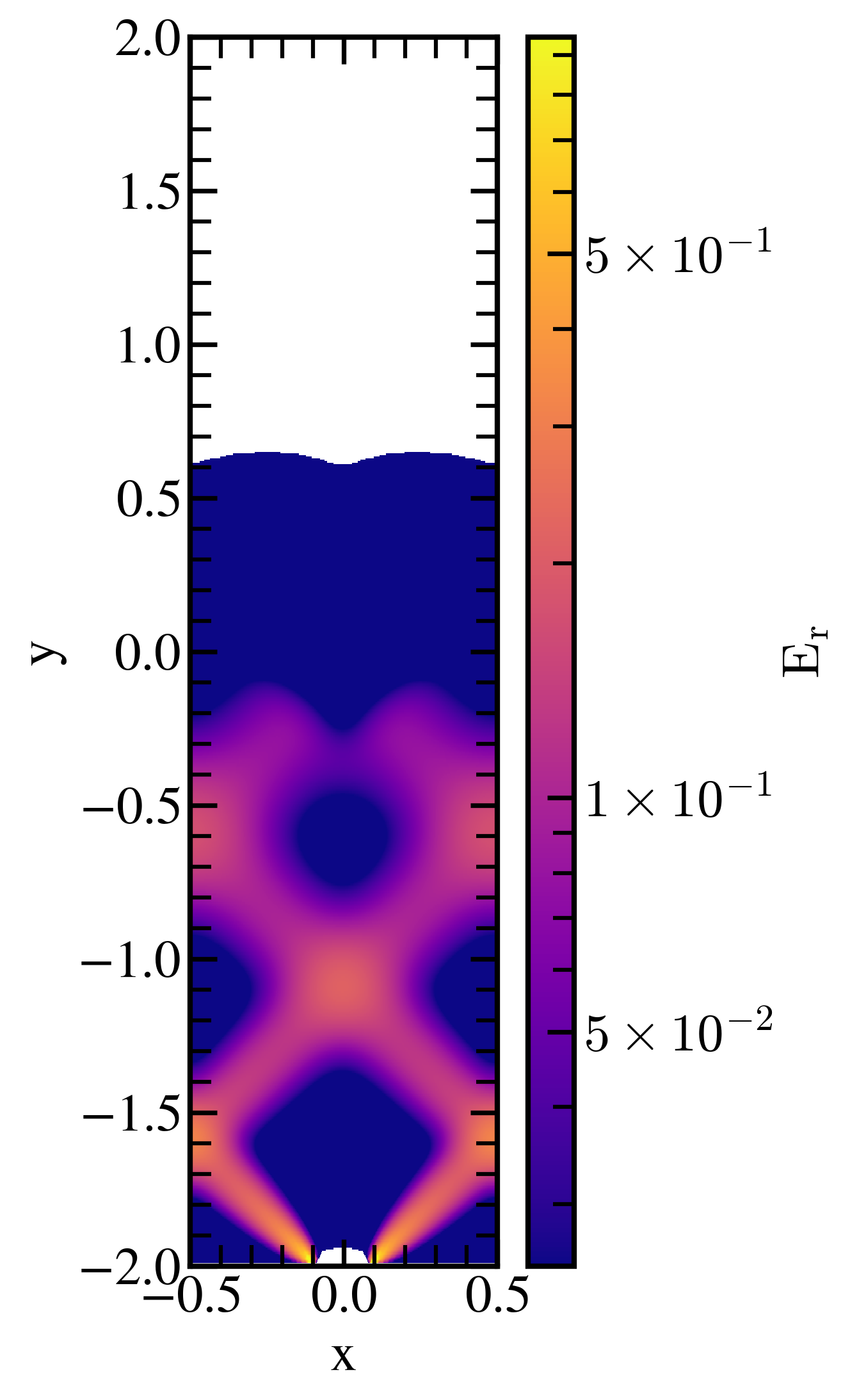}
        \caption{After 200 iterations with resolution\\ $64\times 256$.}
    \end{subfigure}
    \hfill
    \begin{subfigure}{0.3\textwidth}
        \centering
       \includegraphics[width=\linewidth]{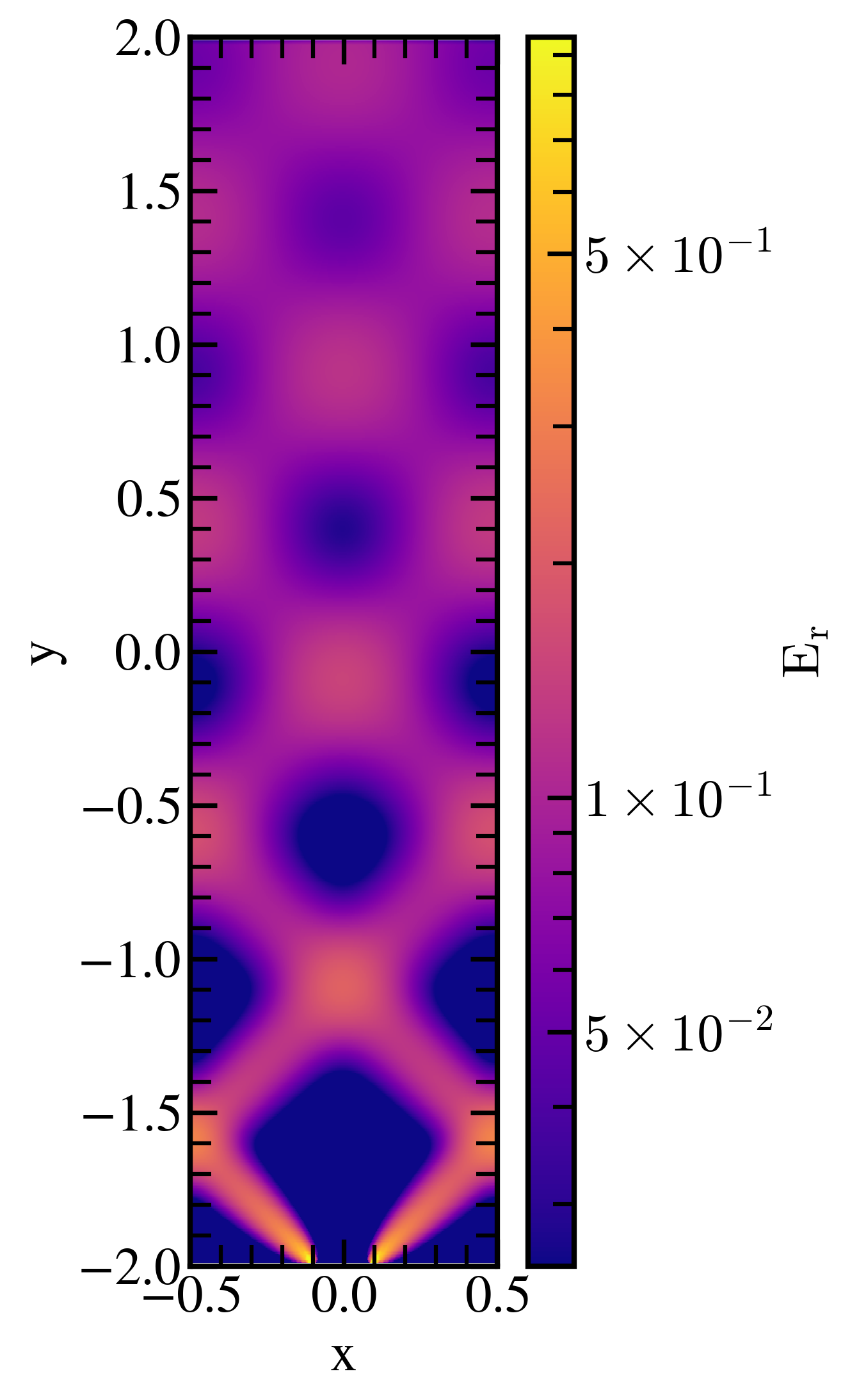}
        \caption{After 2000 iterations with resolution\\ $64\times 256$.}
    \end{subfigure}
    % \vspace{1cm}
    % Second row
    \begin{subfigure}{0.3\textwidth}
        \centering
        \includegraphics[width=\linewidth]{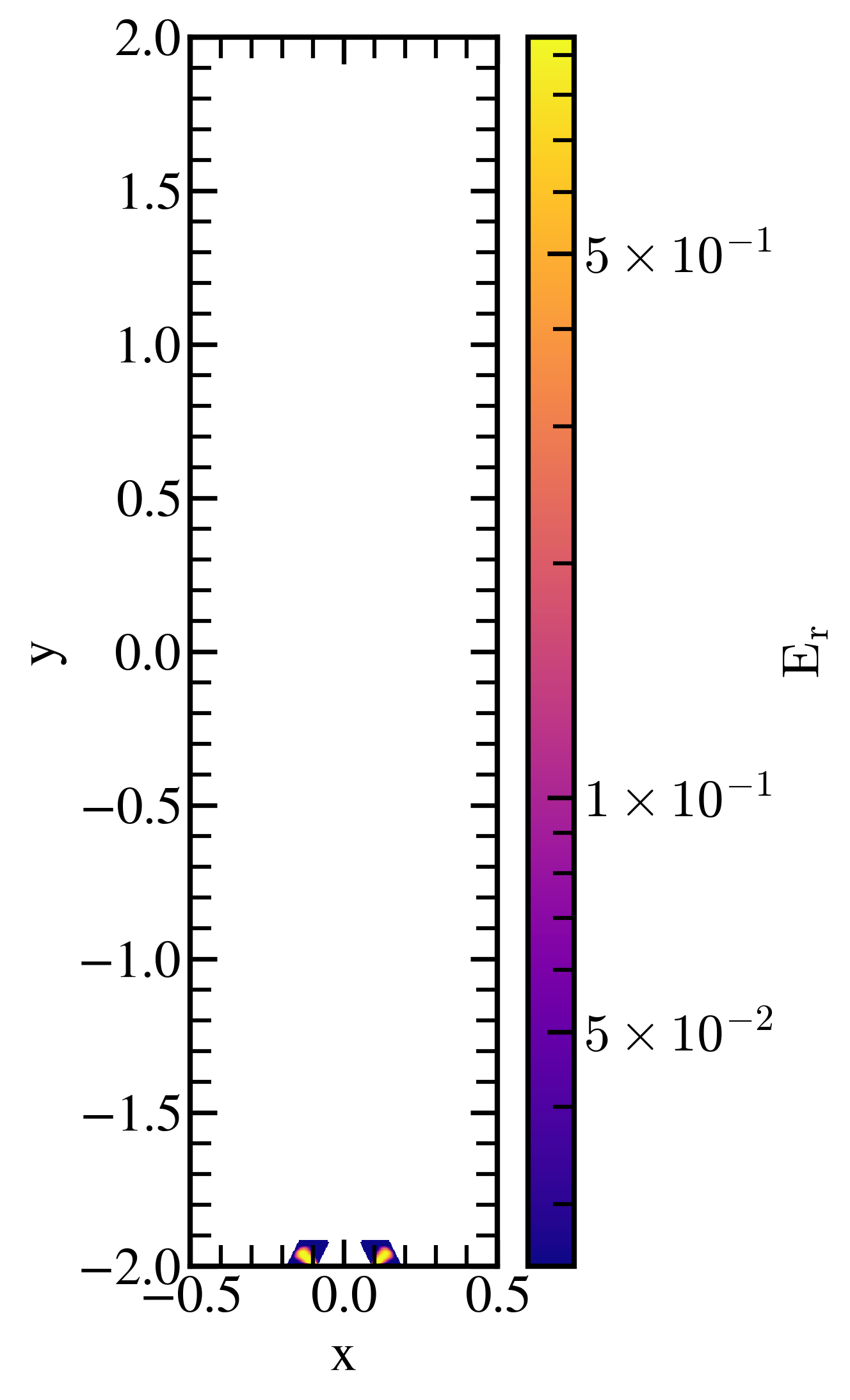}
        \caption{After 20 iterations with resolution\\ $256\times 1024$.}
    \end{subfigure}
    \hfill
    \begin{subfigure}{0.3\textwidth}
        \centering
        \includegraphics[width=\linewidth]{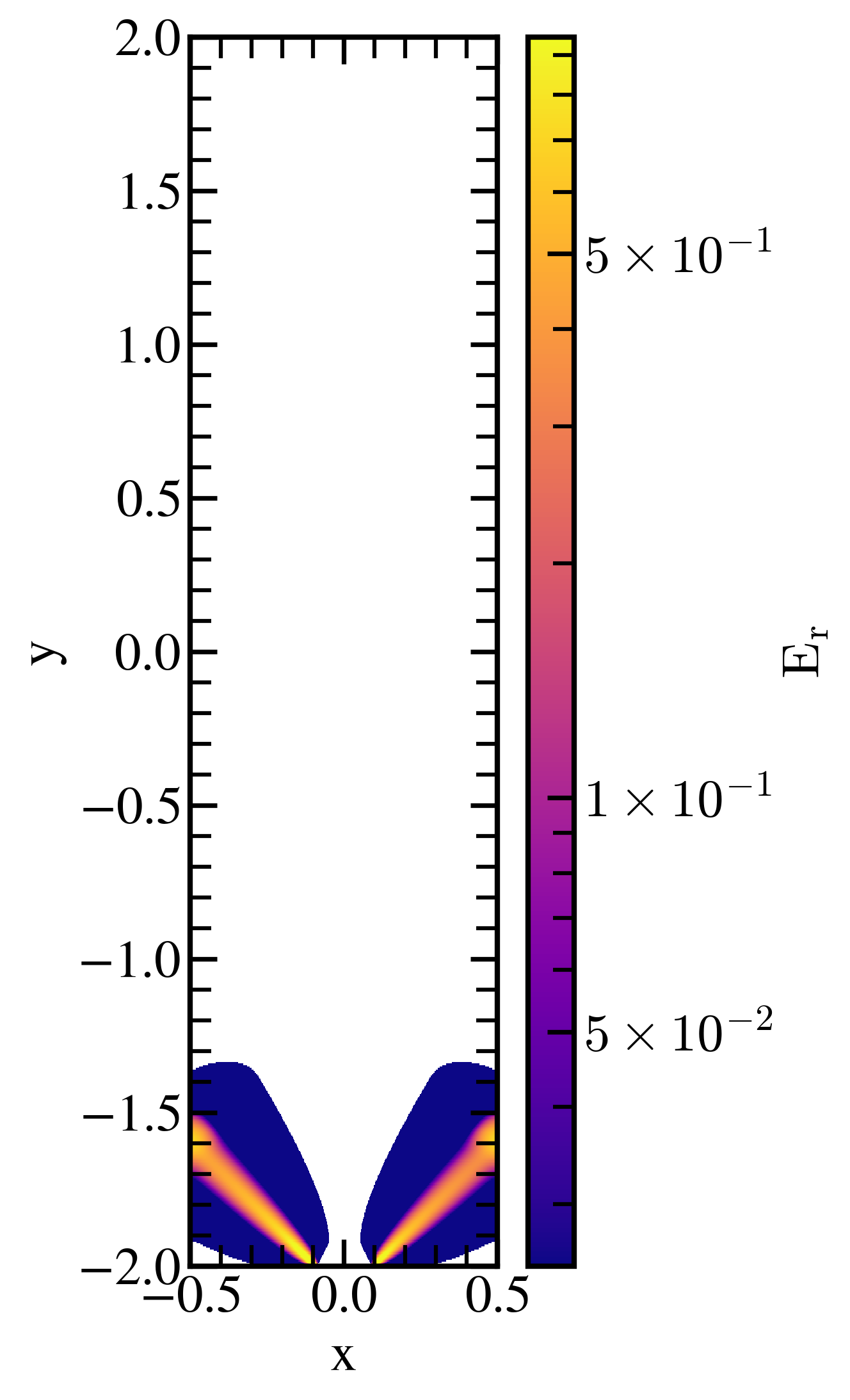}
        \caption{After 200 iterations with resolution\\ $256\times 1024$.}
    \end{subfigure}
    \hfill
    \begin{subfigure}{0.3\textwidth}
        \centering
        \includegraphics[width=\linewidth]{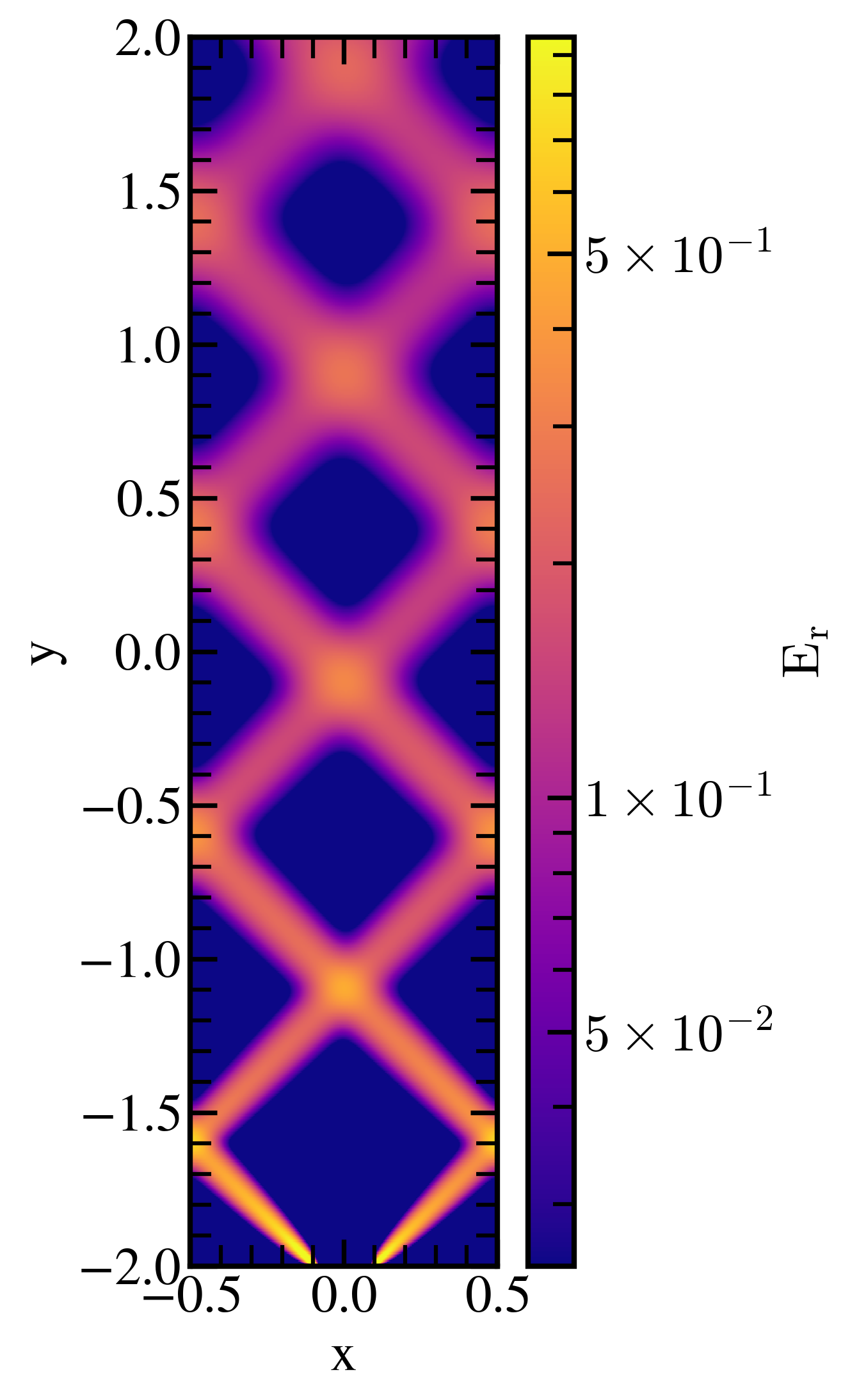}
        \caption{After 2000 iterations with resolution\\ $256\times 1024$.}
    \end{subfigure}
% \begin{figure*}[htb!]
% % \centering
% \gridline{
% \fig{cb_20.png}{0.3\textwidth}{(a) After 20 iterations with resolution $64\times 256$.}
% \fig{cb_200.png}{0.3\textwidth}{(b) After 200 iterations with resolution $64\times 256$.}
% \fig{cb_2000.png}{0.3\textwidth}{(c) After 2000 iterations with resolution $64\times 256$.}
% }
% \gridline{
% \fig{cb_hr_20.png}{0.3\textwidth}{(d) After 20 iterations with resolution $256\times 1024$.}
% \fig{cb_hr_200.png}{0.3\textwidth}{(e) After 200 iterations with resolution $256\times 1024$.}
% \fig{cb_hr_2000.png}{0.3\textwidth}{(f) After 2000 iterations with resolution $256\times 1024$.}
% }
\caption{
2D crossing-beam tests with different spatial resolutions (rows) and different numbers of iterations (columns).
The color shows the radiation energy density distribution when two light beams are injected from the bottom of the simulation box.
White indicates that the radiation energy density is zero as the initial condition.
From left to right, different columns show increasing iterations.
Information of the radiation field approximately propagates through one layer of cells per one or two iterations, so the iteration needed also depends on resolution, as demonstrated in Figure (b) compared to (e).
The light beams spread due to numerical diffusion when propagating through the box.
Figure (a-c) uses the same resolution as figure 3 in \citetalias{jiang2021ApJS} for direct comparison.
Figure (d-f) increases the resolution fourfold in both dimensions, thereby reducing the numerical diffusion.
}
\label{fig:test:steady:thin}
\end{figure*}

This test is designed to show how many iterations it takes for our scheme to converge in the optically thin medium.
As already noted in \citetalias{jiang2021ApJS}, the iterative solver easily converges when the source terms dominate over the transport terms because the matrix of the coefficients is diagonally-dominant, which is typically the case in the optically-thick regime.
However, in the optically-thin regime, we need many iterations for the information of the radiation field to propagate through the computational domain.
Here, we show that if we have approximately $N$ cells per dimension (so that the total cell number is $\mathcal{O}(N^3)$ in 3D simulations), the safe choice would be to run $2N$ iterations to ensure a correct radiation field in the optically-thin regime.

To illustrate this extreme case, we set up a 2D box spanning $[-0.5,0.5]\times[-2.0,2.0]$ with two light beams injected from the bottom.
We choose an angular resolution of 4 rays confined in the 2D plane, and two different spatial resolutions of $64\times 256$ and $256\times 1024$.
All the opacities are set to zero throughout the simulation for a vacuum background, and the intensities are also initialized to be zero.
For the beams at the bottom boundary, we set the specific intensities at $x=-0.1,y=-2.0,n_x=-1/\sqrt{2},n_y=1/\sqrt{2}$ and $x=0.1,y=-2.0,n_x=1/\sqrt{2},n_y=1/\sqrt{2}$ to be $0.8c$.
All the other intensities at the bottom boundary are set to be zero.
We use a periodic boundary in the x direction and an outflow boundary at the top.
The constants are set to be $a=1,c=10^3$, but the values do not matter for the test.

In Figure~\ref{fig:test:steady:thin}, we show the radiation energy density distribution for the crossing-beam tests.
Two different spatial resolutions are shown in two rows, and three columns are for $20,\, 200,\, 2000$ iterations.
Panel (c) adopts the same resolution as \citetalias{jiang2021ApJS}, and thus can be directly compared with their figure 3.
As illustrated in panel (f) compared with (c), increasing the resolution fourfold in both dimensions reduces the numerical diffusion, thereby suppressing the spreading of light beams.
This was also demonstrated in figure 6 in \citet{davis2012ApJS} for the short-characteristic scheme and figure 6 in \citet{jiang2014ApJS} for the semi-implicit discrete ordinates scheme.
Furthermore, we illustrate the propagation of radiation with increasing iterations from left to right.
The sharp boundaries in the left column are likely a mesh effect from the background honeycomb mesh.
Even after $200$ iterations (middle column), the radiation field is still not converged since the information is not propagated to the top of the box.
Our tests suggest that in the optically-thin regime, radiation only propagates through approximately one layer of cells in one to two iterations.
Therefore, for the optically-thin radiation to be correctly captured, simulations with higher spatial resolution would also require more rounds of iterations.

This test shows that our scheme can correctly capture the direction of radiation in the optically-thin regime.
If other moment-based method were applied to this problem, the FLD method will fill the domain with a more uniform radiation field, whilst the M1 method will merge the two crossing beams into one \citep[e.g. figure 3 in ][]{asahina2020ApJ}.
We also illustrate that increasing spatial resolution suppresses the numerical diffusion of our scheme, albeit at the cost of more rounds of iterations in the optically-thin regime.

\subsubsection{1D Absorption-dominated Homogeneous Sphere in Spherical Coordinates}
\label{sec:test:steady:sphere}

With this test, we demonstrate the transition from the optically-thick to the optically-thin region with pure black body emission and absorption.
It also serves as a showcase for our scheme in 1D spherical coordinates for future tests on radiation hydrodynamic simulations of 1D stars, winds, accretion or shock waves, etc.

To construct the initial condition for this test, we adopt exactly the same setup as in \citetalias{jiang2021ApJS}.
We use 1000 mesh points covering $r\in[0.05,7]$ in 1D spherical coordinates and 40 directions for the radiation module.
We setup a homogeneous sphere of radius $R_0=1$, density $\rho_0=1$, temperature $T_0=1$, and absorption opacity $\kappa_\mathrm{R}=\kappa_\mathrm{P}=10$ with zero scattering opacity.
Outside of the sphere, we set zero opacities such that the background is effectively vacuum to radiation.
We initialize the intensities to be the isotropic value $caT_0^4/(4\pi)$.
We fix the intensities as their isotropic value at the left boundary, and use an outflow boundary on the right.
The constants are set to be $a=1,c=100$, but the values do not matter for the test.

\begin{figure}[htb!]
\centering
\includegraphics[width=\columnwidth]{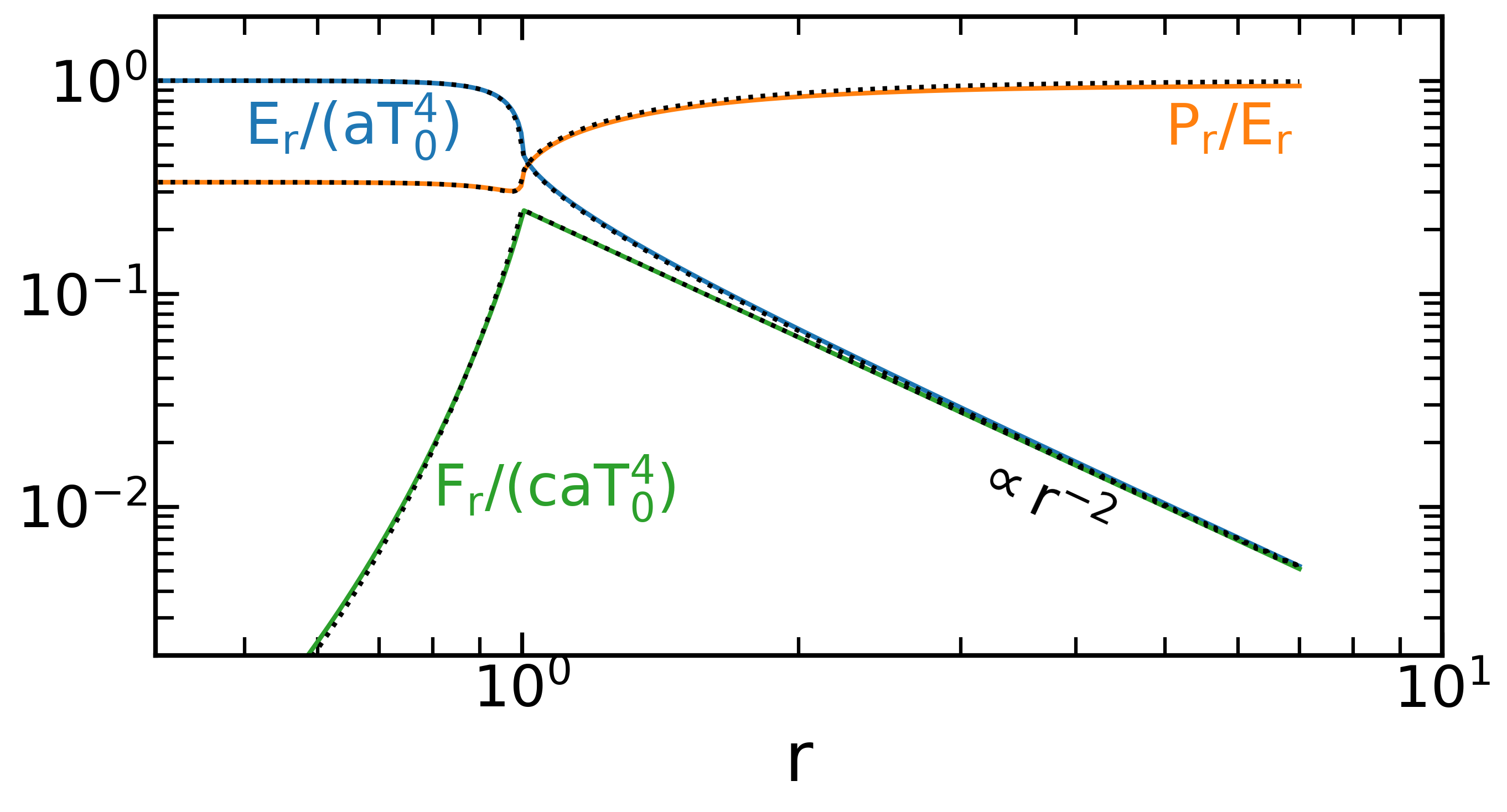}
\caption{
1D absorption-dominated homogeneous sphere test.
Different solid colored lines show the profiles of normalized radiation energy density $E_\mathrm{r}/(aT_0^4)$, normalized radiation flux  $F_\mathrm{r}/(caT_0^4)$, and the radial component of the Eddington tensor $P_\mathrm{r}/E_\mathrm{r}$ as functions of the radius $r$.
All of the simulated results align with the analytical solutions indicated by black dotted lines.
Divided by $r=1$, the radiation field transits from radiative diffusion in LTE inside the optically-thick sphere to free-streaming in the optically-thin vacuum.
}
\label{fig:test:steady:sphere}
\end{figure}

For our scheme to work in 1D spherical coordinates, we use a specific set of angular discretization and take care of the additional transport terms arising from the curvature of the coordinate system, as described in Section 3.2.4 in \citetalias{jiang2021ApJS}.
Since we directly take this part of the code from the public version of \texttt{ATHENA++}\footnote{\href{https://github.com/PrincetonUniversity/athena}{https://github.com/PrincetonUniversity/athena}}, we do not describe the method again, but refer the interested readers to \citetalias{jiang2021ApJS} and the \texttt{ATHENA++} code.

There is an analytical solution to this problem \citep{smit1997A&A}.
The specific intensities depend on the radial coordinate $r$ and the directional unit vector $\vectoraas{n}$ as
\begin{equation}
    I(r,\vectoraas{n}) = \frac{caT_0^4}{4\pi}\left\{1-\exp\left[\rho_0\kappa_\mathrm{R} s(r,n_r)\right]\right\}\, ,
\end{equation}
where $n_r$ is the radial component of $\vectoraas{n}$ and
\begin{align}
    & s(r,n_r) = 
    \begin{cases}
    rn_r+R_0 g(r,n_r), & r<R_0\, ,\\
    2R_0 g(r,n_r), & r\geq R_0\, \&\, \sqrt{1-(R_0/r)^2}\leq n_r\leq 1\, ,\\
    0, & \mathrm{otherwise}\, ,
    \end{cases}\\
    & g(r,n_r)=\sqrt{1-(r/R_0)^2(1-n_r^2)}\, .
\end{align}
In Figure~\ref{fig:test:steady:sphere}, we show the profiles describing the radiation field inside and outside of the homogeneous sphere, which fits well with the analytical solutions (black dotted lines).
Inside the optically-thick sphere, the radiation follows radiative diffusion in LTE, where the radiation temperature is the same as the gas temperature, with the radiation pressure being $1/3$ of the radiation energy density and a radiation flux proportional to the radiation energy density gradient.
Outside of the photosphere at $r>R_0$, the radiation field transits to free-streaming, where the profiles asymptotically approach $P_\mathrm{r}=E_\mathrm{r}$ and $F_\mathrm{r}=cE_\mathrm{r}\propto r^{-2}$, which indicates a constant luminosity.
The radiation field takes 1000 iterations to pass through the 1000 mesh points, which supports the conclusion we reached in Section~\ref{sec:test:steady:thin} that radiation only propagates through one cell in one iteration.

This test illustrates that our scheme correctly simulates the radiation field in the transition zone between optically-thick and optically-thin regions, which is crucial for simulations of stellar atmospheres, transients, and disk photospheres.
For this test, the M1 closure method always overestimates the Eddington tensor near the photosphere \citep{smit1997A&A, oconnor2015ApJS, murchikova2017MNRAS, anninos2020ApJ}, indicating cautions when simulating immediate surroundings of stellar photospheres or transients using M1 radiation transport.

\subsubsection{1D Scattering-dominated Non-LTE Atmosphere}
\label{sec:test:steady:nlte}

This test aims to show the transition from optically-thick to optically-thin region when the medium is dominated by scattering.
When the source term is dominated by scattering, the intensity is also influenced by radiation coming from other directions, such that standard ray-tracing breaks down and many rounds of iterations are needed to capture the radiation field properly.

We use exactly the same setup as in \citetalias{jiang2021ApJS} by adopting 1280 mesh points covering $x\in [-10,10]$ in 1D and 8 directions for the radiation module.
We fix the density profile to be $\rho=10^{-3}\exp(10-x)$ and a uniform temperature field $T=T_0=1$.
The opacities are also set uniformly in the computational domain, with absorption opacity $\kappa_\mathrm{R}=\kappa_\mathrm{P}=\epsilon$ and scattering opacity $\kappa_\mathrm{s}=1-\epsilon$, where $\epsilon$ is a small value for the scattering-dominated atmosphere.
We initialize the intensities to be the isotropic value $caT_0^4/(4\pi)$.
We fix the intensities as their isotropic value at the inner boundary of the atmosphere, and use an outflow boundary for the outer boundary of the atmosphere.
The constants are set to be $a=1,c=1$, but the values do not matter for the test.

\begin{figure}[htb!]
\centering
\includegraphics[width=\columnwidth]{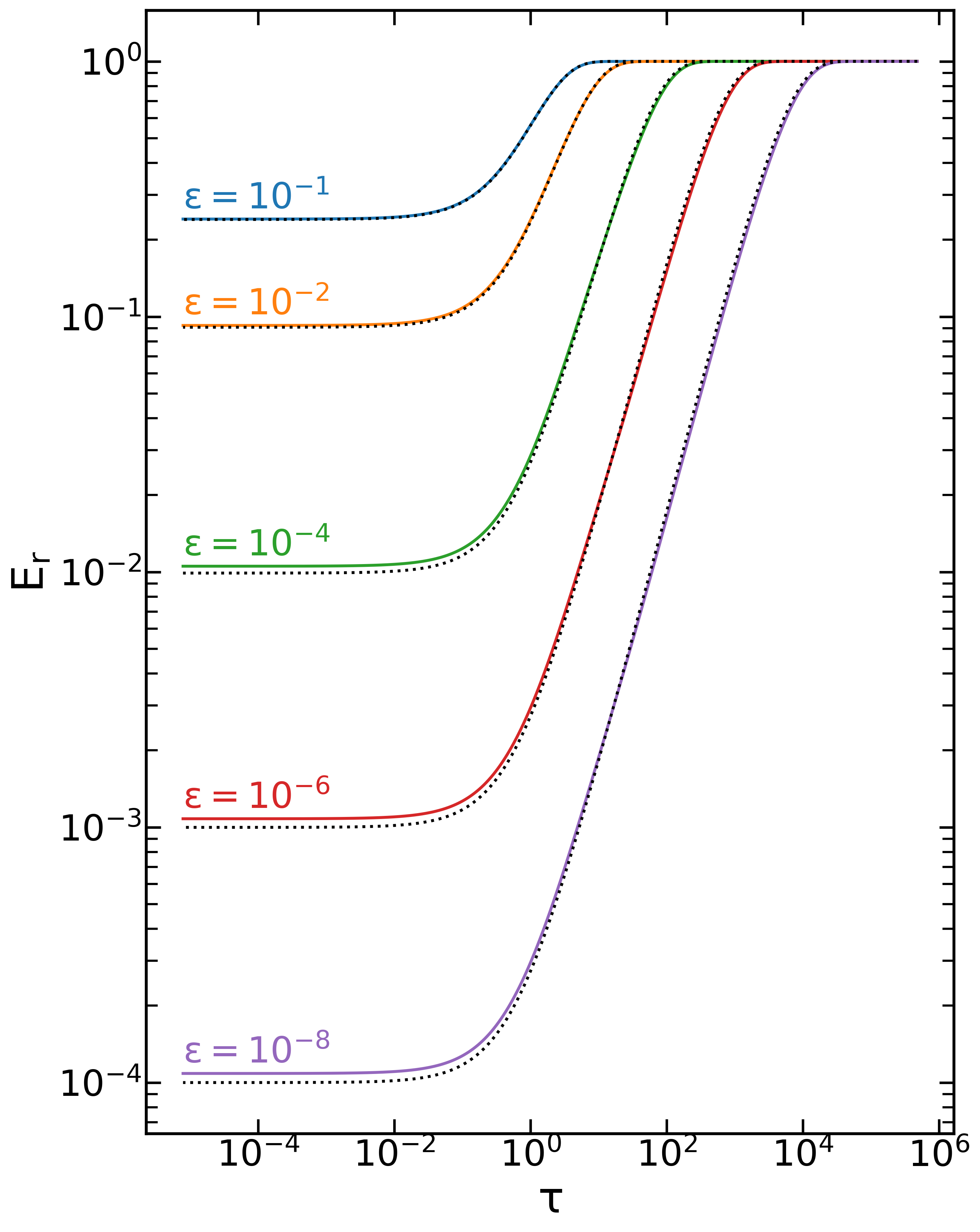}
\caption{
1D scattering-dominated non-LTE atmosphere tests for different absorption opacity $\epsilon$ and scattering opacity $1-\epsilon$.
The solid lines show the radiation energy density $E_\mathrm{r}$ as functions of the optical depth $\tau$, transiting from LTE in the optically-thick region (right) to the non-LTE optically-thin atmosphere (left).
The dotted black lines show the analytical solutions.
As the atmosphere becomes more scattering-dominant (top to bottom indicated by different colors), the radiation temperature deviates from the gas temperature more in a more extended region.
This test also demonstrates that significant scattering can reduce the radiation coming from an optically-thick region.
}
\label{fig:test:steady:nlte}
\end{figure}

The analytical solution to this problem is \citepalias{jiang2021ApJS}
\begin{equation}
    E_\mathrm{r} = aT_0^4\left[1-\frac{\exp(-\sqrt{3\epsilon}\tau)}{1+\sqrt{\epsilon}}\right]\, ,
\end{equation}
where the optical depth from the right boundary is $\tau = \int_x^{10}\rho(\kappa_\mathrm{R}+\kappa_\mathrm{s})dx=\int_x^{10}\rho dx=10^{-3}[\exp(10-x)-1]$.
In Figure~\ref{fig:test:steady:nlte}, we show the radiation energy density as a function of the optical depth.
The solid lines show the simulation results, with different opacities indicated in different colors.
They agree well with the analytical solutions shown in black dotted lines.
The radiation transits from LTE in the optically-thick region (right) to the non-LTE optically-thin atmosphere (left) where the radiation has a much smaller temperature than the gas.
As noted in \citet{davis2012ApJS}, \citet{jiang2014ApJS} and \citetalias{jiang2021ApJS}, greater deviation from LTE requires significantly more rounds of iterations.
For example, our scheme needs about $10^3$ iterations for the $\epsilon=10^{-1}$ case to converge to the analytical solution, while the $\epsilon=10^{-8}$ case needs about $10^6$ iterations.
This is similar to the finding in \citetalias{jiang2021ApJS}, where they showed that the discrete ordinates scheme is slower at convergence for this particular test compared to short characteristics.

With this test, we show that our scheme produces the correct radiation field in scattering-dominated medium regardless of the optical thickness, but is rather expensive in the optically-thin regime.
Such scenario is important for systems involving dust scattering, e.g. wind driving in cool stars \citep{habing1996A&AR} and observational signatures of protoplanetary disks \citep{zhu2019ApJ}.

\subsubsection{2D Shadow Test}
\label{sec:test:steady:shadow}

The shadow test is a classic test to demonstrate the difference between different radiation transport methods, in particular FLD, M1, and other more accurate methods.
FLD and M1 do not give the correct answer to this problem \citep[e.g., figure 13 in ][]{menon2022MNRAS}.
For FLD, the radiation is directed along the radiation energy density gradient, thereby filling the shadow.
M1 treats the radiation like fluid, so the two radiation beams collide and merge into one, and non-stationary radiation vortices are sometimes observed behind the optically-thick object.

We setup the simulation box similar to \citetalias{jiang2021ApJS}.
The box covers $[-0.5,0.5]\times[-0.3,0.3]$ with two parallel rays cast from the left.
We use a spatial resolution of $512\times 256$ and 24 angles consisting of two groups of angular sets to ensure correct Eddington tensors in both optically-thick and optically-thin regime.
We describe the elliptical optically-thick cloud and the near-vacuum background with a density distribution of $\rho(x,y)=1+9/\{1+\exp[10((x/0.1)^2+(y/0.06)^2-1)]\}$, a temperature distribution $T(x,y)$ that creates uniform pressure $P=\rho T=1$, and a pure absorption opacity $\kappa_\mathrm{R}=\kappa_\mathrm{P}=\rho T^{-3.5}$.
The density and temperature background is therefore $\rho_0=1, T_0=1$.
This ensures that the total optical depth is approximately one across the background but reaches more than $10^5$ across the cloud \citepalias{jiang2021ApJS}.
We initialize the intensities to be the isotropic value $caT^4/(4\pi)$.
From the left boundary, we inject two parallel rays of radiation by setting intensities along the directions $\pm 15^\circ$ confined in the 2D plane to be $1031.3ca$ for each ghost cell, set other intensities to be zero, and use an outflow boundary condition for the right boundary.
We adopt periodic boundary conditions in the vertical direction.
The constants are set to be $a=10^{-7},c=1.9\times 10^5$, but the values do not matter for the test.
We set the maximum iteration number to be $2000$.

\begin{figure}[htb!]
\centering
\includegraphics[width=\columnwidth]{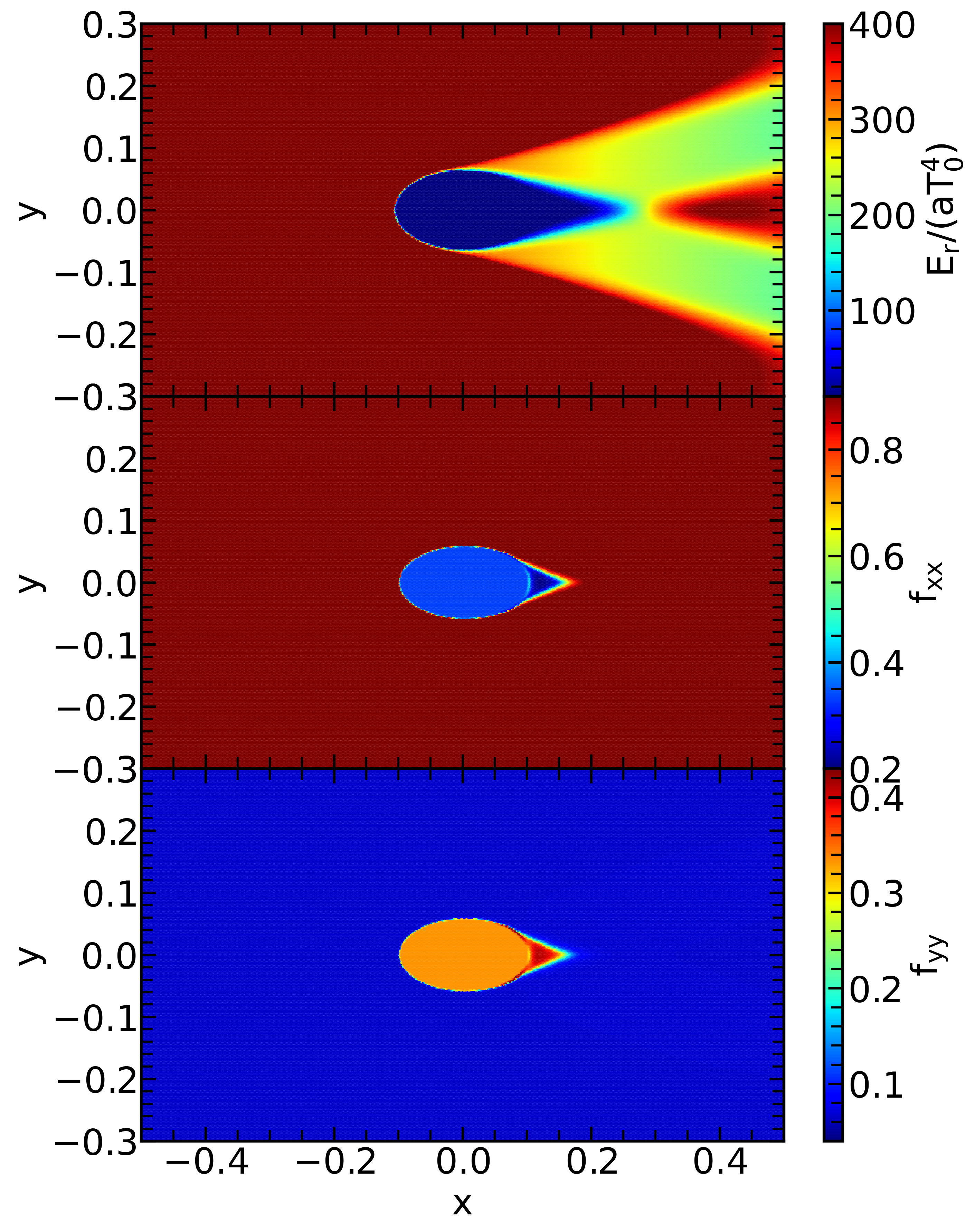}
\caption{
2D shadow test.
We inject two parallel rays of light at $\pm 15^\circ$ from the left, which cast a shadow behind the central optically-thick cloud.
The top, middle, and bottom panels show the nomalized radiation energy density $E_\mathrm{r}/(aT_0^4)$, component of the Eddington tensor along xx direction $f_{xx}$, and component of the Eddington tensor along yy direction $f_{yy}$, respectively.
Inside the optically-thick cloud, the radiation field is nearly isotropic such that the diagonal components of the Eddington tensor are $1/3$.
The radiation field becomes more anisotropic in the shadow.
Outside the shadow in the background, the radiation follows the trajectories of injected rays with a small amount of numerical diffusion.
}
\label{fig:test:steady:shadow}
\end{figure}

In Figure~\ref{fig:test:steady:shadow}, we show the relaxed radiation field when an optically-thick cloud is immersed in two parallel rays of light at $\pm 15^\circ$.
The top, middle, and bottom panels illustrate the normalized radiation energy density, the xx component, and the yy component of the Eddington tensor, respectively.
In the optically-thin background, the radiation energy follows the injected value, and the direction follows the injected rays nearly-parallel to the x axis.
In the optically-thick cloud, the radiation temperature is in equilibrium with the gas temperature of the cloud, and the radiation field is nearly isotropic with an Eddington tensor of $1/3$ on the diagonal.
Immediately behind the cloud, a triangular shadow is formed.
Two extended wings at $\pm 15^\circ$ can also be found behind the cloud with half of the injected radiation energy density, because the injected rays coming from the same direction are blocked by the cloud.
In the middle of the two wings, the injected rays are not blocked by the cloud, and therefore the radiation energy density holds the injected value as the unblocked background.
We illustrate in this test that the shadow behind optically-thick objects is correctly captured in our scheme, and the results are almost identical to \citetalias{jiang2021ApJS}.

\subsection{Evolving-radiation Tests with Fixed Gas Background}
\label{sec:test:rd}

\subsubsection{2D Dynamic Diffusion}
\label{sec:test:rd:diff}

This test is an essential test to demonstrate if the scheme can capture radiative diffusion in the optically-thick regime even in a moving medium.
If the fluxes are not carefully calculated for scattering-dominated regime, the numerical diffusion can easily dominate over the physical diffusion, resulting in a much shorter diffusion time-scale \citep{jiang2013ApJ, jiang2014ApJS, jiang2021ApJS}.
This is especially true for the discrete ordinates method, since the equations are not by design the diffusion equations as solved in FLD.
The test also shows that exploiting local time-stepping does not change the diffusion time-scale or introduce any discontinuous artifacts in the radiation field.

We use the same setup in \citetalias{jiang2021ApJS} but extend the box in 2D, because we aim to perform the test on a moving Voronoi mesh, which can only be constructed in 2D and 3D.
We adopt a slim box with a spatial resolution of $1280\times 16$ covering $[-10,10]\times[-0.125,0.125]$.
We use the default angular discretization not specific to 2D with $8$ directions to reduce the computational cost, since the radiation field is almost isotropic.
The uniform background is fixed with a constant density $\rho=1$, a constant temperature $T=1$, a constant horizontal velocity $v^x=1$, a constant scattering opacity $\kappa_\mathrm{s}=4\times 10^4$ and zero absorption opacities.
The intensities are initialized to be the isotropic values associated with a radiation energy density of Gaussian-peak shape $E_\mathrm{r}(x,y)=\exp(-40x^2),x\in (-0.5,0.5)$ with a floor of $E_\mathrm{r}(x,y)=\exp(-10),|x|\geq 0.5$.
We use periodic boundary conditions for all boundaries.
The constants are set to be $a=1,c=1000$ such that the diffusion time-scale of the peak $t_\mathrm{diff}\sim \rho\kappa_\mathrm{s}/c=40$ is comparable to the gas advection time-scale $t_\mathrm{adv}\sim L^x/v^x=20$ where $L^x=20$ is the horizontal size of the box.
We find that the numerical parameter in the signal speed (Equation~\eqref{eq:alpha}) needs to be increased to $\alpha=20$ to obtain accurate results in 2D.
We manually apply a three-level hierarchy of local time-stepping as illustrated in Figure~\ref{fig:test:2d}.

\begin{figure}[htb!]
\centering
\includegraphics[width=\columnwidth]{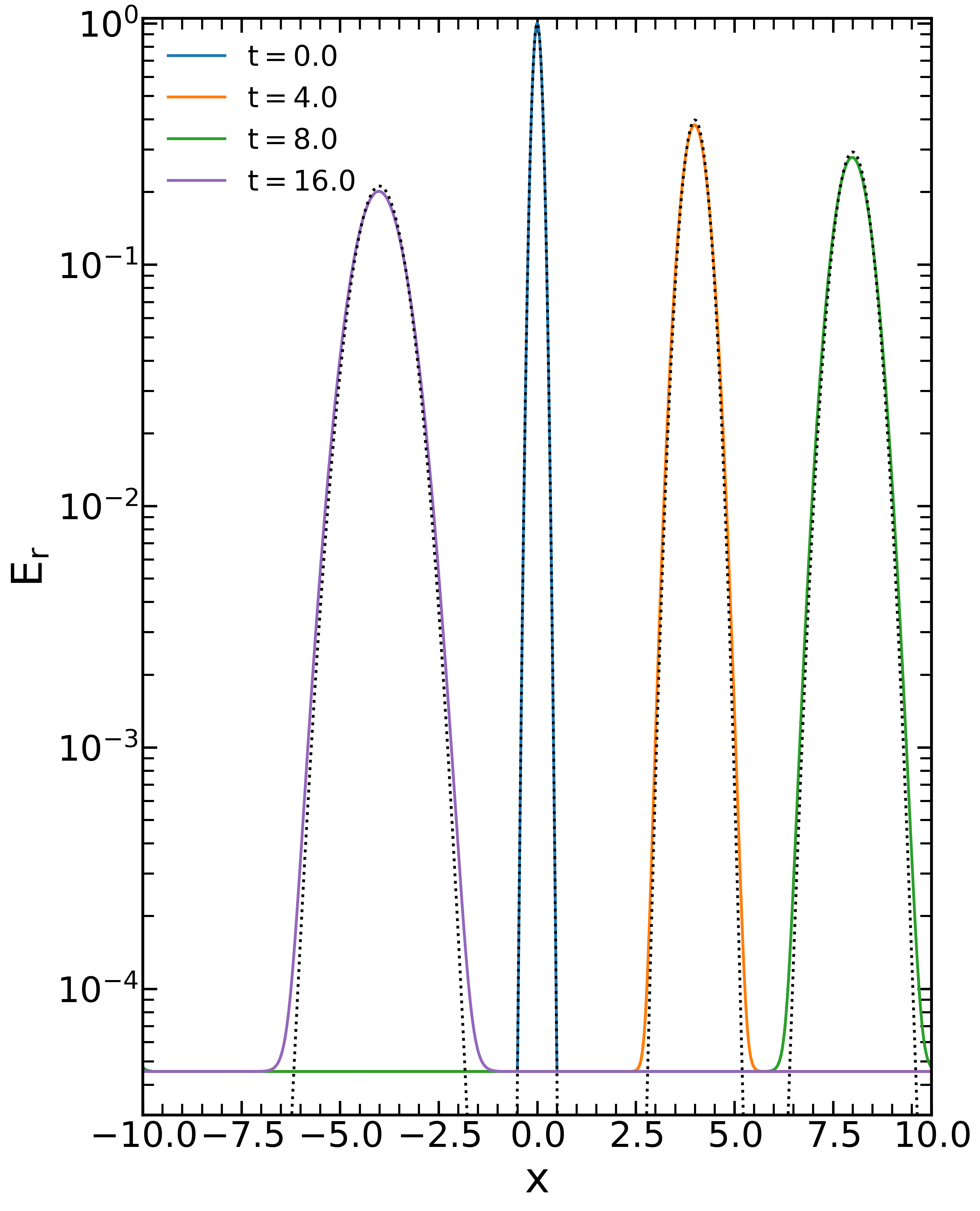}
\caption{
2D dynamic diffusion test.
The plot shows the radiation energy density $E_\mathrm{r}$ as a function of the x coordinate at different times.
We initialize the radiation field as an energy density Gaussian peak at the center (blue).
As the radiation field evolves, the peak is dragged along with the moving fluid, and in the mean time spreads due to radiative diffusion.
The shapes of the peaks agree well with the analytical solutions (black dotted lines), which suggests that the dynamic diffusion is well-captured by our scheme even on an unstructured moving mesh with forced local time-stepping.
}
\label{fig:test:rd:diff}
\end{figure}

The dynamic diffusion of the radiation peak is shown in Figure~\ref{fig:test:rd:diff} for time $t=0,4,8,16$ in different colors.
We sort the 2D data points according to their x coordinates, and plot the radiation energy density $E_\mathrm{r}$ as a function of x.
Initially, the radiation is concentrated at the box center as a narrow Gaussian peak (blue).
As the radiation field evolves in the uniformly moving background gas, the radiation peak is advected with the gas at the gas speed $v^x=1$.
Meanwhile, the peak spreads due to radiative diffusion in the optically-thick medium.
The profiles of the radiation peaks at different times coincide with the analytical solutions (black dotted lines).
For local time-stepping, we do not observe artifacts at the boundaries between regions that hold different timesteps, and the diffused shape of the peak is not changed by forced local time-stepping.
Therefore, we have demonstrated that dynamic diffusion can be correctly captured in our scheme, even on an unstructured moving mesh with forced local time-stepping.

\subsubsection{2D Expanding Multi-source Radiation Bubbles}
\label{sec:test:rd:bubble}

With this test, we aim to show the propagation of radiation in the optically-thin region emitted by multiple optically-thick radiation sources, especially when the radiation only propagates through a small portion of the simulation box in one timestep.
Such setup provides an idealized showcase motivated by the test in \citet{thomas2022MNRAS}, with implications for the ionizing radiation in star-forming regions or in the epoch of reionization.
This test also demonstrates how the numerical diffusion spreads the radiation in optically-thin region, and how local time-stepping does not jeopardize the results.

We construct the initial condition in a 2D box of $[-0.5,0.5]\times[-0.5,0.5]$ with a spatial resolution of $1024\times 1024$ and $200$ directions consisting of two groups of angular sets.
For radiation sources, we setup 20 optically-thick homogeneous spheres with varying radii and temperatures.
Their locations, radii, and temperatures seem to be randomly distributed, but are actually determined by deterministic functions (available at Zenodo: doi: \href{https://doi.org/10.5281/zenodo.15032138}{10.5281/zenodo.15032138}).
% as detailed in Appendix~\ref{apx:bubble:ic}.
This is for the test to be reproducible with other methods, an improvement we made upon the test in \citet{thomas2022MNRAS}.
The whole domain is in pressure balance with a uniform pressure $P=1$, and the density is determined from $\rho=P/T$, where the temperature $T(x,y)$ is different in each sphere and in the background.
The background is fixed to have a temperature of $10^{-3}$ and zero opacity, such that it is effectively vacuum to radiation.
Inside each homogeneous sphere, we set the absorption opacity to satisfy $\kappa_\mathrm{R}=\kappa_\mathrm{P}=10^5/\rho$ with zero scattering opacity, such that the optical depth across each sphere is at least $10^3$.
We initialize the intensities to be the isotropic value $caT^4/(4\pi)$.
All boundaries are set to be outflow for radiation, and we only evolve the radiation field.
We choose the constants to be $a=1,c=10$, such that the light-crossing time for the box is $t_\mathrm{c}=0.1$ and the CFL timestep is approximately $\Delta t_\mathrm{CFL}=4\times 10^{-5}$ for a CFL number of $0.4$ when the signal speed is the speed of light. We set the maximum iteration number to be $200$ per timestep.

\begin{figure*}[htb!]
\centering
% First row
    \begin{subfigure}{0.49\textwidth}
        \centering
        \includegraphics[width=\linewidth]{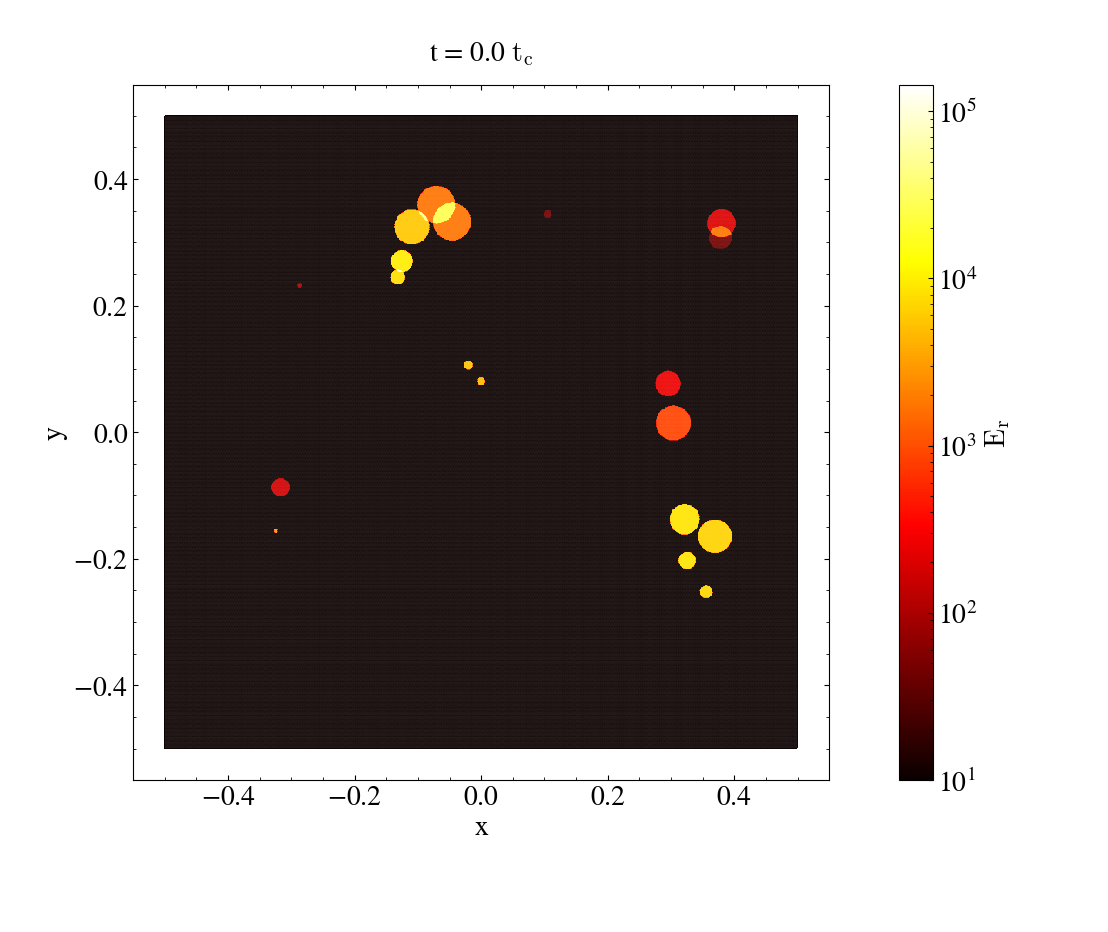}
        \caption{\begin{minipage}{0.7\linewidth} \centering \vspace{0.4cm} Initial condition. \vspace{0.4cm} \end{minipage}}
    \end{subfigure}
    \hfill
    \begin{subfigure}{0.49\textwidth}
        \centering
        \includegraphics[width=\linewidth]{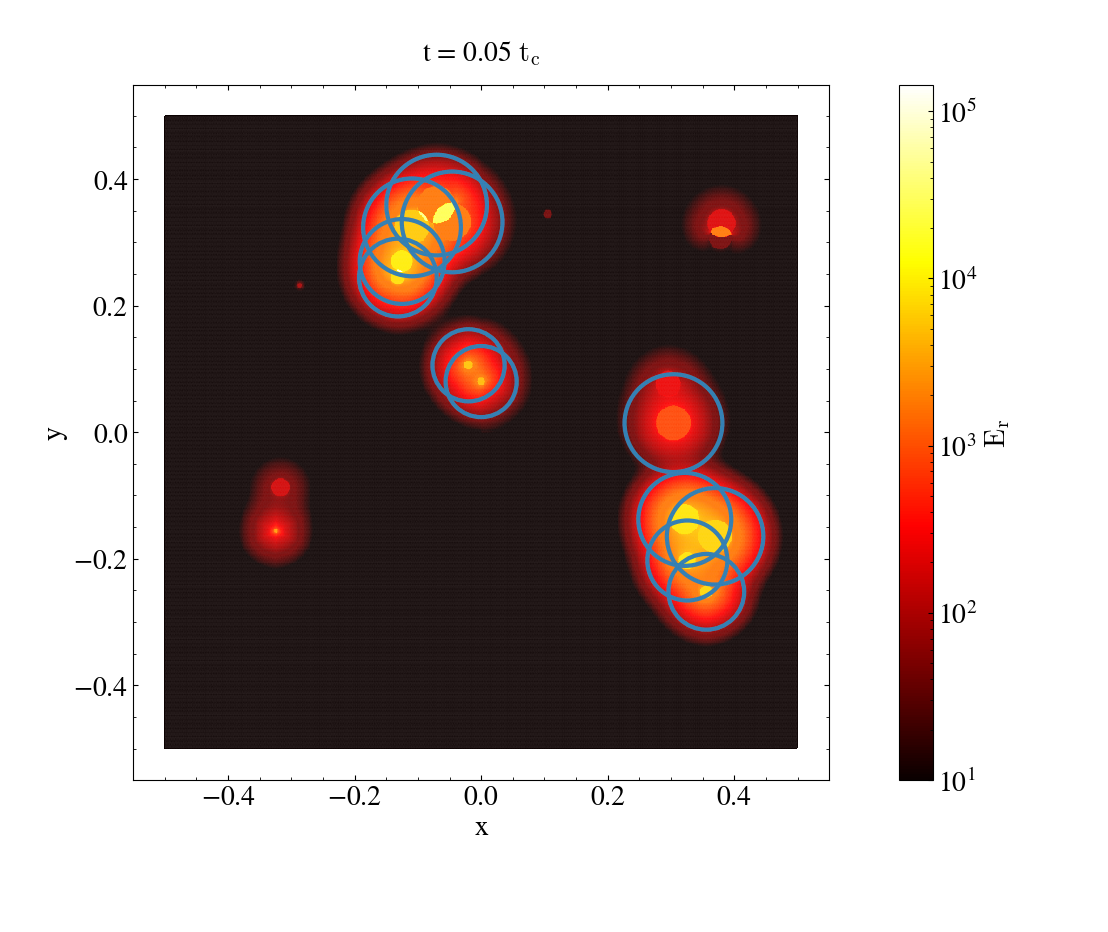}
        \caption{\begin{minipage}{0.7\linewidth} \centering After $0.05$ light-crossing time-scale. Blue circles indicate expected radii of the most luminous expanding radiation bubbles. \end{minipage}}
    \end{subfigure}
    % \vspace{1cm}
    % Second row
    \begin{subfigure}{0.49\textwidth}
        \centering
        \includegraphics[width=\linewidth]{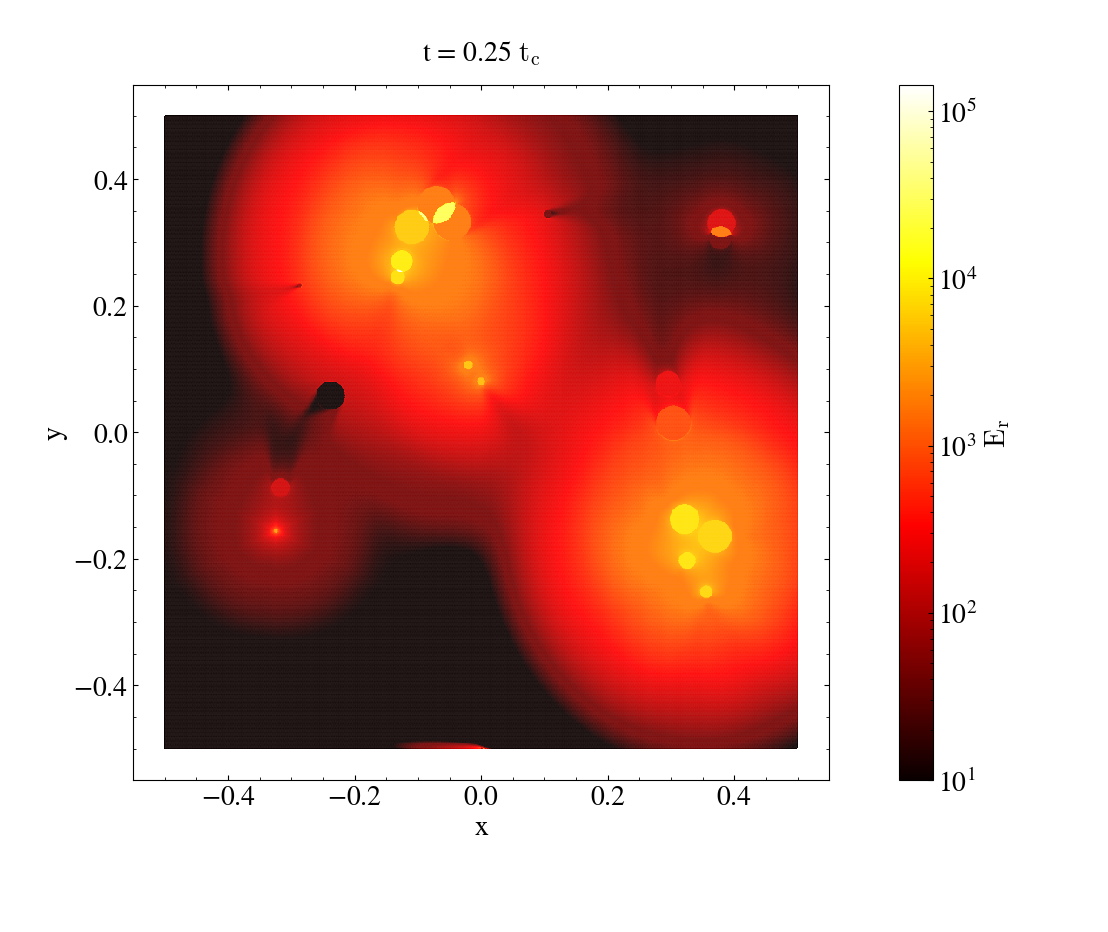}
        \caption{After $0.25$ light-crossing time-scale.}
    \end{subfigure}
    \hfill
    \begin{subfigure}{0.49\textwidth}
        \centering
        \includegraphics[width=\linewidth]{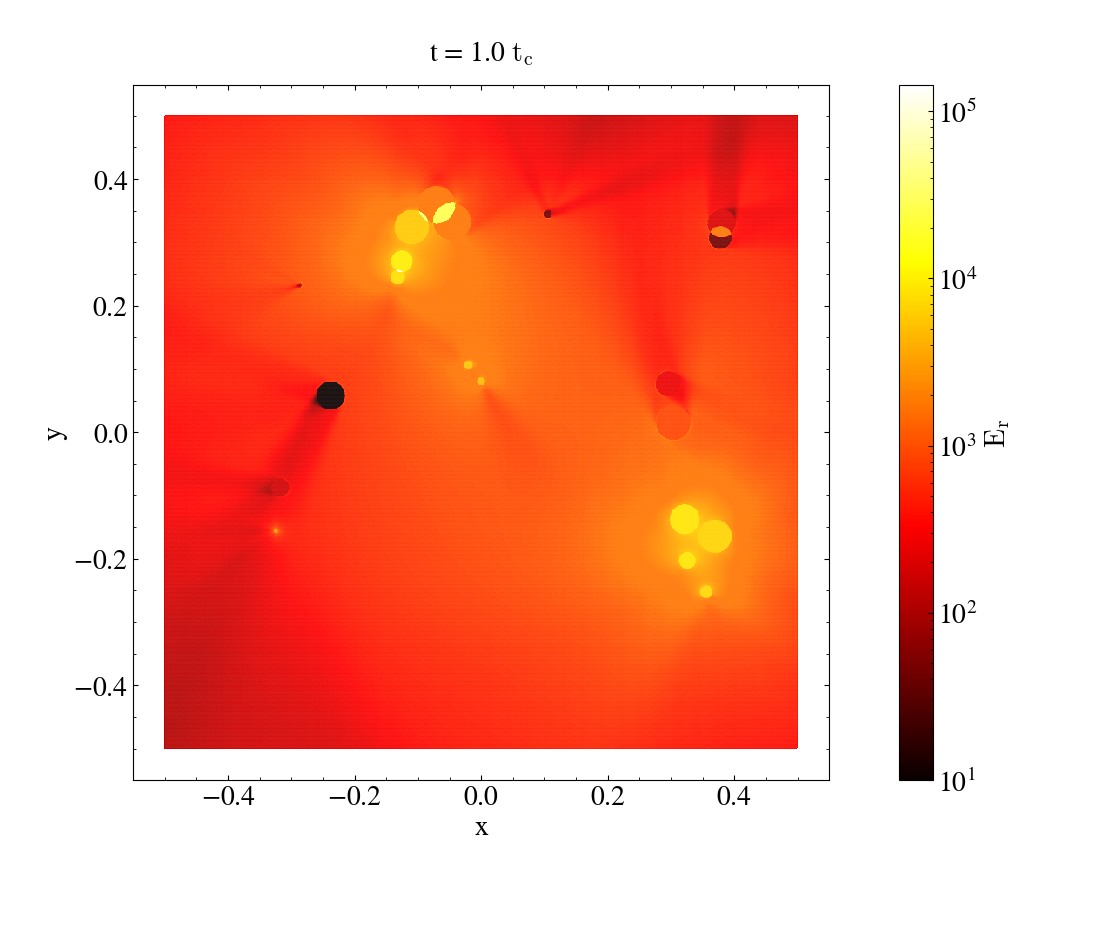}
        \caption{After $1$ light-crossing time-scale.}
    \end{subfigure}
% \begin{figure*}[htb!]
% % \centering
% \gridline{
% \fig{bub_30l_000.png}{0.5\textwidth}{(a) Initial condition.}
% \fig{bub_30l_001.png}{0.5\textwidth}{(b) After $0.05$ light-crossing time-scale. Blue circles indicate expected radii of the most luminous expanding radiation bubbles.}
% }
% \gridline{
% \fig{bub_30l_005.png}{0.5\textwidth}{(c) After $0.25$ light-crossing time-scale.}
% \fig{bub_30l_020.png}{0.5\textwidth}{(d) After $1$ light-crossing time-scale.}
% }
\caption{
2D expanding multi-source radiation bubble test, with a maximum timestep $25$ times larger than the CFL condition among the three levels of local timestep hierarchy.
The colors show the radiation energy density distribution.
We follow the propagation of 20 radiation bubbles in vacuum until the light crosses the whole box.
The bubbles are emitted from 20 optically-thick sources with different luminosities seemingly randomly distributed (panel a).
The radiation does propagate at the speed of light, but the radiation fronts are slightly spread by numerical diffusion compared to the expected fronts (blue circles in panel b).
The radiation bubbles run into each other and merge (panel c).
Eventually, they fill the simulation box, leaving small shadows behind the dim optically-thick sources (panel d).
As shown in the expansion of the bubbles, most of the radiation comes from the luminous sources clustered at the top and bottom right.
}
\label{fig:test:rd:bubble}
\end{figure*}

It is important to demonstrate the power of the implicit scheme to overcome the small timesteps when propagating radiation even if local time-stepping is used.
Therefore, we choose a maximum timestep of $0.001$, about $25$ times larger than the CFL timestep limited by the speed of light.
This results in a three-level timestep hierarchy of $0.00078, 0.00039, 0.000195$ distributed in the 2D plane as illustrated in Figure~\ref{fig:test:2d}.
In Figure~\ref{fig:test:rd:bubble}, we show the expansion and merging process of radiation bubbles.
The 20 radiation sources with varying sizes and luminosities are shown in panel a.
The radiation bubbles expand at the speed of light (panel b), merge (panel c), and eventually fill the simulation box (panel d).
The radiation field is dominated by radiation from the luminous sources clustered at the top and lower-right of the box, which is designed to mimic the clustering of massive stars.
We find no artifact at the boundaries between regions holding different timesteps.
We also compare the figures with another test that adopts global time-stepping, and find almost identical results.
We therefore conclude that local time-stepping does not change the simulation results when the radiation is free-streaming in the optically-thin region, but more complex simulations are needed to confirm this.

In panel b, for each individual expanding bubble, we predict its size as $R_0+ct$ and plot them as blue circles, where $R_0$ is the size of the individual source and $t$ is the time of the simulation.
The radiation energy density falls off as $r^{-2}$ and follows analytical prediction shown in Section~\ref{sec:test:steady:sphere}.
We therefore only plot the predicted bubble sizes for those that are luminous enough to have a radiation energy density at the radiation front larger than the colorbar floor.
The bubble sizes in the simulation generally agree with the predicted ones, indicating that the bubbles indeed expand at the speed of light, even though we take much larger timesteps than the CFL condition.
However, the radiation fronts are not sharp but more extended due to the numerical diffusion.
This seems to suggest that, if our method is applied to more complicated reionization simulations, the surroundings may be ionized sooner than what nature does.
But this may not be an issue, because the radiation energy density in the diffused part falls off quickly with radius, and is generally orders of magnitude smaller than the radiation energy density at the radiation front.

This test also demonstrates the accuracy of our method in simulating a complicated radiation field propagating at the speed of light.
For example, in panel c and d, we show that the intense radiation leaves shadows behind the dim optically-thick sources, and the shadows are pointed towards different directions depending on their relative positions to the luminous clusters.
\citet{thomas2022MNRAS} also illustrated that, due to the fluid-like description of M1 closure, it predicts the collision of radiation bubbles, which generates rarefaction waves and shock fronts that artificially enhance or decrease the radiation energy locally.
But with our method, the radiation bubbles pass each other as light is supposed to behave \citep[see also the discrete ordinates results in][]{thomas2022MNRAS}.

\subsection{Radiation Hydrodynamic Tests for Radiation-gas Coupling}
\label{sec:test:rhd}

\subsubsection{2D Thermal Equilibrium in a Uniform Medium}
\label{sec:test:rhd:te}

This test is a simple demonstration of the coupling between radiation and gas.
Specifically, we test the exchange of energy and momentum between radiation and gas.

We setup the problem the same way as in \citetalias{jiang2021ApJS} with a 2D box covering $[0,1]\times[0,1]$.
We set the spatial resolution to be $32\times 32$, with $24$ directions for the angular resolution consisting of two groups of angular sets.
The box is filled with uniform medium of density $\rho=1$, temperature $T_0$, horizontal velocity $v_0^x$, absorption opacity $\kappa_\mathrm{R}=\kappa_\mathrm{P}=\kappa_\mathrm{a}$ and zero scattering opacity.
The intensities are initialized to be the isotropic value set by the radiation energy density $E_{r,0}$, which is out of thermal equilibrium.
All boundaries are set to be periodic for both fluid and radiation.
We choose $a=1$ with three sets of different parameters: (1) $T_0=1, E_{r,0}=100, v_0^x=0, \kappa_\mathrm{a}=100, c=100$; (2) $T_0=100, E_{r,0}=1, v_0^x=0, \kappa_\mathrm{a}=1, c=100$;  (3) $T_0=1, E_{r,0}=1, v_0^x=3, \kappa_\mathrm{a}=1, c=10$.\footnote{In \citetalias{jiang2021ApJS}, they recorded that the last test is performed with $c=100$. We have checked that in our case, this parameter will not produce the equilibrium $E_\mathrm{r}$ and $v^x$ as reported in \citetalias{jiang2021ApJS}. This is likely a typo, because if we choose $c=10$ the results agree very well.}
The thermal relaxation time-scale can be calculated for different parameter sets as $t_0=1/(c\rho\kappa_\mathrm{a})$.
We manually apply a three-level hierarchy of local time-stepping as illustrated in Figure~\ref{fig:test:2d}.

\begin{figure*}[htb!]
\centering
\includegraphics[width=0.9\textwidth]{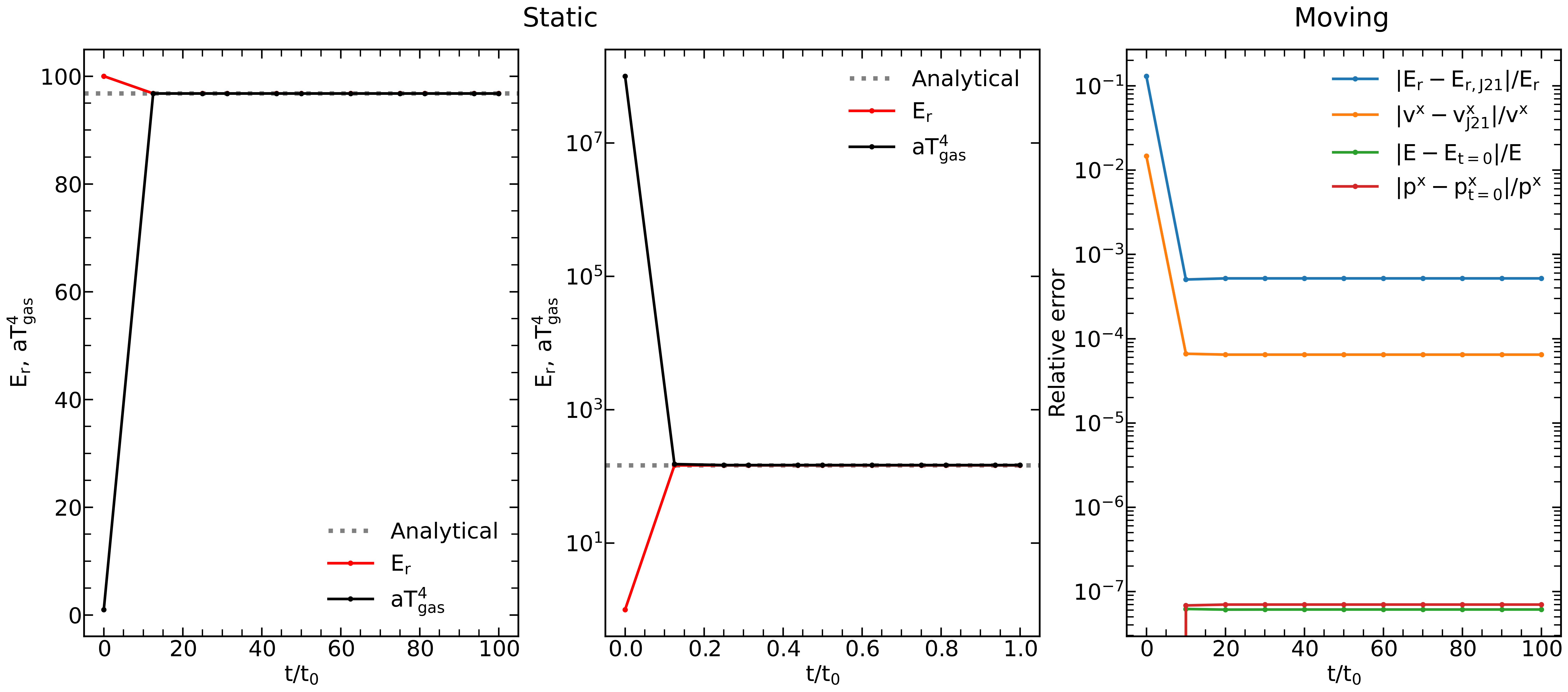}
\caption{
2D thermal equilibrium test.
We start with a radiation field and a temperature field very far from equilibrium with each other, and they quickly reach thermal equilibrium state as the system evolves.
All the x axes indicate the time $t$ of the simulation in unit of the thermalization time-scale $t_0$.
In the first two panels, we explore thermal equilibrium in static media, and show that $E_\mathrm{r}$ (red) and $aT_\mathrm{gas}^4$ (black) become equal at a value consistent with the analytical solutions (grey dotted).
In the last panel, we show the relative error of different values: the radiation energy density compared to the \citetalias{jiang2021ApJS} reported value (blue), the horizontal velocity compared to the values reported in \citetalias{jiang2021ApJS} (orange), the average total energy density compared to its initial value (green), and the average total horizontal momentum density compared to its initial value (red).
The relative errors show that our results are consistent with the values reported in \citetalias{jiang2021ApJS} to the precision they reported, and that the total energy and total momentum are conserved at $10^{-7}$ level even with local timesteps on an unstructured moving mesh.
}
\label{fig:test:rhd:te}
\end{figure*}

In Figure~\ref{fig:test:rhd:te}, we show the results of three tests.
The first two panels illustrate the swift relaxation to thermal equilibrium in static media, either for optically-thick (left panel) or for optically-thin regime (middle panel).
The averaged radiation energy density $E_\mathrm{r}$ (red) and averaged $aT_\mathrm{gas}^4$ (black) reach the same value as the analytical solution (grey dotted).
This proves that the total energy is conserved and the energy is exchanged between gas and radiation.
The time evolution to the equilibrium value is not correctly displayed in the left panel, because the hydro timestep is about $10\; t_0$, and the system reaches thermal equilibrium in one timestep, although physically it should take much less time.
In the middle panel, the system reaches the approximate equilibrium state much faster than $t_0$, because the initial deviation from equilibrium is orders of magnitude larger than the equilibrium value, which accelerates the initial thermal relaxation process.
However, if we plot the temperature in linear scale, it is evident that the system is still approaching the exact equilibrium state at the end of the simulation, because the ultimate thermal relaxation timescale is $t_0$.
In the right panel, we examine the thermal equilibrium in a uniformly moving medium and show the relative errors of different quantities compared to their expected values.
The relaxed radiation energy density $E_\mathrm{r}$ (blue) and the horizontal velocity $v^x$ (orange) fit well with the values reported in \citetalias{jiang2021ApJS} to the reported precision.
The average total energy density (green) and the average total horizontal momentum density (red) also do not change within the relative errors of $10^{-7}$.
Even though this test does not guarantee the conservation of energy and momentum in sophisticated simulations, it shows that the energy and momentum are conserved in the simple test even on an unstructured moving mesh with forced local timesteps.

\subsubsection{2D Linear Waves}
\label{sec:test:rhd:wave}

The linear wave test is a basic test to show the accuracy of the code and the convergence order of the scheme.
It can also expose the errors introduced by the moving-mesh noises that affect the convergence, especially in radiation-dominated optically-thick waves.

We perform the tests similar to \citetalias{jiang2021ApJS} by setting up 2D boxes spanning $[0,1]\times[0,1]$.
We vary the spatial resolution from $16\times 16$ to $256\times 256$, and fix the angular resolution as $8$ directions using the default angular discretization not specific to 2D setup.
We set the uniform background with density $\rho_0=1$, temperature $T_0=1$, gas pressure $P_0$, and absorption opacity $\kappa_\mathrm{R}=\kappa_\mathrm{P}=\kappa_\mathrm{a}$.
The equation of state is chosen to be $P=\rho T$ with an adiabatic index $5/3$.
On top of the uniform background, we add perturbations with an amplitude of $10^{-3}$ via eigenmodes with a wavelength the size of the box, calculated from the dispersion relation detailed in Appendix B of \citet{jiang2012ApJS}.
We initialize the intensities to be the equilibrium values calculated from the time-independent version of our scheme, as we find that this procedure reduces the errors for radiation-dominated optically-thick waves.
All boundaries are set to be periodic.
Two physical parameters to this problem are the optical depth per wavelength $\tau_\mathrm{a}=\rho_0\kappa_\mathrm{a}$, and the ratio between radiation energy density and gas pressure $aT_0^4/P_0$.
We thus fix the speed of light as $c=10$, and vary $a$ and $\kappa_\mathrm{a}$.
We manually apply a three-level hierarchy of local time-stepping as illustrated in Figure~\ref{fig:test:2d}.

\begin{figure}[htb!]
\centering
\includegraphics[width=\columnwidth]{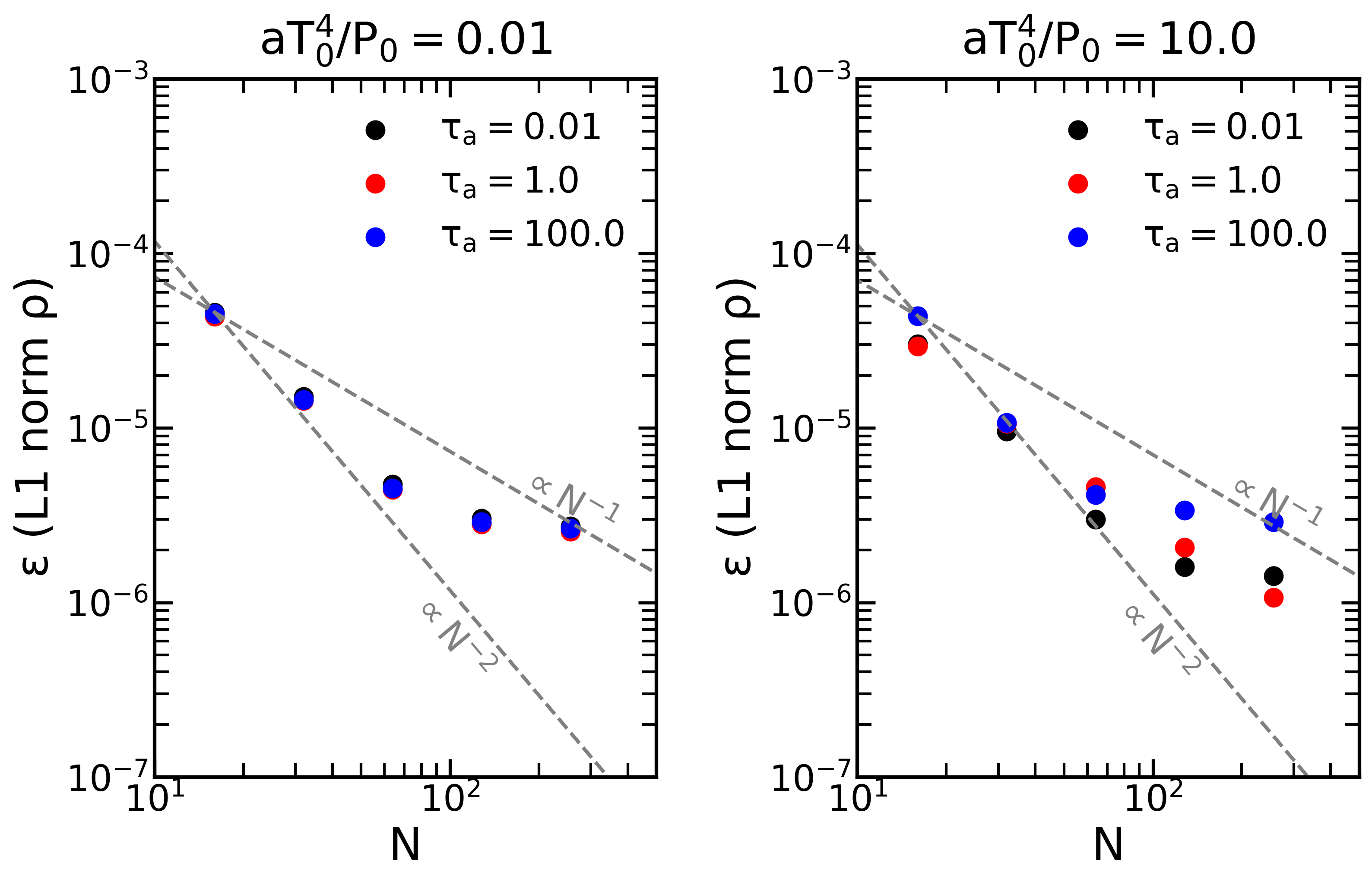}
\caption{
2D linear wave test showing the convergence with increasing spatial resolution.
We explore different physical parameters of $\tau_\mathrm{a}=0.01,1,100$ in different colors and $aT_0^4/P_0=0.01,10$ in different panels.
The L1 norm of the absolute error in density generally decreases with increasing number of cells in the horizontal dimension.
The curves fall off as $N^{-2}$, until they saturate at $10^{-6}$ level, where the errors are dominated by moving-mesh noises as discussed in the main text.
}
\label{fig:test:rhd:wave_conv}
\end{figure}

\begin{figure}[htb!]
\centering
\includegraphics[width=0.4\textwidth]{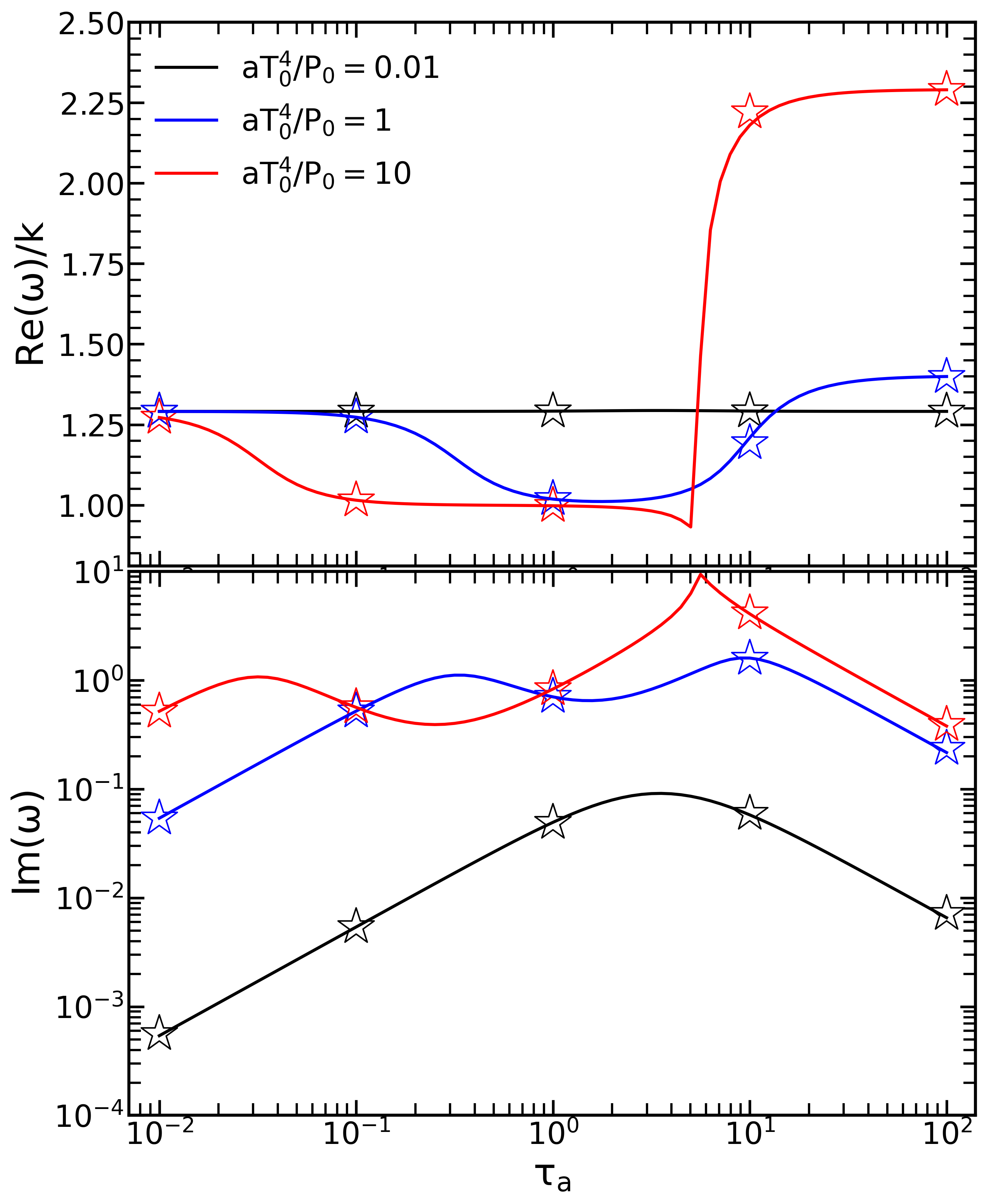}
\caption{
2D Linear wave test showing the phase speed (top) and the damping rate (bottom) of waves.
We use the simulation results with resolution $256\times 256$ (stars), which agree well with the analytical solutions (solid lines).
}
\label{fig:test:rhd:wave_rate}
\end{figure}

We evolve the system for one period, and compare the results with the analytical solutions.
In Figure~\ref{fig:test:rhd:wave_conv}, we measure the L1 norm of the absolute error in density, and plot it against the number of cells in the horizontal dimension.
For gas-dominated waves (left panel), the convergence curve follows $N^{-2}$, as expected for second-order accurate hydro solver in \texttt{AREPO}.
However, we find the same trend for radiation-dominated waves (right panel), even though the errors should be dominated by the first-order accurate radiation solver, as shown in \citet{jiang2012ApJS, jiang2014ApJS} and \citetalias{jiang2021ApJS}.
Such excellent performance is unexpected, and deserves further investigation.
We find that the convergence generally saturates at $10^{-6}$, even for pure hydro tests not shown here.
This is because the moving-mesh noises introduce errors at $10^{-6}$ level, which is also why we set the perturbations with amplitudes $10^{-3}$ instead of $10^{-6}$ as in \citet{jiang2012ApJS, jiang2014ApJS} and \citetalias{jiang2021ApJS}.
However, the moving-mesh noises do not show up in 1D tests, such as those demonstrated in \citet{springel2010MNRAS} and \citet{weinberger2020ApJS}.
In multi-dimensional tests, the moving-mesh noises may be suppressed by using higher-order flux integration \citep{zier2022MNRAS}.
We also find that a numerical parameter of $\alpha=5$ is not enough to obtain sufficiently accurate results for the most optically-thick tests with $\tau_\mathrm{a}=100$.
The results presented here adopt $\alpha=50$ for $aT_0^4/P_0=10,\tau_\mathrm{a}=100$ test, $\alpha=20$ for the rest of $\tau_\mathrm{a}=100$ tests, and $\alpha=5$ for the others.
In Figure~\ref{fig:test:rhd:wave_rate}, we present the phase velocity (top panel) and the damping rate (bottom panel) of the waves for different choices of optical depths $\tau_\mathrm{a}$ and $aT_0^4/P_0$.
We take the last snapshot of each test, fit the density perturbation with a sinusoidal curve, and obtain the phase and the amplitude, by which we calculate the phase velocity and the damping rate.
The results shown (stars) are calculated with resolution $256\times 256$, and agree with analytical solutions (solid lines).
We find no difference between the test results that adopt forced local time-stepping and those that do not (not presented here to avoid duplication).

\subsubsection{1D Radiation Shocks}
\label{sec:test:rhd:shock}

Shock tube tests are standard tests for non-linear numerical schemes for hydrodynamics and MHD.
However, in radiation hydrodynamics or RMHD, the evolution of shocks are more complex, and no analytical solutions can be used to verify numerical results for shock tube tests \citep{jiang2012ApJS}.
Therefore, we resort to the steady-state radiation shock test \citep[e.g. ][]{jiang2012ApJS, jiang2014ApJS, jiang2021ApJS, menon2022MNRAS}.
The test checks the capability of the code to keep the radiation shock structure calculated from analytical solutions.
This is the only time-dependent test where we do not use the moving-mesh or local time-stepping, because we find that the moving-mesh is squeezed near the boundaries such that the mesh construction becomes difficult.

In particular, the setup is almost identical to the radiation shock test in \citet{jiang2014ApJS}.
We use a 1D slab of length $0.115847$ filled by 4096 mesh points and $24$ directions covering the full unit sphere.
We set the pre-shock parameters to be $\rho_0=T_0=E_\mathrm{r,0}=1$ and velocity $v_0={\cal M}$, where we vary the Mach number to be ${\cal M}=1.2,2,3$.
The absorption opacity is chosen to satisfy $\kappa_\mathrm{R}=\kappa_\mathrm{P}=577.4/\rho$.
We set a non-standard equation of state $P=\rho T/\gamma$ to be consistent with \citet{ferguson2017HEDP}, where the adiabatic index is $\gamma=5/3$.
The initial conditions are taken from the analytical solutions \citep{ferguson2017HEDP}, and the intensities are initialized via our time-independent radiation transport calculations.
The left boundaries are set to keep the pre-shock quantities, and we use an outflow boundary condition for the right boundary.
The constants are set to be $a=10^{-4},c=1.73\times 10^3$.

\begin{figure}[htb!]
\centering
\includegraphics[width=\columnwidth]{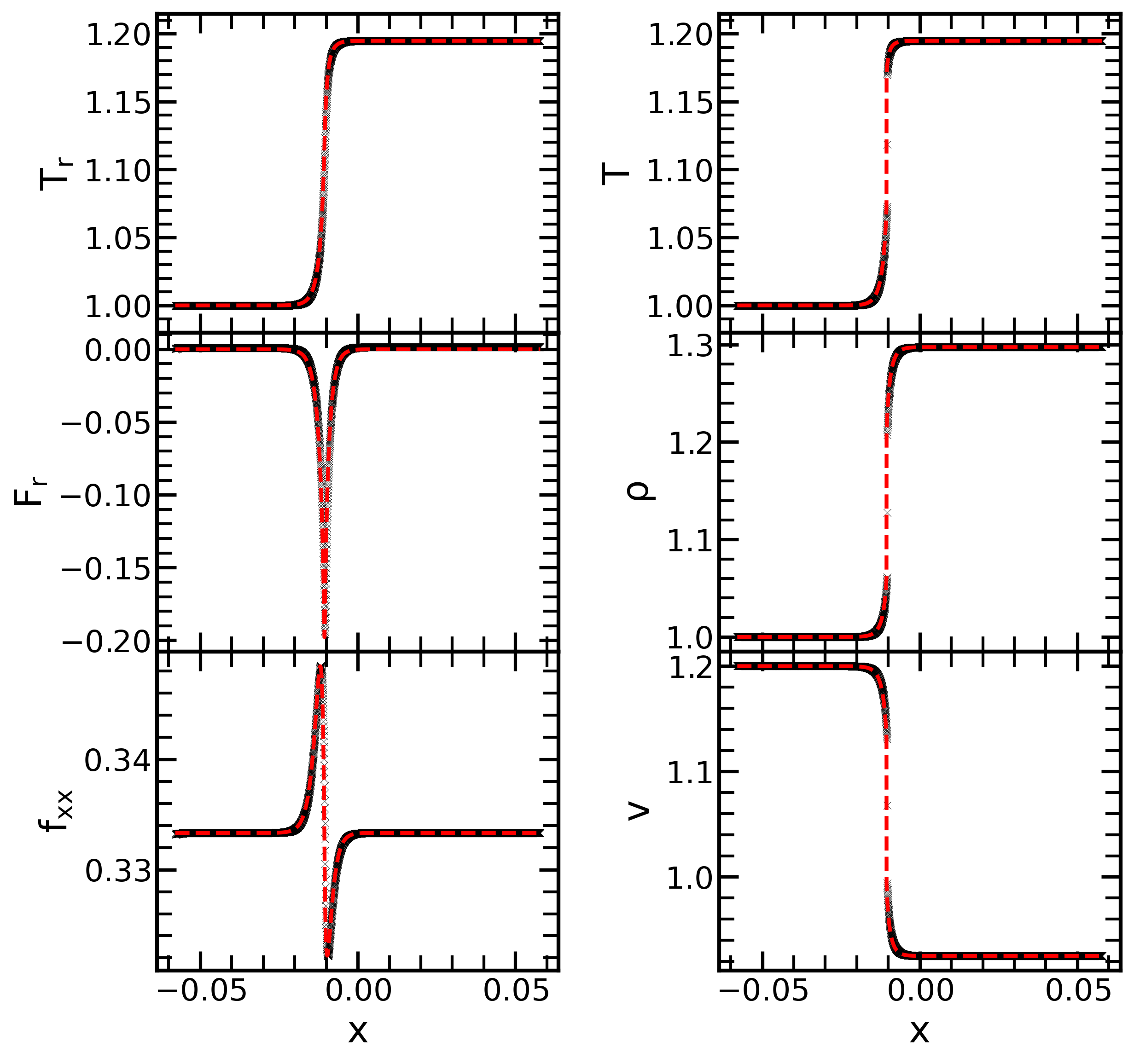}
\caption{
1D radiation shock test with Mach number ${\cal M}=1.2$.
We show the profiles of radiation temperature $T_\mathrm{r}=(E_\mathrm{r}/a)^{1/4}$, gas temperature $T$, radiation flux $F_\mathrm{r}$, density $\rho$, xx component of the Eddington tensor $f_{xx}$, and velocity $v$.
The simulation results (black) align with the analytical solutions (red dashed).
}
\label{fig:test:rhd:shock:1.2}
\end{figure}

\begin{figure}[htb!]
\centering
\includegraphics[width=\columnwidth]{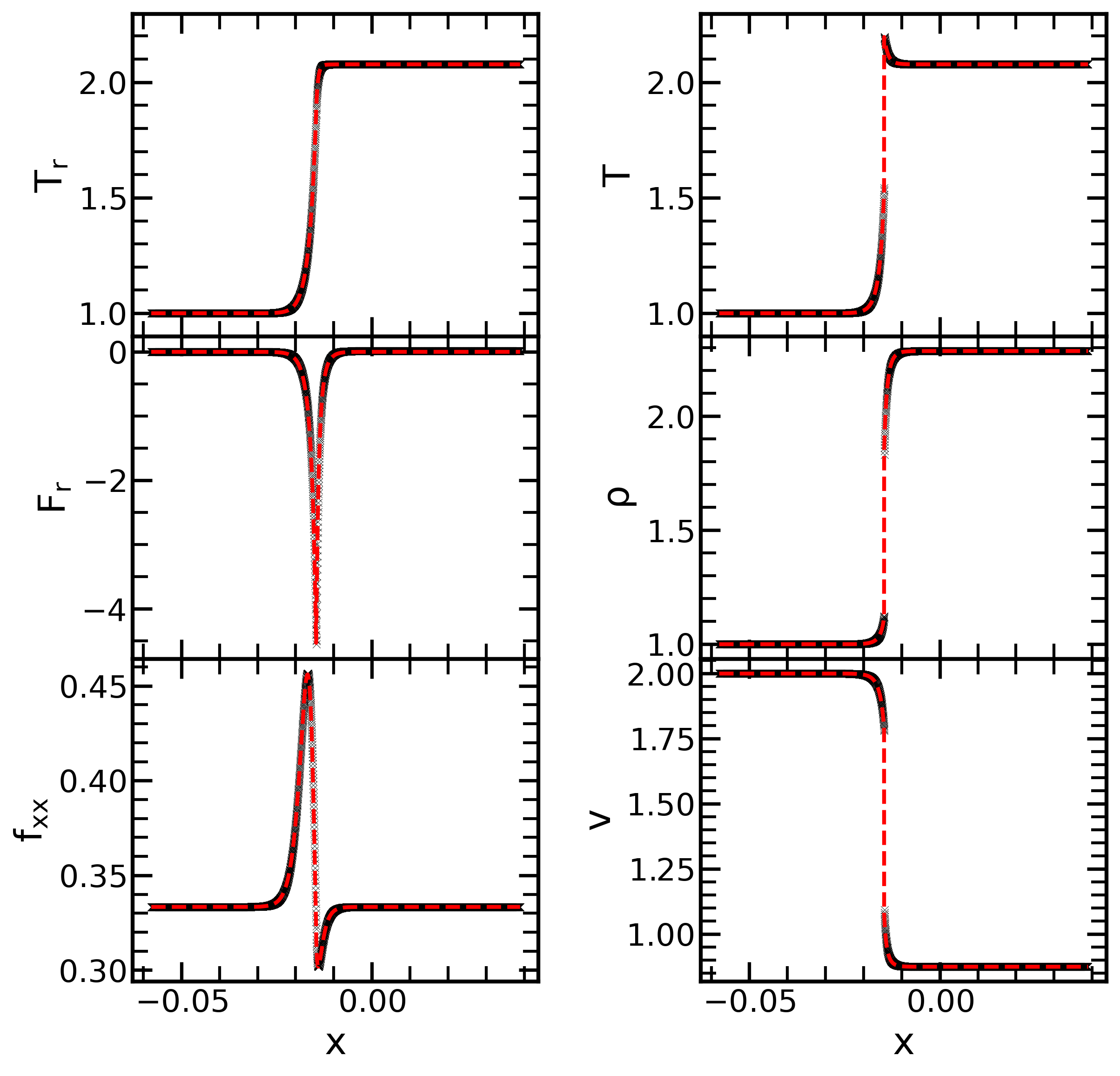}
\caption{
Same as Figure~\ref{fig:test:rhd:shock:1.2} but for Mach number ${\cal M}=2$.
}
\label{fig:test:rhd:shock:2}
\end{figure}

\begin{figure}[htb!]
\centering
\includegraphics[width=\columnwidth]{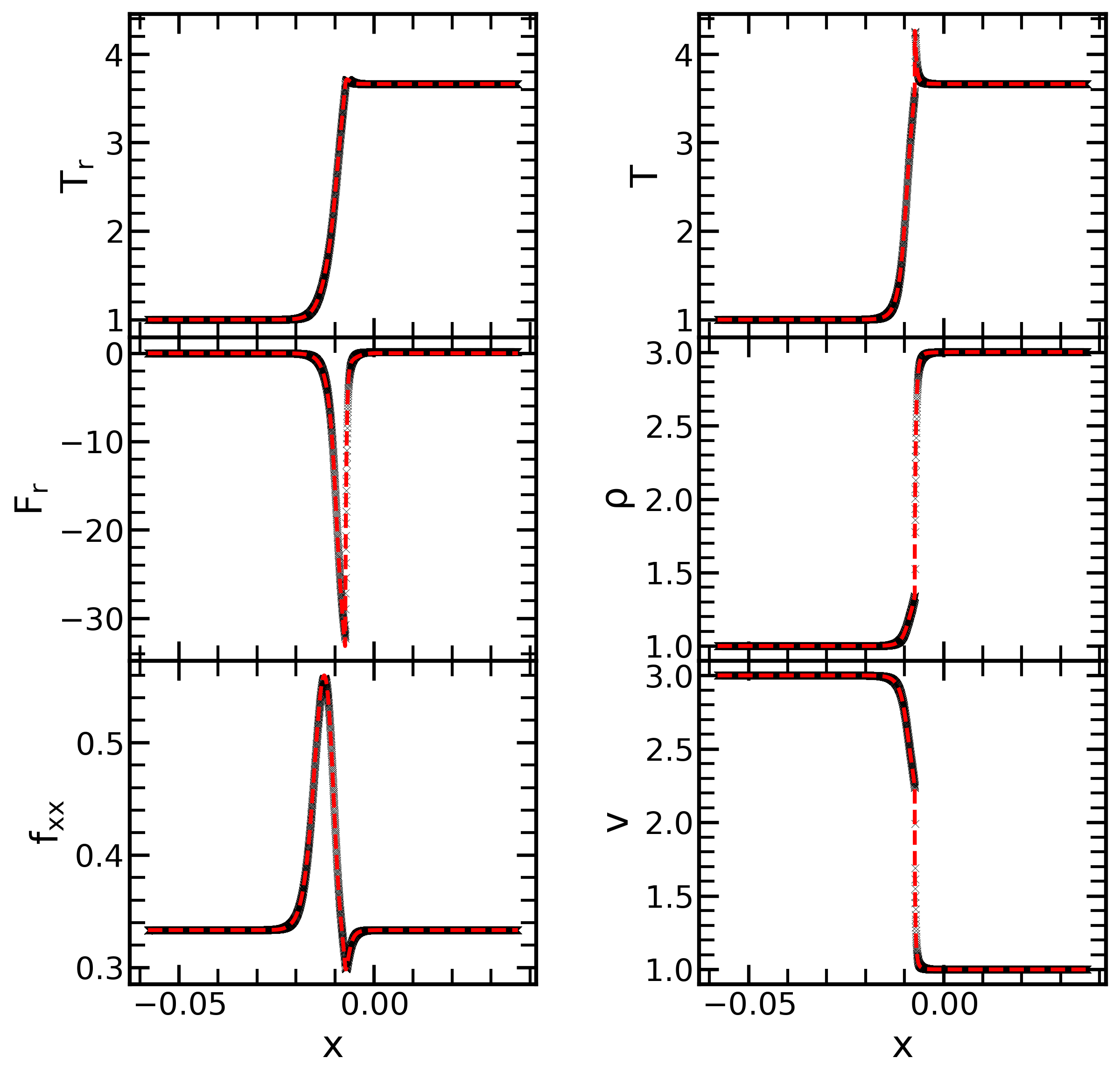}
\caption{
Same as Figure~\ref{fig:test:rhd:shock:1.2} but for Mach number ${\cal M}=3$.
}
\label{fig:test:rhd:shock:3}
\end{figure}

In Figure~\ref{fig:test:rhd:shock:1.2}, \ref{fig:test:rhd:shock:2} and \ref{fig:test:rhd:shock:3}, we show the shock structures for Mach number ${\cal M}=1.2,2,3$, respectively, after three flow-crossing times.
Different panels show the radiation temperature $T_\mathrm{r}=(E_\mathrm{r}/a)^{1/4}$, gas temperature $T$, radiation flux $F_\mathrm{r}$, density $\rho$, xx component of the Eddington tensor $f_{xx}$, and velocity $v$.
The shock structures after three flow-crossing times (black) are identical to the initial profiles (red dashed) calculated from analytical solutions.
A common feature is that the radiation fluxes are always pointing from downstream to upstream, which cool the post-shock downstream gas and heat up the pre-shock upstream gas.
When the upstream Mach number is close to unity (Figure~\ref{fig:test:rhd:shock:1.2}), radiative heating and cooling are not significant so that the gas temperature still increases monotonically from upstream to downstream.
However, for Mach number ${\cal M}=2$ (Figure~\ref{fig:test:rhd:shock:2}), the upstream gas near the shock front is heated up and the downstream is cooled, such that the immediate post-shock gas temperature exceeds the far downstream gas temperature, resulting in a spike in gas temperature typically referred to as the Zel'dovich spike \citep{zeldovich1967}.
For both of those cases, the immediate pre-shock gas temperature is still lower than the far downstream gas temperature, which is called a subcritical shock.
For Mach number ${\cal M}=3$ (Figure~\ref{fig:test:rhd:shock:3}), the immediate pre-shock gas temperature becomes the same as the far downstream gas temperature, termed as the supercritical shock.
Another common feature is that the Eddington tensor is above $1/3$ at the immediate pre-shock region, but below $1/3$ at the immediate post-shock region, which indicate a jump in radiation pressure at the shock front.
The plots can be directly compared with the results in \citet{jiang2014ApJS}, which show identical profiles.

\subsubsection{2D Irradiated Cloud}
\label{sec:test:rhd:cloud}

This simplified test demonstrates the capability of our scheme to perform radiation hydrodynamic simulations, where radiative heating has a significant impact on the gas dynamics.

We use the same setup adopted for the 2D shadow test in Section~\ref{sec:test:steady:shadow}.
The relaxed radiation field in the shadow test is used as the initial condition, and we switch on the time-dependent radiation module coupled to the hydro solver to perform a radiation hydrodynamic simulation.
We switch on the local time-stepping and allow the code itself to decide the local timesteps.
In order to keep the scheme stable, we find it necessary to set a maximum timestep of $0.0003$ to limit the timestep jump in neighboring cells to a factor of two.
This results in a three-level time-stepping hierarchy.

\begin{figure}[htb!]
\centering
% First row
    \begin{subfigure}{\columnwidth}%{0.5\textwidth}
        \centering
        \includegraphics[width=\linewidth]{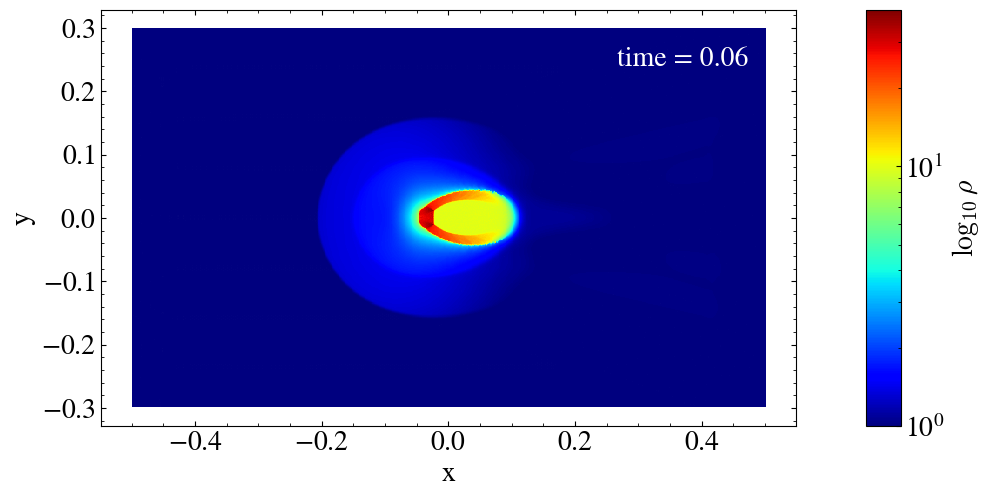}
        \caption{ After $0.06t_0$.}
    \end{subfigure}
    % \vspace{1cm}
    % Second row
    \begin{subfigure}{\columnwidth}%{0.5\textwidth}
        \centering
        \includegraphics[width=\linewidth]{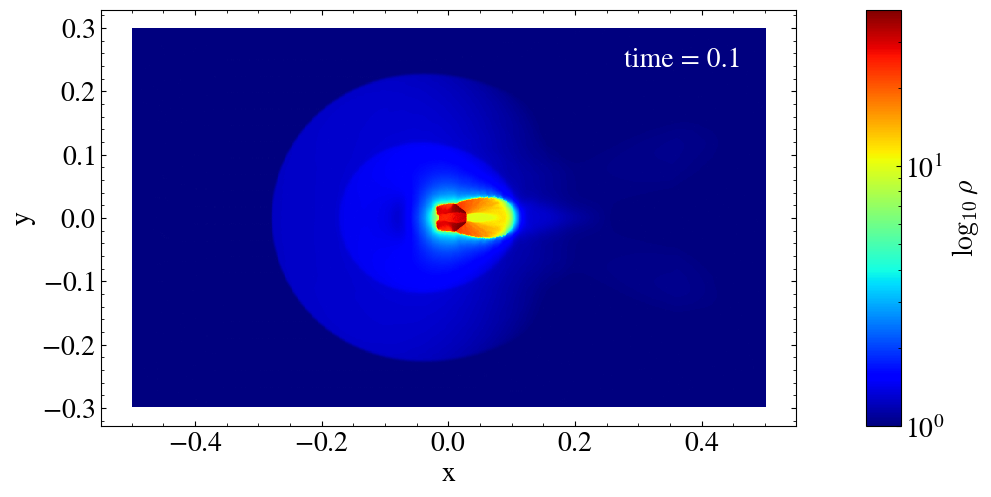}
        \caption{ After $0.1t_0$.}
    \end{subfigure}
    \begin{subfigure}{\columnwidth}%{0.5\textwidth}
        \centering
        \includegraphics[width=\linewidth]{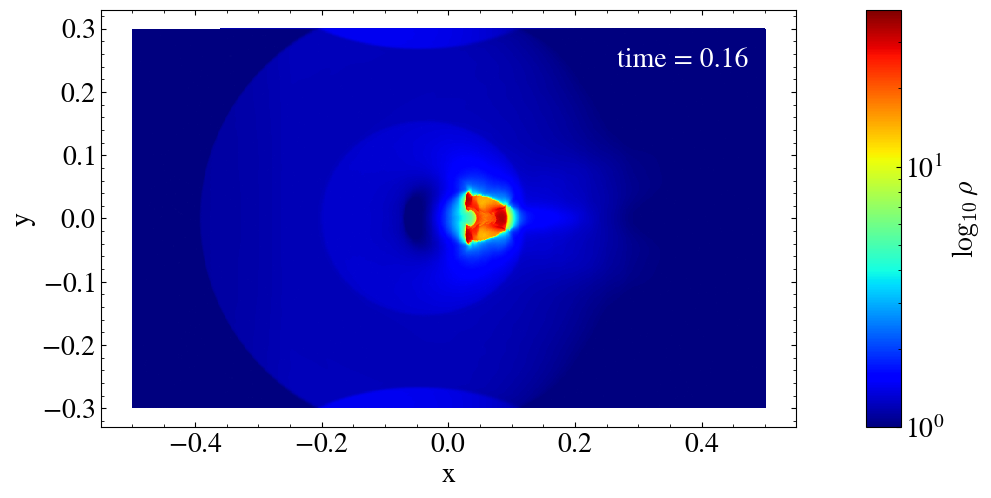}
        \caption{ After $0.16t_0$.}
    \end{subfigure}
% \begin{figure*}[htb!]
% % \centering
% \gridline{
% \fig{irradiation_003.png}{0.5\textwidth}{(a) After $0.06t_0$.}
% }
% \gridline{
% \fig{irradiation_005.png}{0.5\textwidth}{(b) After $0.1t_0$.}
% }
% \gridline{
% \fig{irradiation_008.png}{0.5\textwidth}{(c) After $0.16t_0$.}
% }
\caption{
2D irradiated cloud test.
The plots show the density profile at time $t=0.06t_0,0.1t_0,0.16t_0$, where $t_0=1$ is the sound-crossing time-scale across the box.
We use the same setup in the 2D shadow test (Section~\ref{sec:test:steady:shadow}), where we inject two parallel rays of radiation from the left boundary.
The radiation acts as a pressure that pushes the cloud.
It also heats the cloud on one side, which evaporates the surface and sends out an outflow.
In the meantime, the heated side sends in shocks that propagate inside the cloud, which eventually destroy the cloud.
The figure can be directly compared to figure 5 in \citetalias{jiang2021ApJS}.
}
\label{fig:test:rhd:cloud}
\end{figure}

In Figure~\ref{fig:test:rhd:cloud}, we show the density distribution at simulation time $0.06t_0,0.1t_0,0.16t_0$, where $t_0=1$ is the sound-crossing time-scale across the box.
The optically-thick cloud, which is initially in pressure balance with its surroundings, is compressed by a radiation pressure $0.1\%$ of the gas pressure.
The surface of the cloud is also heated up, driving an outflow from the irradiated side.
Meanwhile, the compressed heated side drives shocks that propagate inwards, which compress the gas further, and eventually destroy the cloud.
Our results can be directly compared with figure 5 in \citetalias{jiang2021ApJS}.
The two results are nearly identical, except that our results show sharper edges because we switch on mesh refinement with a target mass resolution $1/(512\times 256)$ while \citetalias{jiang2021ApJS} fixed the mesh without adaptive mesh refinement.
Similar simulations can also be found in \citet{jiang2012ApJS} and \citet{proga2014ApJ}.

% \subsection{2D Dust-driven Wind}
% \label{sec:test:rhd:wind}
% \revising{To be done. Maybe need to talk to Rahul/Aaron to see how exactly they set up the problem}

\section{Astrophysical Example: Global 3D Convective Envelope of a Red Supergiant Star}
\label{sec:test:rhd:rsg}

To demonstrate the potential of our scheme to simulate complex astrophysical systems, we present a global 3D grey radiation hydrodynamic simulation of a $10\, M_\odot$ red supergiant star that simulates the entire convective envelope and simultaneously marginally resolves the stellar photosphere.
Such simulation is very challenging because it involves physics with orders-of-magnitude span in both length-scale and time-scale.
3D simulations of red supergiants have been performed before by e.g. \citet{freytag2002AstronomischeNachrichten, freytag2024A&A} and \citet{chiavassa2011A&Aa} using the \texttt{CO5BOLD} code \citep{freytag2012JCoPh}, and by \citet{goldberg2022ApJ} using the \texttt{ATHENA++} code \citep{stone2020ApJS}, which are considered the current state-or-the-art 3D models for red supergiants \citep{chiavassa2024LRCA}.

We show that, with our new scheme, we can surpass the current state-of-the-art in the following ways:
% \revising{Begin: Still changing the following part!!}
% \setlist{nolistsep}
\begin{enumerate}%[noitemsep]
    \item We conduct the global 3D simulation for the star that (1) simulates the entire $4\pi$ sphere and (2) retains a large simulation box to capture the atmospheric structure.
    The first point is important for the global convective structure and synthetic interferometric observations \citep[e.g.][]{chiavassa2009A&A}, whilst the second point is essential for studying wind launching, circumstellar material, and shock breakout \citep[e.g.][]{goldberg2022ApJa}.
    The \texttt{CO5BOLD} models achieve the first but not the second, whereas the \texttt{ATHENA++} models achieve the second but not the first.
    We manage to attain both because our scheme (a) achieves fast performance that enables global 3D simulations with reasonable computational cost, and (b) supports Voronoi mesh construction in \texttt{AREPO} which enables very flexible mesh sizes.
    \item Our simulation include the entire convective envelope down to the bottom.\footnote{It is commonly assumed that only simulating the upper part of the convective envelope does not jeopardize the convective structure, because the convection is primarily driven by surface radiative cooling \citep{stein1989ApJ} and the deep thermal timescale is so high that the bottom envelope is not thermally relaxed even if we can simulate it.
    Both of these arguments are true.
    However, if the bottom boundary is not correctly treated, the specific entropy of the efficient convection zone will not be physical, thereby affecting the whole convective envelope.}
    This is to ensure that the specific entropy of the convective envelope matches the internal condition of red supergiants, which is a general challenge that remains to be solved in previous simulations \citep{chiavassa2024LRCA}.
    We manage to solve this by simulating the entire convective envelope.
    This is because our scheme supports local time-stepping in \texttt{AREPO}, which allows us to simulate the deep interior of the star where the local dynamical timescale is orders of magnitude lower than the surface dynamical timescale.
    Local time-stepping in 3D simulations of stellar atmospheres has only been implemented recently \citep[e.g. the \texttt{DISPATCH} framework;][]{nordlund2018MNRAS}, albeit only for local box simulations \citep{eitner2024A&A} or global simulations with simple cooling functions \citep{popovas2022arXive-prints}.
    Here, we illustrate the advantage of local time-stepping in global 3D simulations of stellar surface convection with more accurate radiation transport.
    \item We demonstrate that it is possible to marginally resolve the red supergiant photosphere in global 3D simulations, such that the outgoing luminosity matches the input energy source.
    This is important because it determines the surface cooling rate, which controls the convective structure and the atmospheric temperature.
    We can achieve this because our scheme (1) supports the flexible Voronoi mesh construction of \texttt{AREPO} that can be easily refined in arbitrary regions, and (2) is accurate in the transition region between optically-thick to thin regimes.
    Our surface resolution of $1.5\, R_\odot$ is 5 times higher than the $8\, R_\odot$ surface resolution used in \texttt{ATHENA++} red supergiant models \citep{goldberg2022ApJ}, and also slightly higher than the $2\, R_\odot$ surface resolution used in latest \texttt{CO5BOLD} red supergiant models \citep{ahmad2023A&A}.
    On average, we have 3 cells per photon mean free path at the photosphere, and once in a while 6 cells per photon mean free path during the simulation.
\end{enumerate}

Below we describe the setup and test simulations.
We limit ourselves to a brief demonstration.
In a separate upcoming paper (Ma et al. in prep.), we will discuss the scientific findings in detail.

\subsection{Numerical Setup}

For the initial condition, we take a $10\, M_\odot$ non-rotating red supergiant star at $Z=0.02$ metallicity evolved to the stable core helium burning stage using the 1D stellar evolution code \texttt{MESA} \citep{paxton2011ApJS, paxton2013ApJS, paxton2015ApJS, paxton2018ApJS, paxton2019ApJS, jermyn2023ApJS}.
Since the inner part of the star is too dense and therefore too computationally expensive for 3D simulations, we replace the inner $5\%\, R_\mathrm{bol,MESA}$ with a point mass and a modified polytrope of polytropic index $\gamma=4/3$ that is stably stratified following \citet{ohlmann2017A&A}.
Here, $R_\mathrm{bol,MESA}$ is the \texttt{MESA} bolometric stellar radius.
The star itself is placed at the middle of the simulation box, surrounded by a background filled with low density $\rho_\mathrm{bg}=10^{-17}\, \mathrm{g\, cm^{-3}}$ and low temperature $T_\mathrm{bg}=1800$ K pseudo-vacuum.
We choose the box size to be $300\, R_\mathrm{bol,MESA}$ and use periodic boundary conditions on all sides for the hydrodynamic quantities but not for gravity.
Effectively there is no boundary, because only waves exceeding $60\, \mathrm{km\, s^{-1}}$ can reach the box boundary within $20$ years of simulation time, while the typical outgoing convective speed is below $30\, \mathrm{km\, s^{-1}}$.

To supply an energy source and to keep the inner part stable during the simulation, we introduce an `artificial core' region in our simulation (inner $7\%\, R_\mathrm{bol,MESA}$) that extends beyond the radius where we modified the initial profile.
Within the artificial core, we apply a constant radiative luminosity ($10^{4.6}\, L_\odot$, the surface luminosity of the \texttt{MESA} profile taken as the initial condition) to the radiation transport module.
This is achieved by using a radiation flux profile that follows $F_\mathrm{r}=L_\mathrm{s,MESA}/(4\pi r^2)$, where $r$ is the distance from the center and $F_\mathrm{r}$ is the radial component of the radiation flux.
The intensities are constructed such that the radiation energy density is in equilibrium with local temperature $T$ and gives the correct radiation flux $F_\mathrm{r}$ assuming Eddington approximation, i.e., $I_n = caT^4/(4\pi) + 3F_\mathrm{r}(\vectoraas{n}_n\cdot\vectoraas{\hat{r}})/(4\pi)$, where $\vectoraas{\hat{r}}$ is the unit vector along the radial direction.
Inside the artificial core, we also actively damp the core velocities throughout the simulation, such that $\rho\vectoraas{\dot v}=-\rho\vectoraas{v}/\tau_\mathrm{c}$, where we choose a damping timescale $\tau_\mathrm{c}$ to be $5\%$ of the global sound-crossing timescale of the \texttt{MESA} star.
We find that the core damping is necessary to keep the hydrostatic structure of the core beyond $150$ global dynamical timescales.
We note that this artificial core sits well within the radiative zone (inner $12\%$), below the convective envelope in the 1D profile.
We therefore simulate the entire convective envelope and a part of the radiative zone below in our computational domain.

For the production run shown here, we use the \textit{time-independent} version of our scheme, where we assume the transport term exactly balances the source terms and solve Equation~\eqref{eq:rt:notime}.
This is still valid for non-relativistic simulations such as the simulation presented here.
We find that it is still necessary to implicitly evolve the temperature field simultaneously with the radiation field.
We include the radiation energy and radiation pressure assuming LTE in the equation of state for density $\rho > 10^{-8}\, \mathrm{g\, cm^{-3}}$ and smoothly transit to a pure gas equation of state for density $\rho < 10^{-10}\, \mathrm{g\, cm^{-3}}$, such that the radiation is not included in the equation of state in the optically-thin regime.
In the hydrodynamic solver, we include the radiation source terms in the energy equation for radiative heating/cooling, and only include them in the momentum equation for $\rho < 10^{-8}\, \mathrm{g\, cm^{-3}}$ with a compensating transition region between $10^{-10}-10^{-8}\, \mathrm{g\, cm^{-3}}$ to take into account the possible radiation forces in the optically-thin regime.
Given that the radiative diffusion timescale is orders of magnitude larger than the local timesteps, we only switch on the radiation transport module on the globally synchronised timesteps to save computational time.
At local timesteps that are not synchronised, we only evolve the hydrodynamics and not the radiation field in active cells.
This approach does not guarantee energy or momentum conservation, but as we show in Figure~\ref{fig:test:rsg:conservation}, it does not lead to significant deviations from conservation laws.
We have also attempted to use the time-dependent version of our scheme with local time-stepping on the moving mesh, which did not introduce notable errors by the scheme itself.
However, we find that the inner boundary condition for the time-dependent radiation transport module is difficult to set while keeping the inner core region stable for more than $50$ sound-crossing timescales.
We therefore choose to use the time-independent scheme, and defer the possibility of using the time-dependent scheme to future explorations.

\begin{figure}[htb!]
\centering
\includegraphics[width=\columnwidth]{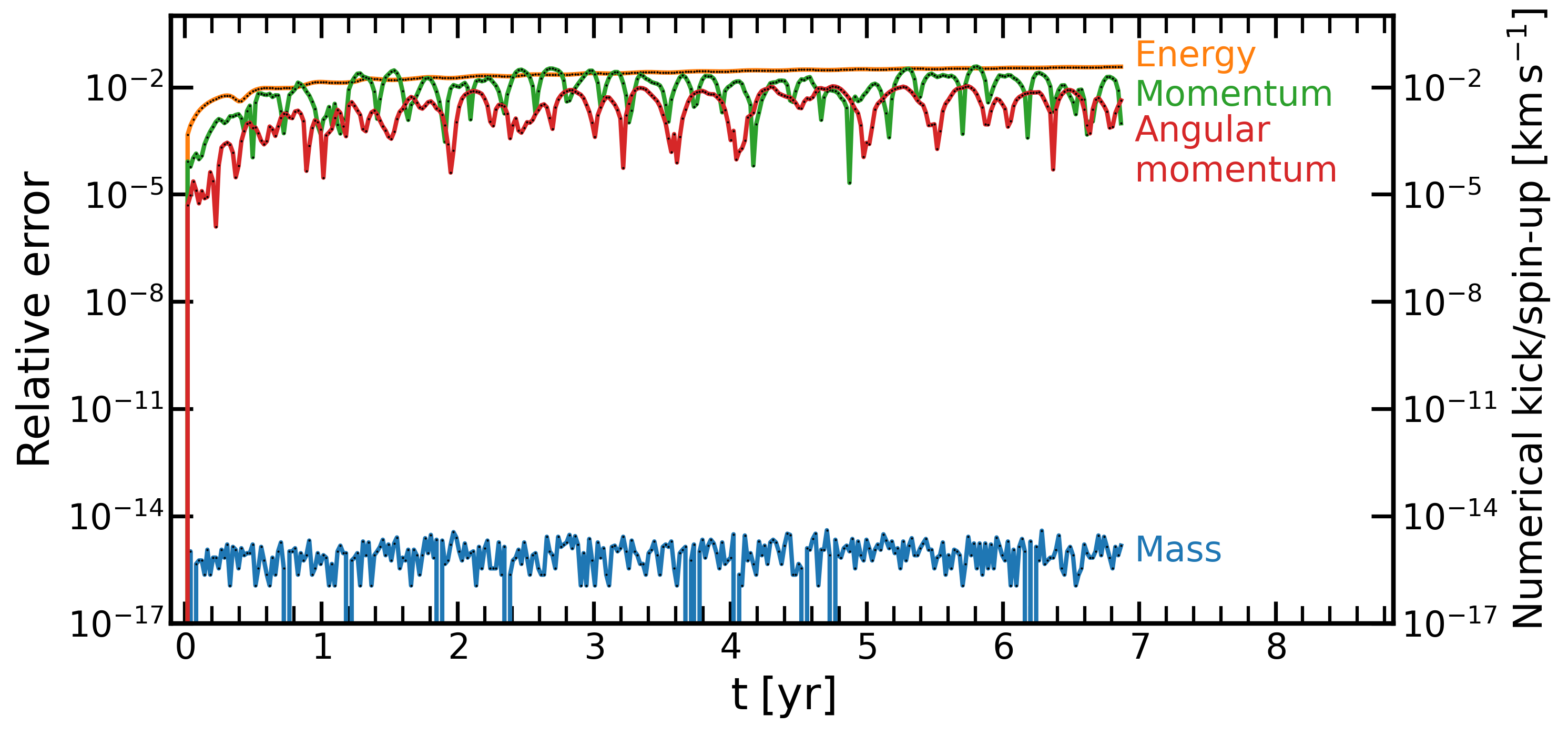}
\caption{
Relative error in conserved global quantities as functions of time for the $10\, M_\odot$ 3D \texttt{AREPO} red supergiant model.
The total mass (blue) is conserved to near machine precision.
As we apply the damping term in the artificial core, the toal energy (thermal + radiation + kinetic + gravitational potential), momentum, and angular momentum are not expected to be conserved.
Therefore, we observe a relative error at the order of $1\%$ in total energy (orange) compared to its initial value.
The momentum and angular momentum are initialized to be zero, so the concept of relative error is not applicable.
Here, we normalize the momentum along the z axis by $M_\mathrm{MESA}v_0=2\times 10^{39}\, \mathrm{g\, cm\, s^{-1}}$, and normalize the angular momentum along the z axis by $0.1 M_\mathrm{MESA}R_\mathrm{bol,MESA}v_0=5.3\times 10^{51}\, \mathrm{erg\, s}$, where we take a reference $v_0=1\, \mathrm{km\, s^{-1}}$.
The green and red curves therefore indicate that the star receives a fluctuating numerical `kick' along the z axis of about $0.01\, \mathrm{km\, s^{-1}}$, and a fluctuating numerical `spin up' along the z axis of about $0.01\, \mathrm{km\, s^{-1}}$ or $10^{-11}\, \mathrm{rad\, s^{-1}}$, which are negligible compared to their normal values of any physical origin for red supergiants.
}
\label{fig:test:rsg:conservation}
\end{figure}

Radiation transport is performed with $80$ directions.
This is slightly smaller than the $120$ directions adopted in the \texttt{ATHENA++} red supergiant simulation \citep{goldberg2022ApJ}, but we find that it does not lead to significant ray-effects.
This is demonstrated in Figure~\ref{fig:test:rsg:ray}, where $24$ or $48$ directions lead to noticeable radiation peaks in the optically-thin atmosphere of the star.
Since the radiation is coupled to the temperature field, this also leads to peaks in temperature field in the stellar atmosphere later on in the simulations.
We thus choose $80$ directions to save computational cost and memory cost while avoiding significant ray-effect.

\begin{figure}[htb!]
\centering
\includegraphics[width=\columnwidth]{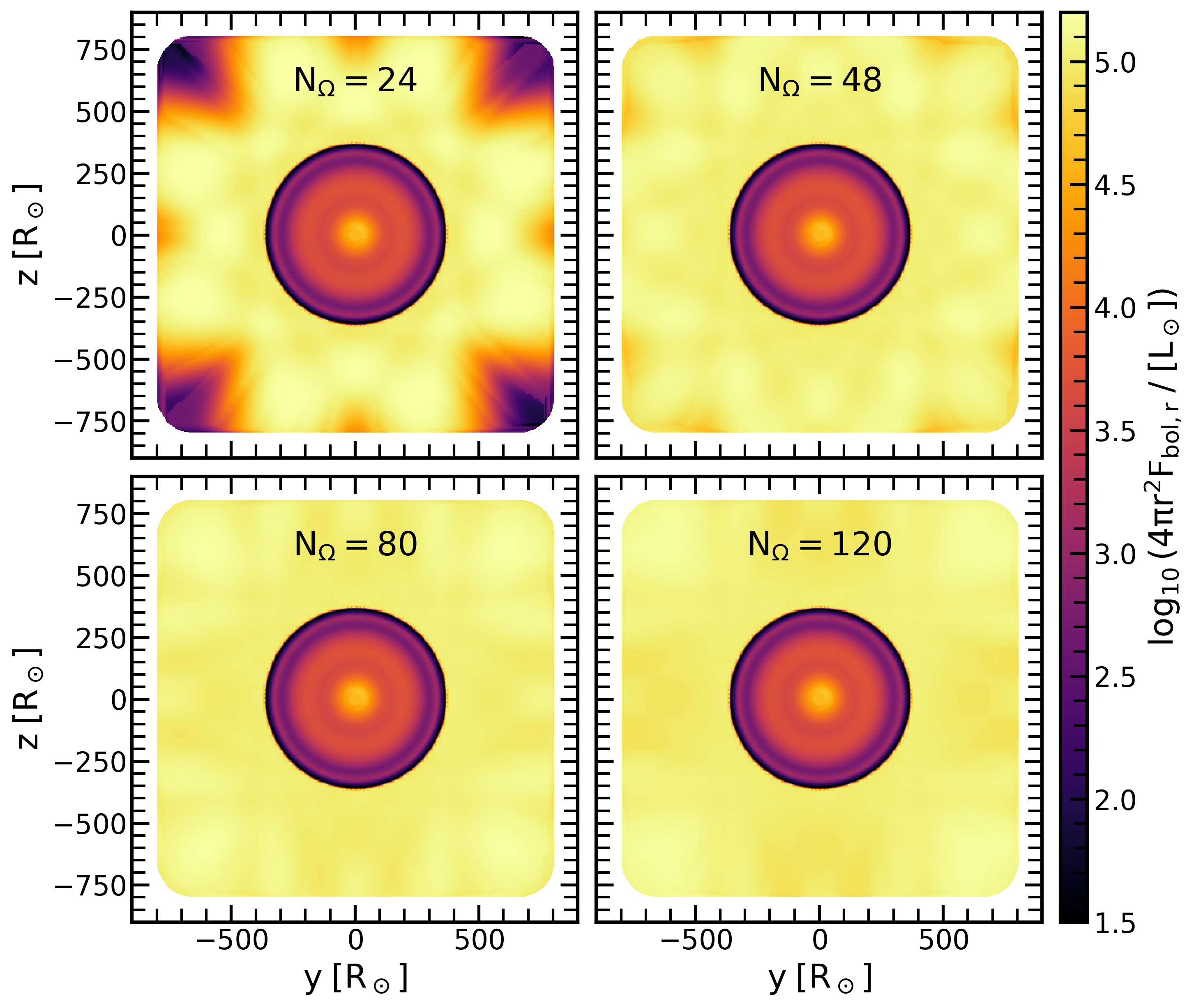}
\caption{
Demonstration of the ray-effect in the $10\, M_\odot$ 3D \texttt{AREPO} red supergiant model.
Different panels show 2D slices of the star when the number of directions $N_\Omega$ in the radiation transport module is taken as $24,48,80,120$.
The color shows the radial radiative luminosity $4\pi r^2 F_\mathrm{bol,r}$.
Inside the star, the radiation is close to the diffusion limit, and the radiative flux is not sensitive to the number of directions chosen.
Across different panels, the radiative fluxes inside the star agree with each other at a relative error below $10^{-10}$.
However, in the optically-thin atmosphere of the star where radiation propagates freely, a small number of directions lead to radiation peaks along the discretized directions.
We generally do not find significant ray-effect for $N_\Omega = 80$ and $120$.
}
\label{fig:test:rsg:ray}
\end{figure}

The simulation includes full self-gravity, uses the \texttt{OPAL} equation of state \citep{rogers2002ApJ}, and Rosseland and Planck opacities $\kappa_\mathrm{R},\kappa_\mathrm{P}$ from high-temperature \texttt{Los Alamos OPLIB} table \citep{colgan2016ApJ}\footnote{\url{https://aphysics2.lanl.gov/apps/}  The opacity due to electron scattering is included in the Rosseland opacity table.} blended with low-temperature \citet{ferguson2005ApJ} table\footnote{\url{https://www.wichita.edu/academics/fairmount\_las/physics/Research/opacity.php}}.
We assume $\kappa_\mathrm{E}\approx \kappa_\mathrm{P}$ and $\kappa_\mathrm{F}\approx \kappa_\mathrm{R}$, which are not good approximations in the optically-thin regime.
But this is the standard approach for 3D grey simulations of stellar atmosphere \citep{freytag2012JCoPh, goldberg2022ApJ}.
We enable local time-stepping for hydrodynamics and gravity on the moving Voronoi mesh in the simulation, which is the main reason that we can afford to run simulations that include the entire convective envelope.

To pass the initial relaxation phase of the simulation, we change the damping terms and the surface resolution during the relaxation.
For the initial $10$ dynamical timescales of the star, we apply a global damping term to the momentum across the simulation box to aid relaxation following \citet{ohlmann2017A&A}.
We drop the global damping afterwards, but we keep the strong damping in the artificial core until the end of the simulation.
We use a target mass resolution of $3.4\times 10^{-6}\, M_\odot$.
We have checked that this leads to sufficient resolution near the artificial core, but not sufficient to resolve the stellar surface.
We therefore adopt several nested layers for the maximum cell volume as the (de-)refinement criterion.
In the spherical shell between $0.2-3\, R_\mathrm{bol,MESA}$, we further constrain the maximum cell size to be $7.4\, R_\odot$.
We find that this is still not enough to resolve the photosphere, in a sense that the outgoing luminosity from the 3D simulation is two times larger than the input luminosity in the artificial core.
Therefore, after about $25$ global dynamical timescale of the \texttt{MESA} star, we further decrease the maximum cell size to be $2\, R_\odot$ between $0.9-1.5\, R_\mathrm{bol,MESA}$, and $1.5\, R_\odot$ between $0.95-1.15\, R_\mathrm{bol,MESA}$.
We find that this is enough to marginally resolve the stellar photosphere, in a sense that the time-averaged outgoing luminosity matches the input.
% On average, we have 3 cells per photon mean free path at the photosphere, and once in a while 6 cells per photon mean free path during the simulation.

\subsection{Results}

\begin{figure}[htb!]
\centering
\includegraphics[width=\columnwidth]{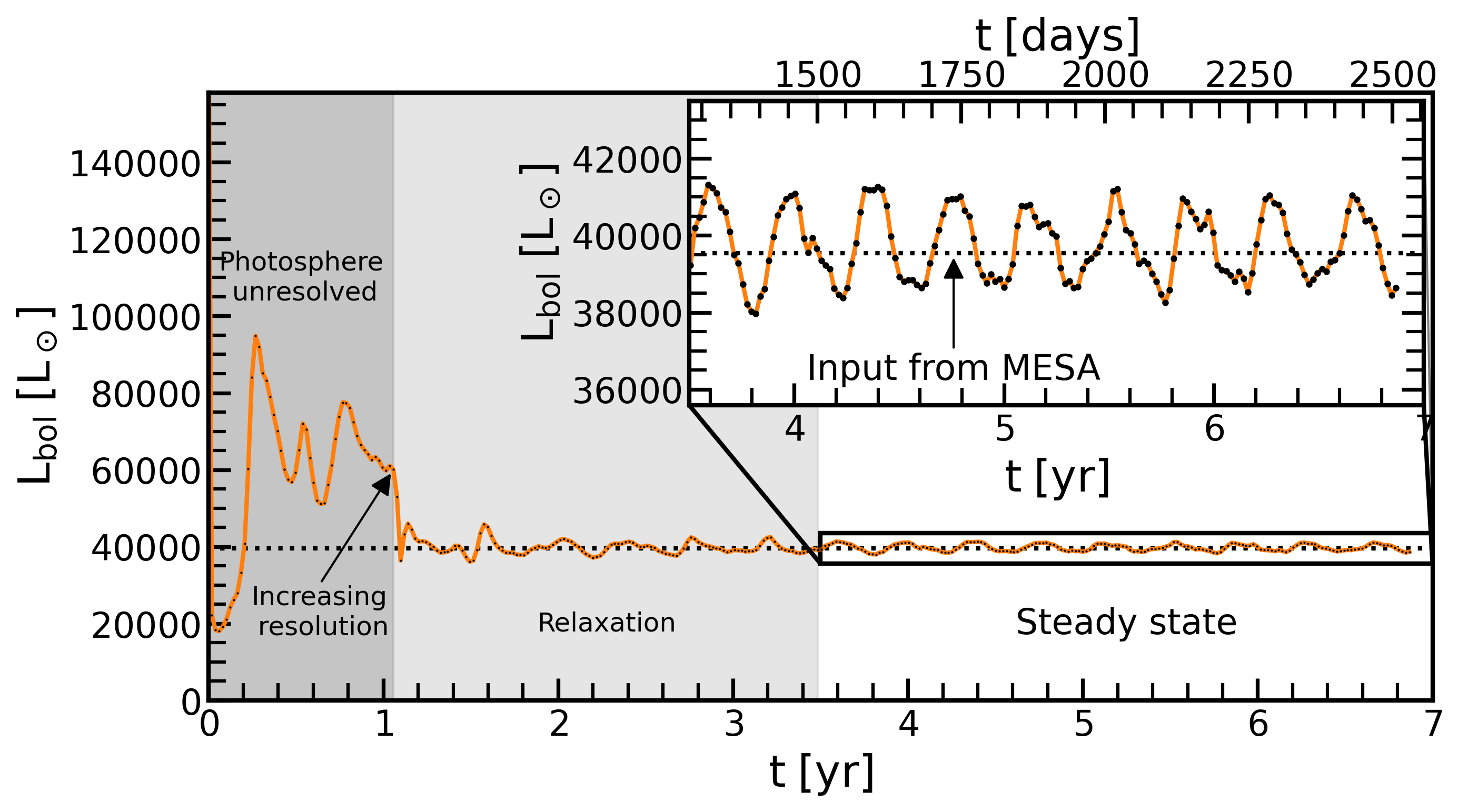}
\caption{
Bolometric lightcurve of the $10\, M_\odot$ 3D \texttt{AREPO} red supergiant model.
We start the simulation at time $t=0$.
During the initial relaxation phase $t=0-1.1$ yr, the resolution near the stellar surface is not enough to resolve the photosphere, and the bolometric luminosity $L_\mathrm{bol}$ does not saturate around the input luminosity (dotted black horizontal line).
Therefore, at $t=1.1$ yr (indicated by the arrow), we suddenly increase the surface resolution from $7.4\, R_\odot$ to $1.5\, R_\odot$.
We find that this is enough for $L_\mathrm{bol}$ to drop to the target value.
After $t>3.5$ yr, the bolometric luminosity settles into a regular variable state around the input luminosity, as shown in the zoom-in plot.
The black dots on the orange line show the snapshots used to draw the lightcurve, which are equally spaced by $7.6$ days.
The overall variability is likely due to the self-excited pulsation of the model, which merits further analyses.
}
\label{fig:test:rsg:lc}
\end{figure}

The relaxation process is shown in the bolometric lightcurve of the 3D \texttt{AREPO} red supergiant model in Figure~\ref{fig:test:rsg:lc}, where we show the bolometric luminosity $L_\mathrm{bol}$ as a function of time.
The bolometric luminosity is calculated by taking the radial component of the radiation flux from the \texttt{AREPO} output (denoted as $F_\mathrm{r,r}$), and averaging the quantity $4\pi r^2 F_\mathrm{r,r}$ over all the cells between $4-5\, R_\mathrm{bol,MESA}$, where $r$ is the distance from each cell to the box center.\footnote{Alternatively, it may seem more intuitive to calculate the bolometric luminosity by integrating the normal radiation flux over all the box boundaries. However, since the box boundary is far away from the star with very low resolution, it is more accurate to calculate the luminosity closer to the star, as we describe in the main text.}
Eventually after $3.5$ yr, the 3D star reaches a steady state, as shown in the zoom-in plot in Figure~\ref{fig:test:rsg:lc}.
On top of other variability, the bolometric luminosity exhibits a regular pulsation pattern around the input luminosity value.
The pulsation is likely driven by the $\kappa$-mechanism \citep{eddington1917TheObservatory, cox1958ApJ, baker1962Z.Astrophys., zhevakin1963ARA&A} where the opacity increases with temperature due to hydrogen recombination \citep{heger1997A&A, joyce2020ApJ}, but detailed analyses are deferred to future work.

Taking a steady-state simulation snapshot at time $3.4$ yr, we show different timescales and optical depth as functions of radial coordinates in Figure~\ref{fig:test:rsg:rt}.
We compute the shell-averaged timescales through radial binning, i.e., taking the average value of quantities in small bins of radial coordinates, weighted by either mass or volume.
The shell-average of quantity $q$ weighted by mass is denoted as $\langle q\rangle_m (r) \equiv \left[\sum_{|r_i-r|<\delta R/2}(q_i\times \Delta m_i)\right]/\left[\sum_{|r_i-r|<\delta R/2}\Delta m_i\right]$, where $\Delta m_i$ is the mass enclosed in each cell, $\delta R=R_\mathrm{bol, MESA}/1000$ is the size of the bin, and the sum is taken over all the cells $i$ whose radial distance $r_i$ from the center fall into the bin.
Similarly, the shell-average of quantity $q$ weighted by volume is denoted as $\langle q\rangle_V (r) \equiv \left[\sum_{|r_i-r|<\delta R/2}(q_i\times \Delta V_i)\right]/\left[\sum_{|r_i-r|<\delta R/2}\Delta V_i\right]$, where $\Delta V_i$ is the volume enclosed in each cell.
We define the shell-averaged convective timescale $\langle t_\mathrm{conv} \rangle$, radial pressure scale height $\langle H_p \rangle$, thermal timescale $\langle t_\mathrm{th} \rangle$, and Rosseland optical depth $\langle \tau_\mathrm{Ross} \rangle$ as follows:
\begin{align}
    & \langle t_\mathrm{conv} \rangle (r)\equiv \frac{\alpha_\mathrm{MLT}\langle H_p \rangle(r)}{\sqrt{\langle v_r^2\rangle_m (r)}}\, , \\
    &  \langle H_p \rangle (r)\equiv -\left(\frac{\langle\vectoraas{\nabla}P\cdot \hat{r} \rangle_V}{\langle P \rangle_V}\right)^{-1}\, ,\\
    & \langle t_\mathrm{th} \rangle (r)\equiv \frac{\sum_{r_i>r}(c_pT\Delta m)_i}{L_\mathrm{bol}}\, ,\\
    & \langle \tau_\mathrm{Ross} \rangle (r)\equiv \sum_{r_i>r}\left[\frac{(\kappa_\mathrm{R}+\kappa_\mathrm{s})\Delta m}{4\pi r^2}\right]_i\, ,
\end{align}
where $c_p$ is the specific heat capacity at constant pressure, $\hat{r}$ is the unit vector in the radial direction and we take $\alpha_\mathrm{MLT}=3$.
The pulsation period is found by taking the Fourier transformation of the relaxed bolometric light curve (Figure~\ref{fig:test:rsg:lc}) and identifying the period of the highest peak in the power spectrum.

In Figure~\ref{fig:test:rsg:rt}, we shade the artificial core in grey, and only analyze the regions outside of the artificial core.
Throughout the convective envelope, the convective timescale (blue) is mostly comparable with the pulsation period (black).
This indicates that convection will interact with pulsation, and 1D treatments of both processes may not be adequate to capture the physics.
The thermal timescale (red) is smaller than the convective timescale at the top of the convective envelope, but is still orders of magnitude larger at the bottom of the envelope.
This is beyond reach for the total simulation time (grey dotted horizontal line).
However, we argue that it does not mean the inner part of the envelope is out of thermal equilibrium in 3D simulations.
As long as we start the simulation from a profile close to the real thermal equilibrium state, it is unlikely for the simulation to deviate too much from reality.
This is why it is essential to take the initial condition from a 1D stellar evolution code, as opposed to relaxing the simulation from a previous 3D simulation of different parameters \citep[as done e.g. in \texttt{CO5BOLD} simulations;][]{ahmad2023A&A}.
Nevertheless, the simulation duration is still an order of magnitude larger than the convective timescale and the pulsation period, and therefore the simulation is dynamically relaxed to the steady state.
We further show the scatter plot of local timesteps taken by all the cells in the simulation.
The smallest timestep is taken by the innermost cells in the artificial core.
The bottom of the convective envelope takes a timestep $16$ times smaller than the largest timestep in the outer envelope.
This is why simulating the entire envelope is beyond reach for 3D codes that only employ global time-stepping.

\begin{figure}[htb!]
\centering
\includegraphics[width=\columnwidth]{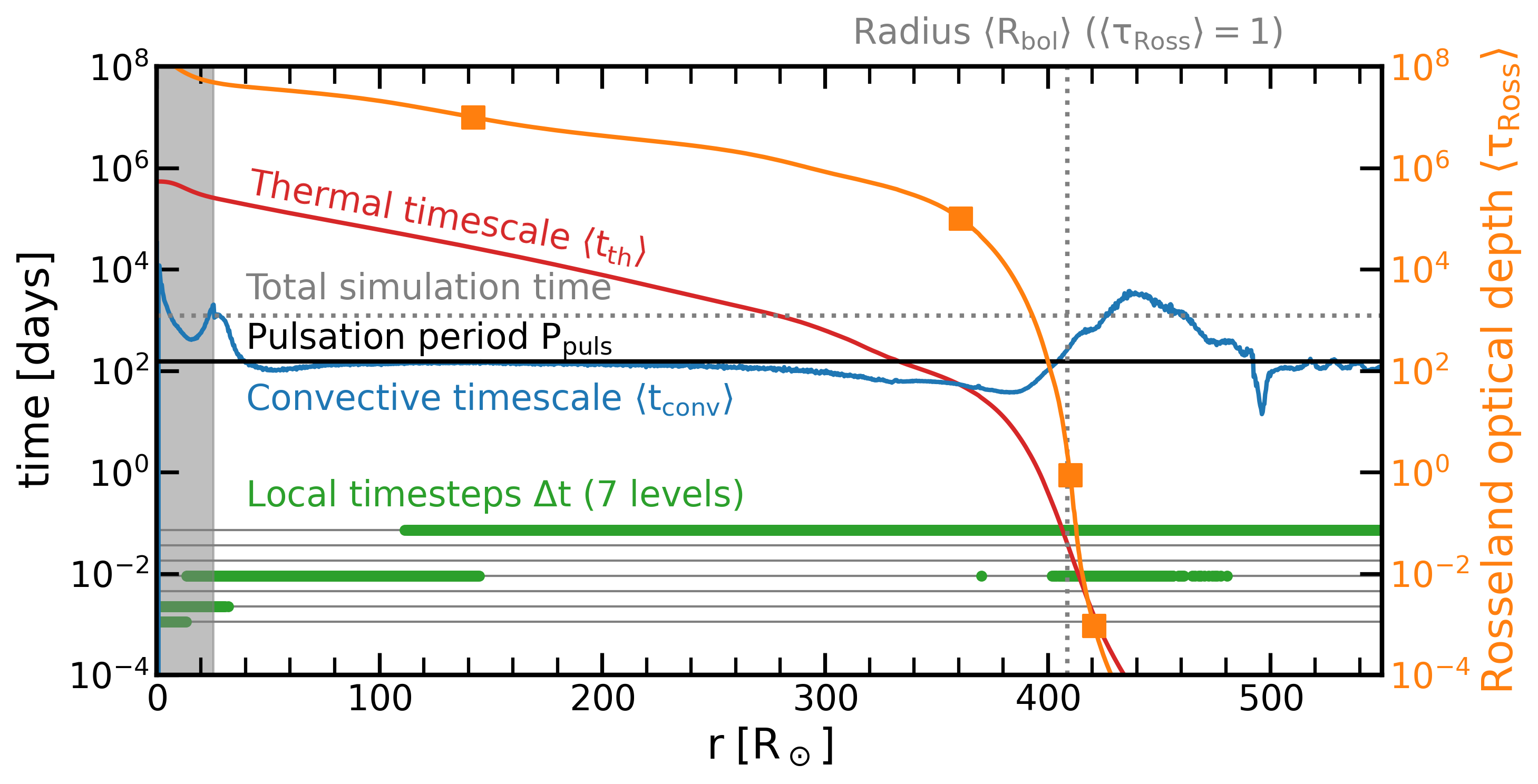}
\caption{
Shell-averaged timescales and Rosseland optical depth profile  of the $10\, M_\odot$ 3D \texttt{AREPO} red supergiant model at time $3.4$ yr.
We show the local timestep distribution in green that occupies 7 levels of hierarchy, which enables us to simulate the entire convective envelope.
The artificial core within $7\%\, R_\mathrm{bol,MESA}$ is shaded in grey (see main text).
We further show different timescales across the envelope: the convective timescale (blue), pulsation period (black), and thermal timescale (red), as defined in the main text.
% The convective timescale and the pulsation period are comparable through most of the envelope, which suggests that convection is likely to interact with pulsation.
% The total simulation time (grey dotted horizontal line) is at least an order of magnitude larger than both timescales, and therefore both processes already reach steady states in our simulation.
% The thermal timescale is orders of magnitude higher than others in the deep envelope, but becomes smaller than others at the outer layer of the envelope.
The smallest timestep is 5 orders of magnitude smaller than the convective timescale and the pulsation period, and 6 orders of magnitude smaller than the total simulation time, which showcases the vast range of timescales covered in the simulation.
In orange, we show the shell-averaged Rosseland optical depth $\langle \tau_\mathrm{Ross}\rangle$, which is a steep function of radial coordinate near the bolometric photosphere (grey dotted vertical line, defined as where $\langle \tau_\mathrm{Ross}\rangle=1$).
We mark different optical depths in orange squares to indicate the approximate locations of the 2D maps shown in Figure~\ref{fig:test:rsg:2d}.
}
\label{fig:test:rsg:rt}
\end{figure}

In Figure~\ref{fig:test:rsg:rt}, we further show the shell-averaged Rosseland optical depth (orange).
We mark the averaged bolometric photosphere (grey dotted vertical line), defined as where the grey optical depth reaches unity.
From outside the grey stellar photosphere to just beneath the photosphere, the optical depth increases steeply by more than $10$ orders of magnitude.
It is therefore essential to have enough resolution near the photosphere to resolve the region where photons escape.
Here, we only marginally resolve this region such that the outgoing luminosity matches the input.
We also mark the location of $\langle \tau_\mathrm{Ross}\rangle =10^7, 10^5, 1, 10^{-3}$ with orange squares, where we take the simulation data to plot the tomography of the star in Figure~\ref{fig:test:rsg:2d}.
Besides the highest optical depth $\langle \tau_\mathrm{Ross}\rangle =10^7$ that is located at about $1/3$ stellar radii, all the others are located rather close to the photosphere.
But as we show in Figure~\ref{fig:test:rsg:2d}, they display distinct flow patterns across different optical depths.

\begin{figure*}[htb!]
\centering
\includegraphics[width=\textwidth]{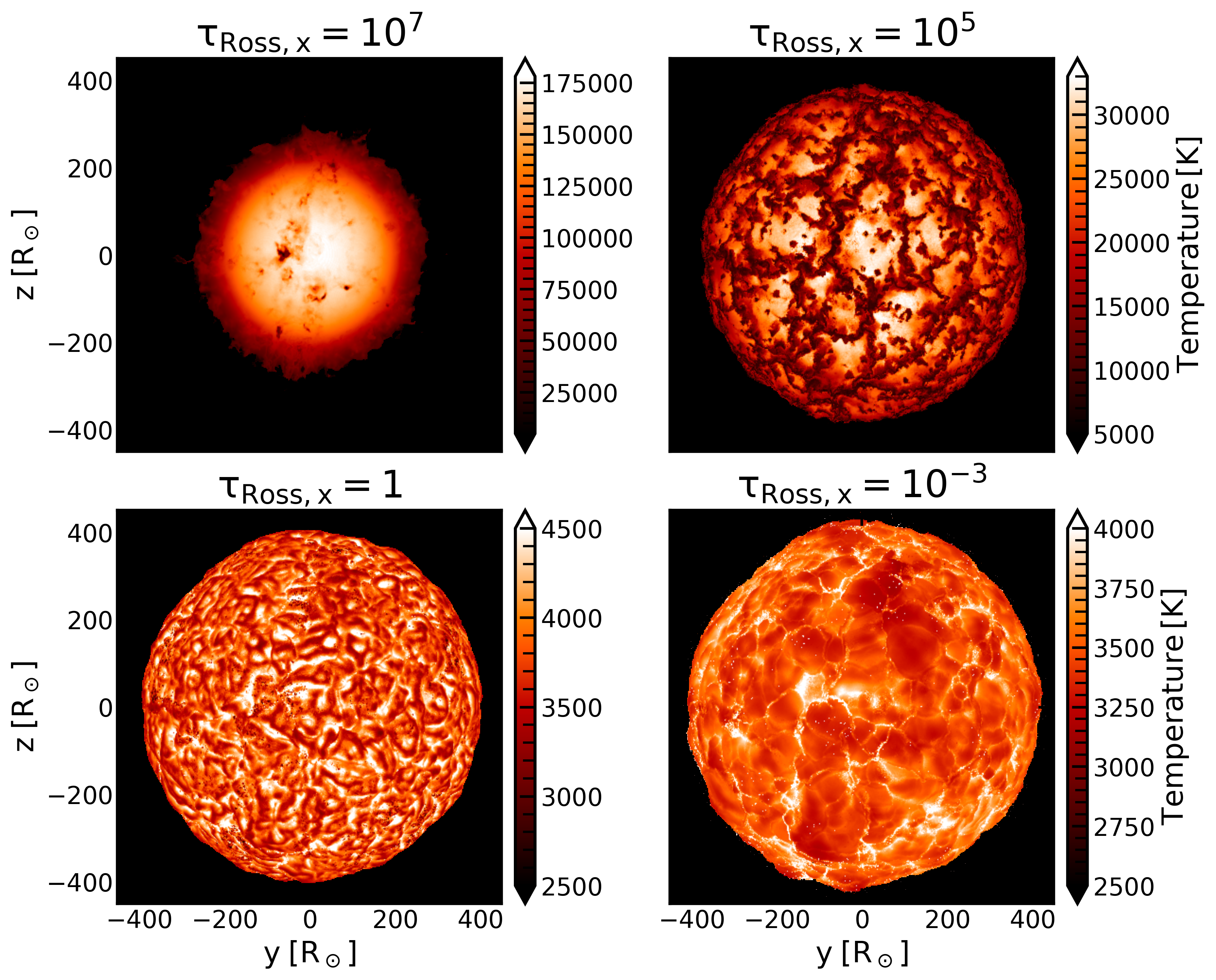}
\caption{
Tomography of the $10\, M_\odot$ 3D \texttt{AREPO} red supergiant model showing 2D temperature maps at different optical depths at time $3.4$ yr.
% The top row shows the temperature, and the bottom row shows the radial component of the velocity.
We scan through the star along the x axis, taking the integrated Rosseland optical depths of $10^7, 10^5, 1, 10^{-3}$.
The optical-depth tomography scans from deep inside the star all the way to the atmosphere outside of the bolometric photosphere.
The convective pattern changes substantially.
This plot highlights the importance of simulating the entire convective envelope and simultaneously marginally resolving the photosphere, which gives us the unprecedented advantage of obtaining the first comprehensive picture of red supergiants from deep inside to high up in the atmosphere, with significant consequences for observational signatures.
}
\label{fig:test:rsg:2d}
\end{figure*}

The fact that we simulate the entire envelope while marginally resolving the photosphere means that we can resolve the flow patterns both in the deep convection envelope and in the atmosphere.
In Figure~\ref{fig:test:rsg:2d}, we present a tomography of our 3D \texttt{AREPO} red supergiant model probed at different layers of optical depths along the x axis.
We take a model snapshot in the steady state at $3.4$ yr, interpolate it onto a Cartesian grid, and calculate the Rosseland optical depth $\tau_\mathrm{Ross,x}(x,y,z)=\int_{-\infty}^x \rho(x,y,z)(\kappa_\mathrm{R}+\kappa_\mathrm{s})(x,y,z)dx$ by integrating along the x axis from infinity to negative infinity.
In Figure~\ref{fig:test:rsg:2d}, we scan through different optical depths from the deep stellar interior to the bolometric photosphere, and eventually into the stellar atmosphere.
% The top row shows the 2D map of the temperature, whereas the bottom row shows the radial component of the velocity.
As illustrated in Figure~\ref{fig:test:rsg:2d}, none of the temperature maps looks alike across different optical depths.
This is in part because radiative cooling and heating gradually becomes important compared to adiabatic expansion as a fluid parcel travels from the deep convective envelope to the upper stellar atmosphere, which will be detailed in the upcoming paper (Ma et al. in prep.).
The distinct morphology across optical depths will have profound implications for interpreting spectroscopic and interferometric observations of red supergiants, which will be explored in the future.

\section{Discussion on Caveats and Directions for Future Development}
\label{sec:disc:code}

\subsection{From Grey to Multigroup}
\label{sec:disc:code:multigroup}

The radiation transport scheme described in this work is currently limited to the grey approximation, where the frequency-integrated Rosseland and Planck opacities are used to calculate the bolometric intensities.
However, in many astrophysical scenarios, different radiation sources emit in distinct frequency bands.
For example, for irradiated exoplanets such as hot Jupiters, the radiation from internal heating is released in infrared (IR), whereas the irradiation from the host star is concentrated in optical \citep{guillot2010A&A}.
For problems involving photo-ionization and photo-dissociation such as protoplanetary disks \citep[e.g.][]{woitke2016A&A}, star formation \citep[e.g.][]{chon2024MNRAS} or the epoch of reionization \citep[e.g.][]{stark2016ARA&A}, strong ultraviolet and X-ray radiation is emitted from (proto)stars which ionize surrounding atoms and dissociate molecules, while in the meantime the dust absorbs the optical and UV radiation and re-emits in IR.
For tidal disruption events, the optical/UV emission and X-rays are observed to have distinct blackbody temperatures and thus emitted from distinct regions \citep{gezari2021ARA&A}.
In general, grey frequency-integrated opacities based on the Planck function are merely crude approximations in the optically-thin regions \citep{malygin2014A&A, jiang2022ApJS, jiang2023Galaxies}, and frequency-dependent opacities are needed for more accurate treatments.

Since the frequency-by-frequency radiation transport is too expensive even in 1D simulations, it is more feasible to adopt a multigroup approach by solving the radiation transport in several frequency bins \citep{mihalas1984}.
The implicit discrete ordinates scheme in \citetalias{jiang2021ApJS} has already been extended to multigroup with Compton scattering in \citet{jiang2022ApJS}.
The multigroup extension in \texttt{ATHENA++} was successfully applied to accretion disks around AGN \citep{secunda2024ApJ} and stellar-mass black holes \citep{mills2024ApJ}, which demonstrates the potential of the scheme.
Given that the multigroup extension does not require a structured static mesh, we expect that it will be straightforward to extend our scheme to multigroup.

\subsection{From Fast to Faster: GPU Acceleration and Node-to-node Communication}
\label{sec:disc:code:faster}

The speed of our scheme depends on the specific problem.
But in general, the computational cost is approximately proportional to $\mathcal{O}(N_\mathrm{p} N_\mathrm{\Omega} N_\mathrm{iter})$, where $N_\mathrm{p}$ is the number of cells, $N_\mathrm{\Omega}$ the number of angles, and $N_\mathrm{iter}$ the number of iterations to reach the convergence criterion.

To demonstrate that our method can be fast for complex 3D simulations, in Figure~\ref{fig:test:rsg:cpu} we show the fraction of CPU hours spent by each module in \texttt{AREPO} for the 3D red supergiant simulation described in Section~\ref{sec:test:rhd:rsg}.
With $N_\mathrm{p} = 2\times 10^7,  N_\mathrm{\Omega} = 80, N_\mathrm{iter} = 10$, we find that the radiation module takes about $47\%$ of the computational cost, whilst $18\%$ is spent on tree-particle-mesh gravity, $22\%$ on Voronoi mesh construction, and $4\%$ on the hydro solver.
% Extrapolating this estimate to other problems, we estimate that for a complex 3D simulation with moving mesh, 
This means that, in this complex 3D example, including radiation only increases the computational cost by a factor of two, and even less if the radiation transport is only applied at the globally synchronised timestep.
If our scheme were to be applied to a simplified 3D simulation on a fixed mesh with a simple treatment of gravity, a RMHD simulation is $10-20$ times as expensive as a pure MHD simulation, which is the typical cost of accurate on-the-fly radiation transport.

\begin{figure*}[htb!]
\centering
% First row
    \begin{subfigure}{0.49\textwidth}
        \centering
        \includegraphics[width=\linewidth]{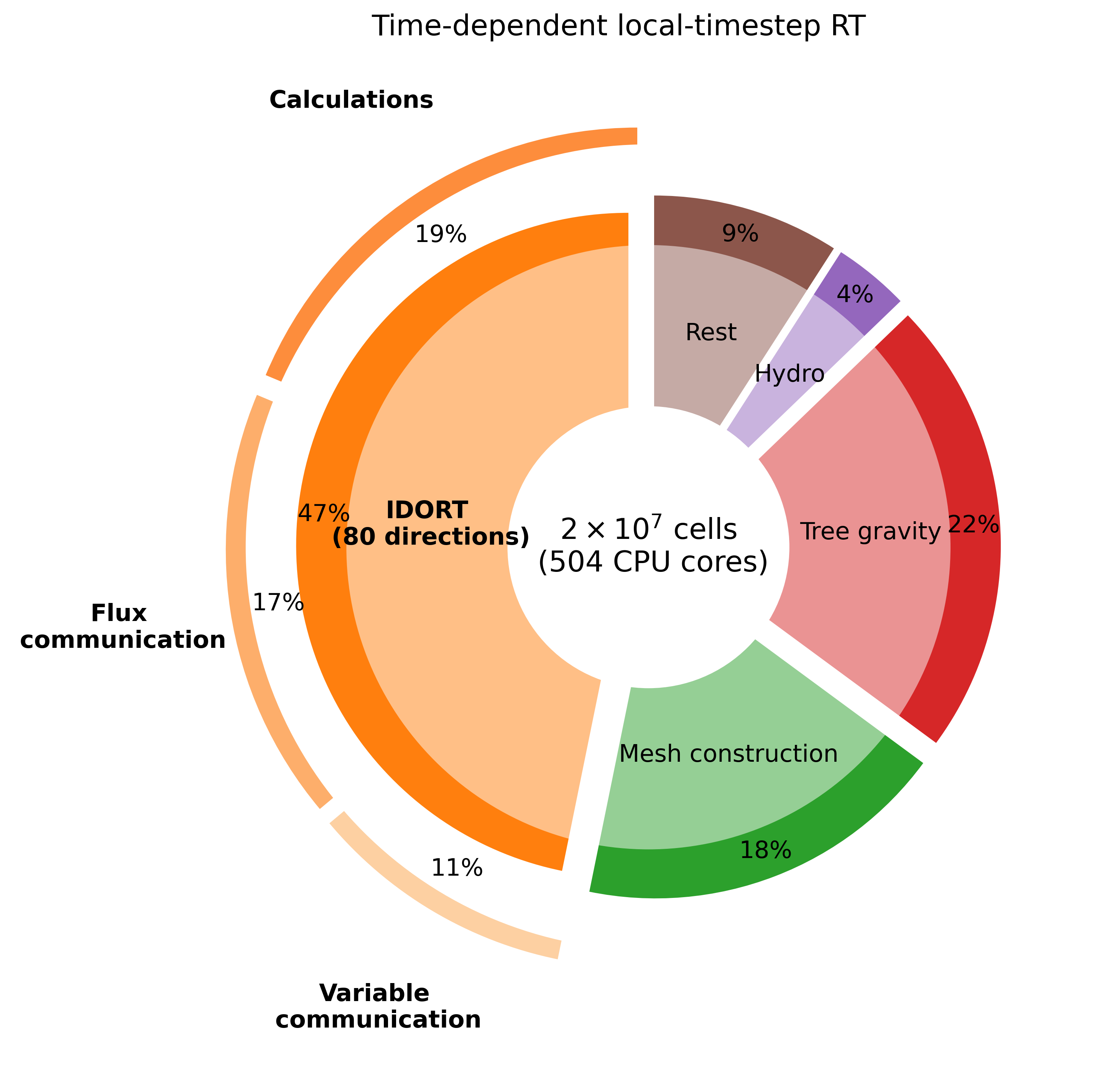}
        \caption{For time-dependent version of the radiation module applied at each local timestep.}
    \end{subfigure}
    \hfill
    \begin{subfigure}{0.49\textwidth}
        \centering
        \includegraphics[width=\linewidth]{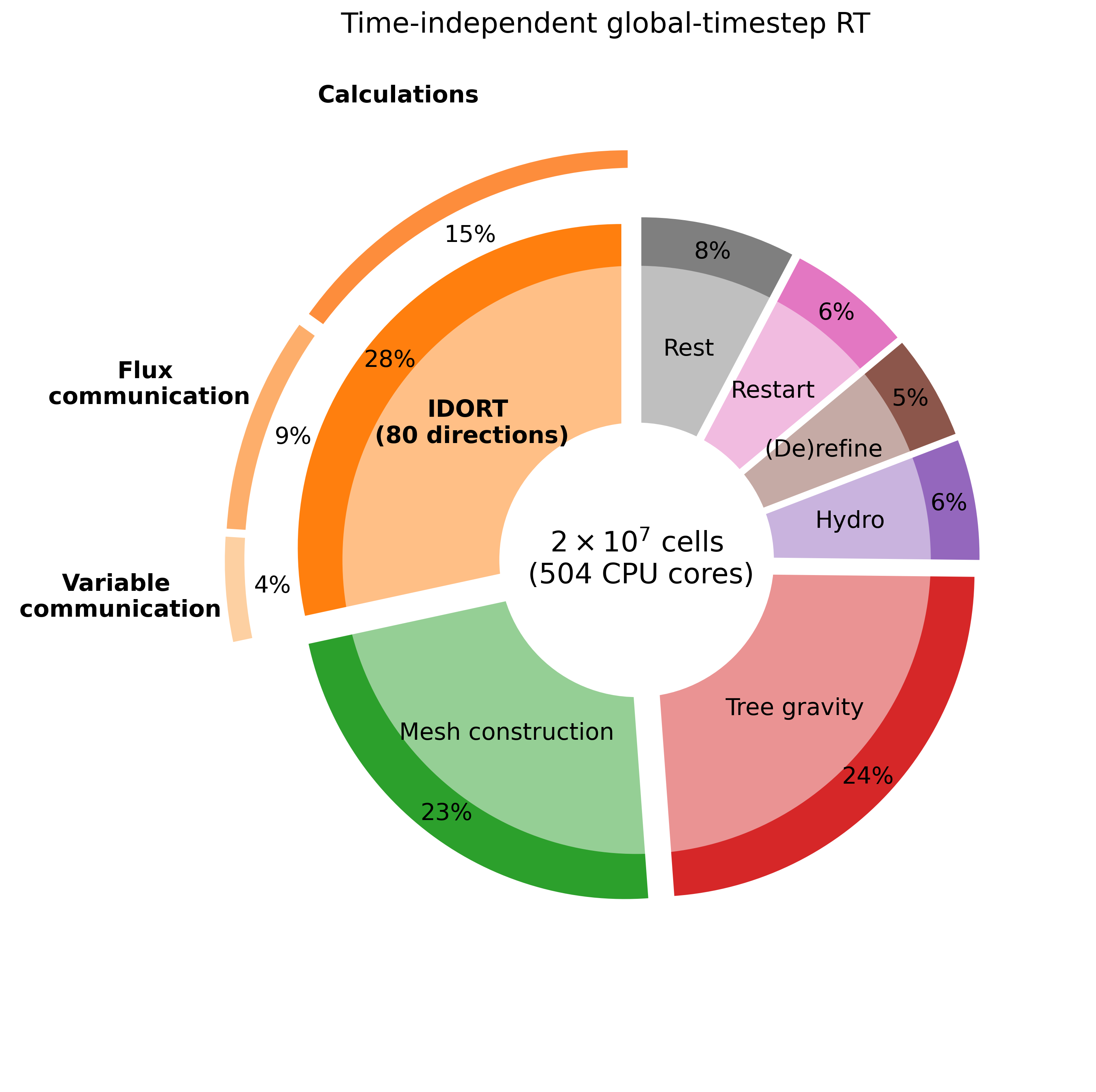}
        \caption{For time-independent version of the radiation module applied only at the globally synchronised timestep.}
    \end{subfigure}
% \begin{figure}[htb!]
% \centering
% \gridline{
% \fig{cpu_250117_real.png}{0.5\textwidth}{(a) For time-dependent version of the radiation module applied at each local timestep.}
% \fig{cpu_250117.png}{0.5\textwidth}{(b) For time-independent version of the radiation module applied only at the highest global timestep.}
% }
% \includegraphics[width=0.5\textwidth]{cpu_250117.png}
\caption{
Percentage of CPU hours spent by each module in \texttt{AREPO} for the 3D red supergiant simulation in Section~\ref{sec:test:rhd:rsg}.
With $80$ directions, the \textit{time-dependent} version of the radiation transport module takes about $1/2$ of the computational cost if applied at each local timestep.
The \textit{time-independent} version of the radiation transport module takes only $1/4$ of the total computational cost if applied only at the highest timestep, and most of the costs are still dominated by the standard gravity solver and mesh construction in \texttt{AREPO}.
}
\label{fig:test:rsg:cpu}
\end{figure*}

Such performance is in fact very fast considering that we use $80$ angles.
For reference, the explicit M1 radiation transport \texttt{AREPO-RT} with $3$ frenquency bins \citep{kannan2019MNRAS} takes $50\%$ of the computing time in the \texttt{THESAN-1} reionization simulation on CPUs \citep{zier2024MNRAS}.
The relative cost of our scheme is similar, but our reported cost adopts a large number of $80$ angles.
It should be noted though that our scheme will be much less efficient for reionization simulations due to the large number of iterations needed for optically-thin regions.
The general relativistic extension of the \citetalias{jiang2021ApJS} scheme was reported to occupy $80\%-96\%$ of the total cost for 3D linear wave tests with $128$ angles \citep[including communications;][]{white2023ApJ}, which is comparable to our scheme.
Using the \citetalias{jiang2021ApJS} scheme to simulate circumbinary disks, \citet{tiwari2025arXive-prints} reported that the RMHD simulation is an order of magnitude more expensive than the MHD simulation, also consistent with our estimate for fixed-mesh simulation without self-gravity.

A useful comparison to make is between the cost of 3D red supergiant simulations in \texttt{ATHENA++} and that of \texttt{AREPO}, where the physical processes are similar despite different setups.
Using the scheme of \citetalias{jiang2021ApJS} in \texttt{ATHENA++}, \citet{goldberg2022ApJ} reported that each 3D red supergiant simulation of $16$-year duration takes two months to finish on 80 skylake nodes on Pleiades supercomputer, or about $10$ million CPU hours.
In comparison, given the current CPU usage, our simulation of $16$-year duration will take one month to finish on 504 CPU cores (7 nodes) on the NHR@FAU Fritz supercomputer, or about $0.4$ million CPU hours, which is $25$ times cheaper than the \texttt{ATHENA++} simulation.
It should be noted that this speed difference is largely due to different setups.
Even though our simulation uses more expensive gravity and mesh construction modules with more mesh points, \citet{goldberg2022ApJ} used a more stringent convergence criterion for the radiation transport.
\footnote{We use full self-gravity and adaptive Voronoi mesh construction with $2\times 10^7$ total mesh points in \texttt{AREPO}.
This setup is supposed to be much more expensive than the \texttt{ATHENA++} setup in \citet{goldberg2022ApJ}, where they used spherically-averaged gravity without adaptive mesh refinement and $10^7$ total mesh points.
On the other hand, we stop the iterative solver after 10 iterations, which typical yields a relative error of $10^{-4}-10^{-5}$. In comparison, \citet{goldberg2022ApJ} stop the iteration after the relative error reaches $10^{-6}$, which sometimes takes $20-50$ iterations (Y.-F. Jiang, priv. comm.).
}
The communication overhead due to a larger number of CPU cores used in \citet{goldberg2022ApJ} may also play a role in the difference of computational cost.
Therefore, we conclude that our red supergiant simulation with \texttt{AREPO} is at least not slower, if not faster, compared to \citet{goldberg2022ApJ} with \texttt{ATHENA++}.
% % Therefore, we conclude that our red supergiant simulation with \texttt{AREPO} is at least $10-50$ times faster compared to \citet{goldberg2022ApJ} with \texttt{ATHENA++}.
% This is due to (1) the communication overhead using a larger number of CPU cores in \citet{goldberg2022ApJ}, (2) the more stringent convergence criterion used in \citet{goldberg2022ApJ}\footnote{We stop the iterative solver after 10 iterations, which typical yields a relative error of $10^{-4}-10^{-5}$. But \citet{goldberg2022ApJ} stop the iteration after the relative error reaches $10^{-6}$, which sometimes takes $20-50$ iterations (Y.-F. Jiang, priv. comm.).}, and (3) the fast performance of our updated radiation transport scheme.

The fast performance of our scheme is largely because of our optimized iterative solver and communication between tasks.
The iterative solver for the intensities effectively solves for a linear system where the intensities can be grouped into a long vector and the coefficients form a large matrix.
Those coefficients do not change during iterations, and thus only need to be calculated once and then saved to be directly used in later iterations, as motivated by \citet{whitehouse2005MNRAS}.\footnote{For specific problems such as fast transients, star formation, or reionization, the opacity also needs to be updated together with radiation and temperature field, and different coefficients are needed. This is to be explored in the future.}
This approach helps reduce the cost for actual calculations without heavily burdening the memory load.
After optimizing the actual calculations, the cost of the radiation module is dominated by the communication between tasks, where we need to exchange primitive variables as well as fluxes.
In the procedure described in Section~\ref{sec:solver:scheme}, we have optimized our scheme such that there is no redundant variable exchange or flux exchange.
Specifically, (1) we only exchange fluxes and not the radiation primitive variables (intensities) when we pre-calculate the coefficients. (2) During each iteration, we exchange the intensities, and then only exchange neighboring fluxes for $I_\mathrm{c}$ and not the conserved variables $IV$. (3) After the iterations, we exchange the intensities, and then only apply fluxes for active-passive faces to update the conserved variables $IV$ for passive cells with neighboring active cells.

Despite the fast performance, the radiation module will dominate the computational cost if we aim for multigroup radiation (Section~\ref{sec:disc:code:multigroup}), since the cost will be at least multiplied by the number of frequency bins $N_\nu$.
The true cost will be even higher because more iterations may be needed to reach convergence.
Although it is possible to reduce the number of angles as a compromise, it is not advisable to reduce the number of angles below $48$ for complex 3D simulations, especially when there are optically thin regions in the simulation \citep[see e.g. the convergence tests in][]{huang2023ApJ, zhang2024ApJ}.
We therefore expect that radiation module will dominate the computational cost for more than $3$ frequency bins.

A possible solution was given in \citet{zier2024MNRAS}, where they ported the M1 radiation transport \texttt{AREPO-RT} \citep{kannan2019MNRAS} onto GPUs with node-to-node communications and achieved a factor of $10-20$ speedup.
Typically, GPUs are much more efficient than CPUs when dealing with millions of simple and identical operations that can be performed independently in parallel.
This is especially relevant for our scheme, since we need to calculate the intensity per cell per angle per frequency bin, which yields $\mathcal{O}(N_\mathrm{p} N_\mathrm{\Omega} N_\nu)$ almost identical operations.
Furthermore, the cost of our implementation is still dominated by task-to-task communication, which exchanges information between different cores.
With modern computing nodes with over $100$ cores per node, using shared memory within each node and only performing node-to-node communication can significantly reduce the computational cost spent in communication.
Since the subcycling approach in explicit M1 \texttt{AREPO-RT} is in its essence very similar to the iterative solver in our scheme, the data structure and workflow in \citet{zier2024MNRAS} can be mostly adopted for our scheme.
Such acceleration techniques will be explored along with the multigroup extension.

\subsection{Convergence Criterion and Iterative Solver}
\label{sec:disc:code:criterion}

The convergence criterion used in this work
\begin{equation}
    \frac{\sum_{i=0}^{N_\mathrm{p}-1}\sum_{n=0}^{N_\mathrm{\Omega}-1}|I_{n,i}^{m+1,l+1} - I_{n,i}^{m+1,l}|}{\sum_{i=0}^{N_\mathrm{p}-1}\sum_{n=0}^{N_\mathrm{\Omega}-1}|I_{n,i}^{m+1,l+1}|} < \epsilon
\end{equation}
is directly inherited from \citetalias{jiang2021ApJS}.
However, it has two limitations.
First, it only constrains part of the simulation domain where the intensities are large, which typically happens in the high-density optically-thick regions.
Second, it only checks whether the relative error between two sequential iterations $l$ and $l+1$ is small, i.e., whether the iterations have converged to a solution, but it does not check how well the converged solution satisfies the original radiation transport equation~\eqref{eq:discretizedrt}.
Therefore, it is important to further check the residual of the radiation transport equation
\begin{equation}
\begin{split}
    \delta (IV)_{n,i}^{m+1} \equiv & [I_n^{m+1}V^m - (IV)_n^m]_i + \sum_{j=0}^{N_i-1}\left[I_n (c\vectoraas{n}_n-\vectoraas{w})\right]_{ij}^{m+1}\cdot \vectoraas{A}_{ij}^m \Delta t_{ij}^m \\
    & - \left[(S_I)_{n}^{m+1} V^m c\Delta t^m\right]_i\, .
\end{split}
\end{equation}
Specifically, since the radiation source terms are the relevant quantities for 3D simulations in practice, we propose to check the relative residuals of the zeroth (radiation energy) and first moments (radiation flux) of the radiation transport equation
\begin{align}
    & \frac{\sum_{i=0}^{N_\mathrm{p}-1}|\delta (E_\mathrm{r}V)_i^{m+1}|}{\sum_{i=0}^{N_\mathrm{p}-1}|(E_\mathrm{r}V)_{n,i}^m|} < \epsilon_{E_\mathrm{r}V}\, ,\\
    & \frac{\sum_{i=0}^{N_\mathrm{p}-1}|\delta (\vectoraas{F}_\mathrm{r}V)_{n,i}^{m+1}|}{\sum_{i=0}^{N_\mathrm{p}-1}|(\vectoraas{F}_\mathrm{r}V)_i^m|} < \epsilon_{F_\mathrm{r}V}\, ,
\end{align}
where
\begin{align}
    & \delta (E_\mathrm{r}V)_i^{m+1} \equiv \frac{4\pi}{c}\sum_{n=0}^{N_\Omega-1} \delta (IV)_{n,i}^{m+1} w_n \, ,\\
    & \delta (\vectoraas{F}_\mathrm{r}V)_i^{m+1} \equiv 4\pi\sum_{n=0}^{N_\Omega-1} \delta (IV)_{n,i}^{m+1} \vectoraas{n}_n w_n\, .
\end{align}
For the 3D simulation presented in Section~\ref{sec:test:rhd:rsg}, we typically find relative errors at the order of $10^{-3}$ in radiation energy and $10^{-2}$ in radiation flux after $10$ iterations.
This is because we have chosen a robust iterative solver with a relatively slow convergence speed as the default solver and that the radiation field does not change significantly.
As already discussed in the Appendix of \citetalias{jiang2021ApJS}, if we do not separate the positive coefficients from negative ones for the diagonal terms (see Section~\ref{sec:solver:derivation}), the convergence speed will be much faster.
If we adopt this alternative solver that converges faster, the relative errors are at the order of $10^{-7}$ in radiation energy and $10^{-3}$ in radiation flux after $10$ iterations.
However, we do find that this alternative solver is prone to divergence for the 3D simulation in Section~\ref{sec:test:rhd:rsg}.
Therefore, we have chosen to keep the original iterative solver as the default one, but bear in mind that this leads to slow convergence and relatively larger errors in radiation energy and flux.
We have also chosen to keep the original convergence criterion (maximum 10 iterations on top of Equation~\eqref{eq:criterion}) as the default one, but print the relative residuals in the terminal output for debugging purposes.
One future direction is to find an alternative implicit solver that converges faster, which is further discussed in the next subsection.

\subsection{Faster Convergence and Higher-order Accuracy: Alternative Implicit Solvers}

The computational cost of our scheme almost scales with the number of iterations $N_\mathrm{iter}$.
To correctly capture the radiation field in the optically-thin regions, a large $N_\mathrm{iter}$ is typically needed (see e.g. test problems~\ref{sec:test:steady:thin}, \ref{sec:test:steady:nlte}, \ref{sec:test:rd:bubble}).
It is thus desirable to accelerate the convergence of the implicit scheme.
Furthermore, our scheme is only first-order accurate both in time and space, as directly inherited from the \citetalias{jiang2021ApJS} scheme.
As explained in \citetalias{jiang2021ApJS}, this is to avoid oscillatory behaviors when the source terms are stiff and when flux-limiters for higher-order spatial reconstruction are difficult to treat implicitly.
However, the low-order accuracy renders the scheme diffusive (see e.g. test problems~\ref{sec:test:steady:thin} and \ref{sec:test:rd:bubble}).

A potential direction for the future is to look for alternative implicit solvers that have been successfully applied to other astrophysical problems.
For example, implicit methods are also adopted to simulate compressible low-Mach stellar convection, e.g., the \texttt{SEVEN-LEAGUE HYDRO} code \citep{miczek2015A&A, leidi2022A&A} or the \texttt{MUSIC} code \citep{viallet2016A&A, goffrey2017A&A}.
The Crank-Nicolson time-integration used in those codes yields second-order accuracy in time, but it remains to be explored how to incorporate it with the moving-mesh construction in \texttt{AREPO} \citep{pakmor2016MNRAS}.
Second-order spatial reconstruction was also shown to be possible with implicit schemes adopting Newton-Raphson methods with pre-conditioners \citep{viallet2016A&A, leidi2022A&A}, but saving the gradients will pose a serious challenge to memory load.
Furthermore, since the Jacobian matrix for our radiation transport problem is huge, it is attractive to adopt a Jacobian-free method to avoid overloading the memory \citep[e.g.][]{viallet2016A&A}, although the construction of a pre-conditioner and massive parallelization are challenging to realize compared to the iterative solver adopted in this work.

\subsection{Heavy Memory Load for Large Simulations}

Large simulations such as cosmological simulations are usually bound by physical memory \citep{weinberger2020ApJS, kannan2022MNRAS}.
As our first attempt to overcome the memory issue, we avoid saving gradients for radiation quantities by using a first-order spatial reconstruction, and free most of the saved radiation quantities after radiation transport calculations except for the conserved radiation quantity $IV$ and primitive radiation variable $I$.
However, when the radiation module is active, we still temporarily save at the order of $\mathcal{O}(10 N_\mathrm{p} N_\mathrm{\Omega} N_\nu)$ quantities.
A minor improvement is to avoid saving the angle-dependent coefficients (Section~\ref{sec:solver:scheme}), which reduces the temporarily saved quantities to $\mathcal{O}(N_\mathrm{p} N_\mathrm{\Omega} N_\nu)$.
This is especially relevant if we attempt to move the scheme to GPUs, since the coefficients are supposed to be computed fast and we aim to limit the amount of data exchanged between tasks.
Another way to explore is to save the radiation quantities as single precision instead of double precision, as mentioned in \citet{kannan2022MNRAS}.
Nevertheless, more aggressive approaches will be needed if we aim to reduce the number of saved radiation quantities by an order of magnitude.
It is foreseeable that, given the memory limit, a VET method closed either by our time-independent discrete-ordinates method or by ray-tracing may be a more practical choice for large simulations, in particular when the simulation is dominated by optically-thin regimes (e.g. reionization).

\section{Conclusions}
\label{sec:conc}

In this paper, we describe an accurate implicit discrete ordinates radiation transport scheme for general-purpose multi-dimensional RMHD simulations on an unstructured moving mesh.
We extend and optimize the original finite-volume fixed-mesh scheme developed by \citetalias{jiang2021ApJS}, and implement the scheme in the moving-mesh MHD code \texttt{AREPO} \citep{springel2010MNRAS, pakmor2016MNRAS, weinberger2020ApJS}.
Compared to \citetalias{jiang2021ApJS}, our new scheme supports
\begin{enumerate}%[noitemsep]
    \item An unstructured moving mesh;
    \item Local time-stepping;
    \item General equations of state.
\end{enumerate}
% \begin{itemize}
%     \item[$\bullet$] Unstructured moving mesh;
%     \item[$\bullet$] Local time-stepping;
%     \item[$\bullet$] General equation of state;
%     \item[$\bullet$] Fast performance (for a 3D red supergiant simulation in Section~\ref{sec:test:rhd:rsg}, we reach a relative computational cost of RMHD : MHD = 2 : 1 for moving-mesh with full gravity, and an estimated cost of RMHD : MHD = 10 : 1 for fixed-mesh with simple gravity).
% \end{itemize}
Our scheme also achieves fast performance.
For a 3D red supergiant simulation in Section~\ref{sec:test:rhd:rsg}, we reach a relative computational cost of RMHD : MHD < 2 : 1 for moving-mesh with full gravity, and an estimated cost of RMHD : MHD = 10 : 1 for fixed-mesh with simple gravity.
We perform an extensive list of tests to demonstrate these new features and the accuracy of our scheme in optically-thick, optically-thin, or the transition region in between (Section~\ref{sec:test}).
Most of those tests were also performed in \citetalias{jiang2021ApJS} and some in \citet{davis2012ApJS} and \citet{jiang2012ApJS, jiang2014ApJS}, and the results show clear consistencies, which demonstrates that our scheme performs well even on an unstructured moving mesh with local time-stepping.
% In particular, we show that local time-stepping in our scheme does not jeopardize significantly the energy, momentum, or angular momentum conservation even for a complex 3D simulation (Section~\ref{sec:test:rhd:rsg}).

We collect our thoughts on future code developments in Section~\ref{sec:disc:code}.
The immediate follow-ups are to extend our scheme to multigroup based on \citet{jiang2022ApJS} and further speedup with GPUs and node-to-node communication based on \citet{zier2024MNRAS}.
We expect both of these extensions will be relatively straightforward, since both our scheme and the \citet{jiang2022ApJS} multigroup scheme are direct decendants of \citetalias{jiang2021ApJS}, and the \citet{zier2024MNRAS} speedup was developed for another radiation transport module in \texttt{AREPO} \citep{kannan2019MNRAS}.
Further uncharted territories include improving the convergence, order of accuracy, and memory load, for which we discuss alternative implicit solvers and trimming the unnecessary storage.
% We also discuss the potential to modify our radiation transport scheme for SPH simulations.

By combining the unique features of \texttt{AREPO} with an accurate radiation transport, we hope that our scheme will open up possibilities to perform moving-mesh RMHD simulations for a vast variety of astrophysical problems where radiation plays an important role.
The flexible mesh sizes and local time-stepping provide the rare ability to explore a large dynamic range in both space and time.
This is particularly relevant for star formation \citep{grudic2021MNRAS, chon2024MNRAS, mayer2024arXive-prints}, tidal disruption events \citep{ryu2023ApJ, price2024ApJ, steinberg2024Nature}, accretion around supermassive black holes \citep{hopkins2024OJAp, guo2023ApJ, jiang2024arXive-prints}, and galaxy formation and reionization in the early Universe \citep{vogelsberger2020MNRAS, kannan2022MNRAS, kannan2023MNRAS}, where radiative feedback and/or radiation force are dynamically important.
In addition, the moving mesh offers an excellent tool to simulate systems with bulk supersonic motions.
This typically happens in gravitationally-bound multi-body systems, such as binary star interactions \citep{schneider2019Nature, ropke2023LRCA}, astrophysical disks and their interactions with planets, stars, black holes \citep{benitez-llambay2016ApJS, munoz2019ApJ, zhu2021MNRAS, chen2023ApJ, yao2024arXive-prints}, where the interactions can take the form of radiative heating/cooling or radiation pressure.
Supplemented by recent \texttt{AREPO} implementations of non-ideal MHD \citep{zier2024MNRASb, zier2024MNRASa} and unique adaptive mesh refinement in shearing box \citep{zier2022MNRAS}, we expect that many difficult astrophysical problems can now be tackled using moving-mesh RMHD simulations with \texttt{AREPO}.
As the `first light' of this prospective endeavor, we are exploring the trans-sonic convection and pulsation in global 3D simulations of red supergiant stars with our new scheme (see Section~\ref{sec:test:rhd:rsg} for a preview), which will be detailed in a forthcoming paper (Ma et al. in prep.).

\begin{acknowledgements}
      % We thank the referee for the dedicated comments that helped improve this work.
We thank Yan-Fei Jiang for sharing some test problems and helpful discussions.
We thank Sebastian Ohlmann for sharing the module to construct stable 3D AREPO giant stars.
We thank Andrea Chiavassa for insights on red supergiant simulations.
We thank Volker Springel for helpful discussions and giving us access to AREPO and to the computational resources.
We thank Paul Ricker, Rahul Kannan, Aaron Smith, Oliver Zier, Thomas Guillet, Daniel Price, Xue-Ning Bai, Robert Andrassy, and Giovanni Leidi for discussions on imlicit methods and radiation transport with local timesteps.
We thank Mike Lau, Jared Goldberg, Taeho Ryu, Sunmyon Chon, Aniket Bhagwat for discussions on 3D radiation hydrodynamic simulations in different fields of astrophysics.
This research project was partly conducted using computational resources (and/or scientific computing services) at the Max-Planck Computing \& Data Facility. 
The authors gratefully acknowledge the scientific support and HPC resources provided by the Erlangen National High Performance Computing Center (NHR@FAU) of the Friedrich-Alexander-Universität Erlangen-Nürnberg (FAU) under the NHR project b234dd.
NHR funding is provided by federal and Bavarian state authorities.
NHR@FAU hardware is partially funded by the German Research Foundation (DFG) – 440719683.
The idea for this work was conceived when the author was attending the Interacting Supernovae workshop at the Munich Institute for Astro-, Particle and BioPhysics (MIAPbP) which is funded by the Deutsche Forschungsgemeinschaft (DFG, German Research Foundation) under Germany´s Excellence Strategy – EXC-2094 – 390783311.
This work was initiated at the 2023 MPA-Kavli Summer Program in Astrophysics supported by the Max Planck Institute for Astrophysics (MPA) and the Kavli Foundation.
Part of this work was done when the author was attending the TDE24 program at Kavli Institute for Theoretical Physics (KITP), which is supported in part by grant NSF PHY-2309135.
\end{acknowledgements}

% \vspace{5mm}

\textit{Software:}
\texttt{AREPO} \citep{springel2010MNRAS, pakmor2016MNRAS, weinberger2020ApJS}, 
\texttt{MESA} \citep{paxton2011ApJS, paxton2013ApJS, paxton2015ApJS, paxton2018ApJS, paxton2019ApJS, jermyn2023ApJS},
\texttt{Astropy} \citep{astropycollaboration2013A&A, astropycollaboration2018AJ, astropycollaboration2022ApJ}, 
\texttt{NumPy} \citep{harris2020Nature}, 
\texttt{SciPy} \citep{virtanen2020NatureMethods}, 
\texttt{Matplotlib} \citep{hunter2007Comput.Sci.Eng.}, 
\texttt{Jupyter} \citep{kluyver2016PositioningandPowerinAcademicPublishing:PlayersAgentsandAgendas}

\textit{Data availability:}
All the test problems and \texttt{MESA} inlists are available at Zenodo: doi: \href{https://doi.org/10.5281/zenodo.15032138}{10.5281/zenodo.15032138}. The radiation module is merged into the master branch or the IDORT branch of \texttt{AREPO}. To use the code, please see the description in Zenodo and contact the first author.

\bibliographystyle{aa}
\bibliography{RSG}

\end{nolinenumbers}

% \end{CJK*}
\end{document}